\documentclass{elsarticle}
\usepackage{lineno,hyperref}
\usepackage{amsmath}
\usepackage{amssymb}
\usepackage{bm}
\usepackage{graphicx}
\usepackage{color}
\usepackage{float}
\usepackage{caption}
\usepackage{subcaption}
\usepackage{subeqnarray}
\usepackage{geometry}
\usepackage[normalem]{ulem}

\journal{Journal of Computational Physics}

\begin{document}

\begin{frontmatter}

\title{A gradient-based framework for maximizing mixing in binary fluids}

\author{M.F. Eggl\fnref{myfootnote}}
\author{P.J. Schmid}
\address{Department of Mathematics, Imperial College London, London SW7 2AZ, United Kingdom}
\fntext[myfootnote]{Corresponding author: maximilian.eggl11@imperial.ac.uk}

\begin{abstract}
  A computational framework based on nonlinear direct-adjoint looping
  is presented for optimizing mixing strategies for binary fluid
  systems. The governing equations are the nonlinear Navier-Stokes
  equations, augmented by an evolution equation for a passive scalar,
  which are solved by a spectral Fourier-based method. The stirrers
  are embedded in the computational domain by a Brinkman-penalization
  technique, and shape and path gradients for the stirrers are
  computed from the adjoint solution. Four cases of increasing
  complexity are considered, which demonstrate the efficiency and
  effectiveness of the computational approach and
  algorithm. Significant improvements in mixing efficiency, within the
  externally imposed bounds, are achieved in all cases.
\end{abstract}


\end{frontmatter}


\section{\label{sec:intro}Introduction}

The mixing of binary fluid is a problem of fundamental concern in
fluids dynamics, as its mechanisms play an important role in a wide
variety of industrial fields and in many fluid processes encountered
in daily life. The spreading and mixing of pollutants in the
ocean~\cite{Lekien2005}, the ventilation of a
building~\cite{Hunt1999,Linden1999}, the mixing of air and fuel for
subsequent combustion~\cite{Annaswamy1995} or the mixing in
microfluidic devices~\cite{Hessel2005, Nguyen2005} are but a few
examples where mixing processes play an important role. The food
processing, pharmaceutical and consumer-product industry are further
sectors where an improvement of mixing efficiency would translate into
immediate profits as well as a more consistent quality of the end
product.

Mixing in industrial applications is often accomplished by stirrers,
i.e., moving bodies of a given shape embedded in a mixing vessel whose
task it is to produce long filaments~\cite{Aref1984} which are
subsequently diffused. This two-step process, which is principally at
the core of binary fluid mixing by stirrers, has been recognized as a
fundamental mechanism and has been studied extensively to gain insight
and to guide control strategies. We will consider the case of
stirrer-induced mixing and its optimization in this article.

From a computational point of view, one of the difficulties of
stirrer-based mixing stems from the treatment of the fluid-structure
interaction between the embedded, moving stirrers and the binary
fluid. Various options exist and have been pursued by previous
studies, such as body-fitted meshes~\cite{Mattiussi2000} or the method
of fictitious domains~\cite{Glowinski1996}. These techniques are,
however, often restricted to simpler configurations or are excessively
costly due to the need to remesh at each time step (after the solid
object has been advanced). An attractive alternative to the above
methods is the penalization method which will be adopted in this
article. Introduced in reference~\cite{arquis1984}, solutions obtained
by the penalization method have been rigorously shown to converge to
the corresponding solution of the Navier Stokes
equation~\cite{Angot1999} for the respective complex
domain. Supporting studies, including an asymptotic analysis, are
summarized in~\cite{Liu2007}, and applications to high Mach-number
flows~\cite{Boiron2009} and turbulent flow past
cylinders~\cite{Kevlahan2001}, among many other examples, have
demonstrated the effectiveness and flexibility of the approach. A
penalization approach has also been taken by~\cite{Chantalat2009},
coupled with a level-set technique to express the geometry, and
applied by~\cite{Bruneau2013} to optimize the shape of actuators. The
appeal of the penalization method lies in its simple derivation and
straightforward numerical implementation.

A further complication of the mixing problem is the nonlinear nature
of its underlying governing equations. Optimizing stirring strategies
using a gradient-based approach, as will be pursued in this article,
will have to deal with the solution of nonlinear equations and the
checkpointing problem for the dual/adjoint problem (see details
below). The complexity of the flow does not furnish equilibrium points
about which to linearize, not even over a limited horizon of
validity. Instead, the full nonlinear problem has to be tackled, and
nonlinear adjoint looping techniques~\cite{Juniper} have to be
employed for the stirrer geometry and/or stirrer path.

Rather than investigating merely the effects and mechanisms of mixing,
several previous studies have attempted the control and optimization
of mixing processes. These include, among others, optimal control of
mixing via entropy maximization of a flow governed by two orthogonal
shear flows ~\cite{DAlessandro1999}, and optimal mixing of a steady
Stokes flow by optimizing the mix-norm~\cite{Mathew2007, Lin2011}. A
concise review of measuring mixing as well as mixing optimization is
given in~\cite{Ottino1990} as well as in~\cite{Thiffeault2012}.

We wish to blend the advantages of a penalization method for
describing the motion of stirrers through the binary fluid with the
direct-adjoint methodology for gradient-based optimization. The
challenges of this approach lie in the extraction of path-derivatives
and shape-derivatives from the forward-backward integration of the
direct and dual problem (the Karush-Kuhn-Tucker system). The full
computational framework inherits the flexibility, efficiency and
accuracy of the fluid-structure treatment by penalization and the
effective convergence of the PDE-constrained optimization method to
reach a minimum in mixing variance (or any other mixing norm), and
thus a better mixed state of the binary system.

The test cases and their underlying geometry, namely a circular dish
with one or more embedded rotating stirrers, are taken with a view
towards industrial configurations, where cylindrical mixing containers
with rotating stirrers are commonplace~\cite{Handbook6}. Within this
setup the algorithm can prove its mettle and provide significant
improvements in mixing efficiency.

The rest of this article is organized in the following manner. In
section \S~\ref{sec:govequ} we present the governing equations of the
system we wish to analyze as well as an in-depth review of the
penalization method incorporated into these equations. In section
\S~\ref{sec:NumMethod} we then turn our attention to the numerical
algorithm and implementation, covering the discretization of the
system and introducing a formulation that is particular amenable to
deriving the dual/adjoint system. We then present, in section
\S~\ref{sec:optim}, the methodology and algorithm used to achieve
optimality. This includes the introduction of an augmented Lagrangian
of the system and the derivation of the optimality system
(KKT-system), with special emphasis on the penalization notion
established earlier. Furthermore, we also present a step-by-step
summary of the optimal-mixing algorithm. Lastly, in section
\S~\ref{sec:validation}, we demonstrate the efficacy of the algorithm
by presenting test cases of progressing complexity. These test cases
have been chosen to probe and assess the convergence and optimization
behavior of the algorithm, and results range from simple (and
anticipated) to more difficult (and less intuitive). They are meant to
gain experience with the optimization strategy, which in turn will
guide future efforts and more complex setups. Conclusions are offered
in section~\ref{sec:concl}. The appendices will provide details on
various derivations in the text.

\section{\label{sec:govequ}Governing equations and general assumptions}

\subsection{Governing equations}

We will consider mixing processes of a binary, miscible fluid in a
parameter regime where inertial effect cannot be ignored, yet
turbulent fluid motion has not developed. In addition, we assume for
simplicity that both liquids behave as Newtonian fluids. This
parameter regime is commonly observed in a wide range of industrial
applications and shall serve here as a basis for establishing a
mathematical and computational framework for the optimization of
mixing efficiency.

The underlying equations governing the motion of the fluid is then
given by the incompressible Navier-Stokes equations which read in
primitive form
\begin{subeqnarray}
 \partial_t \bm{u} + {\bm{u}} \cdot \nabla \bm{u} +
 \nabla p- Re^{-1}\nabla^2\bm{u} &=& 0, \\
 \nabla \cdot \bm{u} &=& 0
  \label{eq:NavierStokes}
\end{subeqnarray}
with $\bm{u}$ as the velocity vector and $p$ as the pressure
field. The equations have been stated in non-dimensional form, where
appropriate characteristic length ($L_0$) and velocity ($u_0$) scales
have been chosen to render the equations dimensionless. This
introduces the Reynolds number $Re$ based on these characteristic
scales, i.e., $Re = \displaystyle{\frac{u_0 L_0}{\nu}}$ with $\nu$ as
the kinematic viscosity.

The above equations have to be augmented by a scalar field, denoted by
$\theta,$ to describe the mixing process. This is accomplished by
considering

\begin{equation}
  \partial_t \theta + \bm{u}\cdot \nabla \theta -
  Pe^{-1}\nabla^2\theta = 0
  \label{eq:PassiveScalar}
\end{equation}
as the evolution equation that transports $\theta$, where $\theta$ is
defined between 0 and 1. This advection-diffusion equation for the
passive scalar is parameterized by the P\'{e}clet number $Pe,$ defined
as $Pe=\displaystyle{\frac{u_0 L_0}{\kappa}}$, with $u_0$ and $L_0$
again as the characteristic velocity and length scales, respectively,
and $\kappa$ denoting the diffusion coefficient of the mixing fluid.

As mentioned above, we target the parameter regime beyond Stokes
mixing, where inertial effects become relevant, but below the onset of
turbulence, where mixing by turbulent fluctuations dominates. To this
end, we consider the parameter setting $Re=Pe=1000,$ which represents
inertial, yet laminar flow. In this regime, we retain two important
physical mixing processes: advection and diffusion. The complex
interplay between these two mechanisms shall be the focus of our
analysis and the target of our optimization algorithm; advection in a
generated shear field will be exploited to produce small-scale
structures, which are subsequently subjected to diffusion and
mixing~\cite{Lin2011}.

\subsection{Measuring mixedness}

In anticipation of our stated goal of enhancing mixing efficiency, we
have to introduce a measure that quantifies the degree of mixedness of
a particular configuration. This measure shall be based solely on the
passive scalar field $\theta.$

In general, mixing is defined as the reduction of inhomogeneities of a
given indicator field~\cite{Handbook0}, which still leaves open a
precise mathematical definition to be used in our framework. Several
norms of the passive scalar $\theta$ that attempt to mathematically
define the measure of mixedness have been proposed and used in
the past~\cite{Mathew2005}, among them the variance or the more complex
negative-index and fractional-index Sobolev norms~\cite{Foures2013}.

As the choice of norm may influence the outcome of the optimization,
but will not affect the design of our computational optimization
platform, we will focus, for simplicity, on the variance norm of the
passive scalar $\theta.$ This rather intuitive measure attains higher
values for an unmixed field (with high levels of inhomogeneities)
and decreases as the scalar field $\theta$ becomes more
mixed. Mathematically, the variance is defined as

\begin{equation}
  \text{Var } \theta = \frac{1}{\vert \Omega
    \vert}\int_{\Omega} \theta(\bm{x},t)^2 \text{ d}\Omega,
    \label{eq:Variance}
\end{equation}
where $\Omega$ is our computational domain, and $\vert \Omega \vert$
denotes the size (volume or area) of our domain. In the above
definition, we have assumed, without loss of generality, a zero mean
of the passive scalar field $\theta.$ Throughout this paper we will be
optimizing with respect to this quantity, but we stress again that
other norms can be employed without conceptual changes in the
optimization procedures.

\subsection{Complex geometry via penalization}

Stirrers will be used to achieve fluid mixing, categorizing our
problem as a fluid-structure interaction problem. We will use the the
penalization method~\cite{arquis1984} to approach this problem. Its
appeal lies in the simple modification of the governing Navier-Stokes
equations by adding external forcing terms. These terms model our
solid bodies as Brinkman-style porous media with vanishing
permeability $C_\eta.$ This method has been shown to converge to the
exact solid-fluid solution as $C_\eta$ tends to zero~\cite{Angot1999};
furthermore, it is able to enforce Dirichlet as well as Neumann
boundary conditions on the respective flow variables. The advantage of
this method lies in its rather simple implementation, its flexibility
in imposing complex boundary conditions, and its numerical
efficiency. Moreover, moving solids are straightforwardly treated by
remapping masks to a new position, without any need for remeshing or
sophisticated grid operations. We will provide a brief overview here;
for alternative applications of this method, or a more in-depth
coverage of this method, the reader is referred
to~\cite{Schneider2005}.

In preparation for the fact that we will deal with multiple embedded
solids with independent characteristics, we will introduce $\chi_i$,
a mask function of the $i$-th solid by defining
\begin{equation}
  \chi_i({\bm{x}},t) =
  \begin{cases}
    1, \hspace{12pt} \hbox{if } {\bm{x}} \in \Omega_{s,i}\\
    0, \hspace{12pt} \hbox{if } {\bm{x}} \in \Omega_f \\
    0, \hspace{12pt} \hbox{if } {\bm{x}} \in \Omega_{s,j\neq i}
  \end{cases}
\end{equation}
with $\Omega_{s,i}$ denoting the $i$-th solid domain, while $\Omega_f$
stands for the fluid domain. The global mask $\chi$ for our
computational domain is then given as

\begin{equation}
  \chi({\bm{x}},t) = \sum_i \chi_i({\bm{x}},t).
\end{equation}
The mask $\chi$ acts as an indicator function which distinguished
between the solid part ($\chi=1$) and the fluid part ($\chi=0$) of the
computational domain. This indicator function then allows us to
supplement the Navier-Stokes equations~(\ref{eq:NavierStokes}) by
external driving terms that impose a given velocity of the $i$-th
solid, denoted by ${\bm{u}}_{s,i},$ on the fluid and thus model the
motion of individual bodies through the fluid.

Assuming Einstein summation over identical indices, we can then state
the penalized Navier-Stokes equations as

\begin{subeqnarray}
 \partial_t \bm{u} + \bm{u}\cdot \nabla \bm{u} +
 \frac{\chi}{C_{\eta}}\bm{u} -\frac{\chi_i}{C_{\eta}}\bm{u}_{s,i} +
 \nabla p- Re^{-1}\nabla^2\bm{u} &=& 0, \\ \nabla \cdot \bm{u} &=& 0,
  \label{govEqEnd}
\end{subeqnarray}
where $C_{\eta}$ is the permeability of the solids (which, for
simplicity, we take identical for all solids). It has been
shown~\cite{Engels2015} that an optimal value of $C_{\eta}$ is
proportional to $(\Delta x)^2$, and this leads to accurate numerical
results. Recalling the definition of $\chi_i,$ we note that the
equations above reduce to the Navier-Stokes
equations~(\ref{eq:NavierStokes}) in the fluid domain ($\chi=0$).

Proceeding to the governing equations for the passive scalar $\theta,$
we also have to apply penalization terms to enforce no-flux boundary
conditions at the various solids. The scalar field
equation~(\ref{eq:PassiveScalar}) then becomes~\cite{Kadoch2012}
\begin{eqnarray}
  \partial_t \theta + \left( 1-\chi \right)\bm{u}\cdot\nabla\theta +
  \chi_i \left( \bm{u}_{s,i}\cdot\nabla\theta \right) - \nabla\cdot
  \left( \left[ Pe^{-1} \left( 1-\chi \right) + \frac{\chi}{C_{\eta}}
    \right] \nabla\theta \right) = 0.
\end{eqnarray}
The terms $\left( 1-\chi \right) \bm{u}\cdot\nabla\theta$ and
$\nabla\cdot \left( \left[ Pe^{-1} \left( 1-\chi \right) +
  \chi/C_{\eta} \right] \nabla\theta \right)$ prevent the passive
scalar field $\theta$ advecting or diffusing, respectively, into any
of the solids; the term $\chi_i \left( \bm{u}_{s,i}\cdot\nabla\theta
\right)$ transports the scalar field with the velocity of the $i$-th
solid.

\section{\label{sec:NumMethod}Numerical method}

The starting point for the discretization of the penalized governing
equations is the open-source software {\tt{FluSI}}~\cite{Engels2015},
a Fourier pseudo-spectral code for fluid-structure interactions. It
solves the three-dimensional, incompressible Navier-Stokes equations on
equispaced grids using a spectral formulation, adaptive time-stepping
and a pressure-projection approach. The inclusion of solid bodies, or
complex computational domains, is treated by a Brinkmann-type
penalization method; sponge layers are utilized to handle open and
outflow boundaries.

Below we give a brief outline of the key features of the numerical
methodology and introduce special details that had to be added to suit
our goal of mixing enhancement. In particular, we put in place a
spatially discrete formulation which will build the basis for an
efficient derivation and implementation of an adjoint solver.

\subsection{Discretization of the governing equations}

Following the discretization strategy of the original {\tt{FluSI}} code, we
replace the continuous spatial derivatives with multiplications of the
discretized velocity, pressure and passive scalar fields by a Fourier
discretization matrix. We introduce the discrete analog of the
continuous derivative according to
\begin{equation}
  \frac{\partial}{\partial x_i} \quad \to \quad {\mathsf{A}}_i.
\end{equation}
with ${\mathsf{A}}_i$ as an $n\times n$ matrix where $n$ is the number
of grid points in a single dimension. The associated discrete gradient
operator, consisting of ${\mathsf{A}}_i$ for the two coordinate
directions, is represented by ${\mathbf{{\bm{A}}}}.$ Furthermore,
discretization on an equispaced two-dimensional mesh results in the
following variables: ${\bm{u}}$ represents a $3 \times n$ vector
containing the three velocity components, and $\chi, \theta$ and $p$
are $1 \times n$ vectors representing the mask, passive scalar and
pressure field, respectively. In addition, we introduce as
${\bm{u}}_{s,i}$ the $3 \times n$ velocity vector of the $i$-th solid,
as well as the corresponding mask for the same solid, denoted by
$\chi_i$. We note that $({\bm{u}}_{s,i})_j$ is a $1 \times n$ vector
containing the velocity component in the $j$-th coordinate direction
of the $i$-th solid.

With these notations and assuming the Einstein summation convention,
we can then state the spatially discretized governing equations in the
form

\begin{subeqnarray}
  \partial_t \bm{u}+ \bm{u}_j \circ \left[ {\mathsf{A}}_j\bm{u}
    \right]+ \frac{\chi}{C_{\eta}} \circ \bm{u} -
  \frac{\chi_i}{C_{\eta}} \circ \bm{u}_{s,i} +
  \bm{\mathsf{A}} p - Re^{-1} {\mathsf{A}}_i {\mathsf{A}}_i \bm{u} &=&
  0, \label{FLUSI:Eq1}\\
  {\mathsf{A}}_i \bm{u}_i &=& 0, \\
  \partial_t \theta - {\mathsf{A}}_i \left( \left[ Pe^{-1} \left(
    \bm{1}-\chi \right) + \kappa\chi \right] \circ {\mathsf{A}}_i
  \theta \right) + \left( \bm{1}-\chi \right) \circ \bm{u}_j \circ
  \left[ {\mathsf{A}}_j \theta \right] & &\nonumber \\
  + \chi_i \circ \left( \bm{u}_{s,i} \right)_j \circ \left[
    {\mathsf{A}}_j \theta \right] &=& 0 \label{FLUSI:Eq2}
\end{subeqnarray}
where we used the Hadamard (element-wise) product
$\circ$ (see~\cite{Horn2012}). We invoke an operator-splitting
approach and enforce the continuity equation ${\mathsf{A}}_i \bm{u}_i
= 0$ via a pressure Poisson equation which reads
\begin{equation}
  {\mathsf{A}}_j {\mathsf{A}}_j p + {\mathsf{A}}_i \left( \bm{u}_j
  \circ \left[ {\mathsf{A}}_j \bm{u}_i \right] \right) +
        {\mathsf{A}}_i \left[ \frac{\chi}{C_{\eta}} \circ
          \bm{u}- \frac{\chi_i}{C_{\eta}} \circ
          \bm{u}_{s,i} \right] = 0.\label{FLUSI:Eq3}
\end{equation}
We will consider the masks $\chi_i$ (representing the shape of the
solid objects) and the velocities ${\bm{u}}_{s,i}$ (representing their
speed) as the control variables of the governing set of equations
which need to be adjusted to influence and optimize mixing
efficiency. The velocities ${\bm{u}}_{s,i}$ can further be specified
in the form
\begin{eqnarray}
  \bm{u}_{s,i} &=\bm{u}_{C_i}(t) +\omega_i(t)
  \bm{r}_i(\bm{x}), \label{FLUSI:US}
\end{eqnarray}
where ${\bm{u}}_{C_i}(t)$ is the time-dependent velocity of the center
of the solid, $\omega_i(t)$ is the rotational speed of the solid about
its center, and $\bm{r}_i$ denotes the distance from the same center.

The above spectrally discretized governing equations are nonlinear,
and special care has to be exercised to avoid numerical instabilities
due to aliasing errors stemming from the quadratic nonlinearities. In
the original approach ({\tt{FluSI}}), aliasing errors have been
eliminated by using the $\frac{2}{3}$-rule: higher-resolution spectral
transforms in combination with zero-padding and downsampling are used
to circumvent contamination of the lower wavenumbers by mapped higher
ones. In our implementation, a spectral cut-off filter (suggested
in~\cite{Hou2007}) has been used instead, as it has been found more
efficient for our problem.

\subsection{Numerical implementation of masks}

The representation of solid bodies on an underlying Cartesian grid
calls for a transfer of geometric information onto the background
mesh. This transfer is accomplished by a mollified delta-function,
smoothing the otherwise discontinuous mask onto the grid and thus
avoiding numerical inaccuracies and
instabilities~\cite{Kolomenskiy2009}. We choose the widely used,
piece-wise defined function
\begin{equation}
  \chi_i({\bf{x}},t) =
  \begin{cases}
    1, & \hbox{for } \quad |f|<r_i \\
    \displaystyle{\frac{1}{2}} \left(1+\cos \left( \displaystyle{\frac{\pi
        (f-r_i)}{2h}}\right)\right) & \hbox{for } \quad r_i <|f|<r_i
    + 2h \\
    0, & \hbox{otherwise}
  \end{cases}
  \label{Mask:ChiEq}
\end{equation}
where $f$ is some parametric form for the solids we wish to study. We
note that $h$ is proportional to the grid size, $\Delta x$, and as
$\Delta x \to 0$ the function $\chi_i$ tends to a Heaviside
function~\cite{Engels2015}.

\section{\label{sec:optim} Optimization using adjoint methodology}

The effectiveness of mixing in binary fluids can be quantified by a
variety of measures. In this article, we concentrate on the variance
of the passive scalar defined in section \emph{2.2}, while being fully
aware that fractional Sobolev norms of the same quantity, as used,
e.g., in~\cite{Foures2014}, are mathematically more suited for mixing
problems. The choice of norm, however, does not markedly alter the
computational framework introduced below. The choice of the variance
is thus for convenience and for the sake of a less cluttered
notation. With this choice, we determine the quantity we seek to
minimize as

\begin{eqnarray}
  \mathcal{J} = \frac{1}{V_\Omega}\int_{\Omega} \theta^2
  \ \text{d}\Omega\bigg |_{T^{F}}
\end{eqnarray}
which constitutes the cost functional ${\mathcal{J}}$ for our
optimization. In the above expression, $T^{F}$ denotes
the final time. We note that this particular choice of ${\mathcal{J}}$
would lead to unconstrained optimization, as there is no bound on the
energy we are able to inject into the system. Therefore, we modify our
cost functional with the addition of a term that constrains the energy
we supply for the optimal strategy. This enhanced cost functional then
takes the form
\begin{eqnarray}
  \mathcal{J} = \frac{1}{V_\Omega}\int_{\Omega} \theta^2
  \ \text{d}\Omega\bigg |_{T^{F}} + \lambda
  \int_0^{T^{F}} \sum_i [(\bm{u}_{s,i})_j
    \chi_i]^{H} R_i [(\bm{u}_{s,i})_j \chi_i] \ \text{d}t
\end{eqnarray}
where we choose $R_i$ such that the energy penalization constitutes a
valid norm. In line with the semi-discretized formulation above, we
express the discretized cost functional in the form
\begin{eqnarray}
  \mathcal{J} = \frac{\theta^{H} {\mathsf{M}}
    \theta}{V_\Omega}\bigg|_{T^{F}}+ \lambda \int_0^{T^{F}} \sum_i
          [(\bm{u}_{s,i})_j \chi_i]^{H} \mathsf{R}_i [(\bm{u}_{s,i})_j
            \chi_i] \ \text{d}t
\end{eqnarray}
where ${\mathsf{M}}$ is a symmetric, positive definite weight matrix,
taking into account the grid resolution and (possible) spatial
weightings, $\mathsf{R}_i$ denotes a positive definite diagonal matrix
and the $^H$ refers to the conjugate transpose of the relevant
matrix/vector. We note that for real quantities this simply reduces to
the transpose. The user-specified parameter $\lambda$ controls the
weight of the penalization term. With no spatial preference and our
uniform spatial mesh of $n^2$ grid points, the weight matrix
${\mathsf{M}}$ is simply a diagonal matrix of the form
\begin{equation}
  {\mathsf{M}} = \left(\frac{1}{n}\right)^2 {\mathsf{I}}
\end{equation}
with ${\mathsf{I}}$ as the identity matrix.

At this point, we will briefly elaborate on the control parameters we
use to minimize the variance. The framework is sufficiently flexible
to manipulate various internal or external parameters, but we will
concentrate -- in view of possible industrial applications -- on
time-independent variables, specifically the shape $\chi_i$ of
(multiple) stirrers and their rotational speed $\omega_i.$ These
quantities can be altered between iterations using the optimality
conditions, but during the forward solution they remain constant in
time.

A final observation is the fact that our energy penalization is not a
proactive measure, but is instead reactive. When a new optimized
solution is applied, the energy penalization refers to the energy of
the unoptimized system. Therefore, a situation could arise where the
optimization step ventures beyond physical constraints and introduces
singularities that prevent the energy penalization from properly
taking effect. For this reason, it is imperative to achieve a proper
balance between imposing too stringent an energy penalization and thus
stifling the system, and too weak a penalization, yielding potentially
unphysical scenarios.

\subsection{Introducing Lagrange multipliers or adjoint variables}

We note that the cost functional $\mathcal{J}$ is a function of
$\theta$, which is implicitly influenced by $\bm{u}$, $\bm{u}_s$
and $\chi$. To be able to optimize our control variables, we need to
embed these in an augmented cost functional $\mathcal{L}$ which
explicitly expresses these various dependencies. This is achieved by
including the governing equations as well as $\mathcal{J}$ in a single
functional, given by

\begin{eqnarray}
  \mathcal{L} = \mathcal{J} -& \int_0^{T^{F}} & (\bm{u}^{\dag})_k^{{H}}
          {\mathsf{M}} \biggl\{\partial_t \bm{u}+ \bm{u}_j \circ
          [{\mathsf{A}}_j\bm{u} ]+ \frac{\chi}{C_{\eta}}\circ\bm{u} -
          \frac{\chi_i}{C_{\eta}}\circ\bm{u}_{s,i} + {\mathbf{\bm{A}}} p -
          Re^{-1}[{\mathsf{A}}_i{\mathsf{A}}_i\bm{u}] \biggr\}_k
            \nonumber \\
            &+&
            p^{\dag,{H}}{\mathsf{M}}\biggl\{[{\mathsf{A}}_i{\mathsf{A}}_i]p
            + {\mathsf{A}}_i(\bm{u}_j \circ[ {\mathsf{A}}_j\bm{u}_i])
            + {\mathsf{A}}_i\left[ \frac{\chi}{C_{\eta}}\circ\bm{u}
              -\frac{\chi_i}{C_{\eta}}\circ\bm{u}_{s,i}\right]  \biggr\} \nonumber\\
            & + & \theta^{\dag,{H}}{\mathsf{M}}\biggl\{\partial_t \theta
            + (\bm{1}-\chi)\circ \bm{u}_j \circ[
              {\mathsf{A}}_j\theta]- \chi_i \circ (\bm{u}_{s,i})_j
            \circ [{\mathsf{A}}_j\theta]\nonumber\\
          &-& {\mathsf{A}}_i([Pe^{ -1}(\bm{1}-\chi) + \kappa\chi]
          \circ {\mathsf{A}}_i\theta)\biggr\} +
          \chi_i^{\dag,{H}}{\mathsf{M}}[\chi_i -
            g_i(\bm{x},t)]\ \text{d} t \nonumber \\
          & - & \omega_i^{\dag,{H}}{\mathsf{M}}[\omega_i - z_i].
          \label{eq:augL}
\end{eqnarray}

Here, we have introduced Lagrange multipliers or adjoint variables,
denoted by the superscript ${}^\dag$. The adjoint variables enforce
the constraints given by the governing equations, i.e.,
$\bm{u}^{\dag}$ enforces the time evolution of $\bm{u}$, $p^{\dag}$
the pressure Poisson equation, and $\theta^{\dag}$ is associated with
the equation governing the passive scalar field. Lastly, $\chi^{\dag}$
and $\omega^{\dag}$ ensure that the conditions which define the
embedded solids are satisfied. We use this semi-discretized framework,
rather than a continuous approach that requires the explicit
enforcement of boundary conditions, for reasons of greater simplicity,
flexibility and functionality when it comes to optimizing mixing by
manipulating $\chi$ and $\bm{u}_s$. The entire information related to
the solids is encapsulated in the penalization terms, and is thus
easily captured by associated Dirichlet, Neumann or Robin boundary
conditions -- without the need to include supplementary terms in the
augmented Lagrangian $\mathcal{L}$. Moreover, in the ensuing
derivation of the adjoint equations, the appropriate boundary
conditions on the solids will be fully encoded in the discretized
penalization terms, making the resulting numerical implementation
substantially simpler and less error-prone.

The key to deriving the optimality conditions is to minimize the
augmented cost functional $\mathcal{L} $ by taking first variations of
$\mathcal{L}$, i.e., enforcing $\delta \mathcal{L} = 0$. Since $\delta
\mathcal{L}$ depends on several independent variables, including the
adjoint variables, we require that each individual variation is in
effect zero. Proceeding along this line, we note that the first
variation with respect to the adjoint variables recovers the original
governing equations. We then continue by focusing on the variation
with respect to the direct variables which will ultimately produce a
governing equation for the adjoint variables. For the sake of clarity,
some explicit parts of this calculation have been relegated to the
appendix. The full adjoint equations defining $u^{\dag}$, $p^{\dag}$
and $\theta^{\dag}$ are as follows:

\begin{subeqnarray}
  \partial_t\bm{u}^{\dag}_i - \Pi^{\dag}_k\circ
          [{\mathsf{A}}_i\bm{u}_k ] - {\mathsf{A}}_j^H[\bm{u}_j \circ
            \Pi^{\dag}_i] \label{Adjoint:EqStart} -
          \frac{\chi}{C_{\eta}}\circ\Pi^{\dag}_i + Re^{-1}{\mathsf{A}}_j^H
               {\mathsf{A}}_j^H \bm{u}_i^{\dag} && \nonumber \\
   -(\bm{1}-\chi)\circ\theta^{\dag}\circ[ {\mathsf{A}}_i\theta] &=&0
  \\
  {\mathsf{A}}_j^H\Pi_j^\dag &=& 0 \\
  \partial_t\theta^{\dag} -{\mathsf{A}}_j^H[(\bm{1}-\chi)\circ\bm{u}_j
    \circ \theta^{\dag}] + {\mathsf{A}}_i^H([Pe^{ -1}(\bm{1}-\chi) +
    \kappa\chi] \circ {\mathsf{A}}_i^H\theta^{\dag}) && \nonumber \\
  -{\mathsf{A}}_j^H[ \chi_i \circ (\bm{u}_{s,i})_j\ \circ
    \theta^{\dag}] &=&0
  \label{Adjoint:EqEnd}
\end{subeqnarray}
with initial conditions
\begin{eqnarray}
  \bm{u}^{\dag}(\bm{x},T^{F}) = 0, \qquad
  \theta^{\dag}(\bm{x},T^{F}) = \frac{2\theta}{\Omega}.
\end{eqnarray}
The optimality conditions, stemming from the first variation with
respect to the control variables, are found to be

\begin{subeqnarray}
  \chi^{\dag}_i &= &
  \left[(2 \lambda \mathsf{R}_i)((\bm{u}_{s,i})_j\circ
    \chi_i)\right]\circ (\bm{u}_{s,i})_j \nonumber \\
  &+&\left[\theta^{\dag}\circ[
      \mathsf{A}_j\theta]-\frac{\Pi^{\dag}_j}{C_{\eta}}\right]\circ(\bm{u}_j
  -(\bm{u}_{s,i})_j) +
  (\kappa-Pe^{-1})\mathsf{A}_j^{{H}}\theta^{\dag}\circ \mathsf{A}_j\theta
 \label{Optimal:Chi} \\
  \omega_i^{\dag}&= &\int_0^{T^{F}} \left(\chi_i\circ
  \bm{h}_j(\phi)\right)^H\left((2 \lambda \mathsf{R}_i)((\bm{u}_{s,i})_j\circ
  \chi_i)+\frac{\Pi^{\dag}_j}{C_{\eta}}-\theta^{\dag}\circ[\mathsf{A}_j\theta]\right)
  \ \text{d}t \label{Optimal:Om}
\end{subeqnarray}
where $\Pi^{\dag}_i = \bm{u}^{\dag}_i + {\mathsf{A}}^{H}_ip^{\dag}$.

We note that since rotation is more conveniently defined in polar
coordinates (yet we work in Cartesian coordinates), we introduce the
vector-valued function $\bm{h},$ which transforms the radial
velocity from one coordinate system to the other as follows
\begin{subeqnarray}
  \bm{h}_1(\phi) &=& -\sqrt{(\bm{x}^2-\bm{x}^2_0)}\sin \phi, \\
  \bm{h}_2(\phi) &=&  \sqrt{(\bm{x}^2-\bm{x}^2_0)}\cos \phi,
\end{subeqnarray}
where $\phi$ is the angle of the point $\bm{x}$ with respect to the
horizontal axis.

When considering the above system of equations, we note that we have
to simultaneously solve the direct and adjoint equations, as well as
the optimality condition. Rather than following this procedure, it is
customary to solve the direct and adjoint equations exactly, and to
iterate on the optimality condition, until a user-specified criterion
is satisfied. With this approach, we use the gradient with respect to
the control variables to advance the solution towards an optimum. We
recall that our system is nonlinear, which necessarily implies that we
may not achieve convergence to a global optimum; instead, only a local
optimum may be guaranteed.

\subsection{Summary of algorithm}

The system of equations, reformulated as an iterative scheme, coupled
with an optimization strategy, completes the full algorithm for
computing optimal mixing strategies.

The step-by-step procedure advances along the following
lines. Starting with an initial (guessed) mixing strategy, we solve
the direct (forward) problem over a chosen time horizon from $t=0$ to
$t=T^{F}.$ In a second step, we turn to the adjoint
equations, which advance the adjoint variables backwards in time, from
$t=T^{F}$ to $t=0.$ During this step, it is important to
notice that, owing to the nonlinearity of the direct problem, there is
an explicit dependence of the adjoint equations on the direct
variables. For this reason, we have to store the direct variables
during the forward sweep and inject them, at the appropriate time
steps, into the adjoint equations. For high-resolution cases and large
time horizons, we cannot afford to store all necessary direct
variables. In this case, we refer to checkpointing, where we store
relevant direct information at specific checkpoints in time. From
these checkpoints, we then reconstruct the necessary solutions as they
are needed in the adjoint equations. In this manner, we trade memory
limitations for a (minor) decline in run-time efficiency.

\medskip

The algorithmic steps of the full optimization scheme are then:
\begin{enumerate}
\item We begin by running our forward solution from $t=0$ to
  $t=T^{F}$, solving our discretized and penalized
  governing equations~(\ref{FLUSI:Eq1}) and~(\ref{FLUSI:Eq3}). At this
  point we have specifically chosen checkpoints, at which we save our
  state variables to disk. We aim to have sufficient checkpoints such
  that (i) the memory required to save all state variables to RAM does
  not exceed our resources, and (ii) efficiency of reading and writing
  to disk is ensured.
\item Once we have reached the endpoint of our simulation at
  $t=T^{F},$ we run our program forward from our last
  checkpoint, say $t_n,$ starting from the state that was saved at
  this point, to $T^{F}$. During this forward solution,
  we now ensure that we save the required state variables in RAM for
  each time step.
\item When we reach $T^{F},$ we begin to run the adjoint
  simulation backwards in time from $T^{F}$ to $t_n.$ We
  have all the relevant forward variables in RAM, and so can feed them
  into the adjoint equations~(\ref{Adjoint:EqStart}) at the correct
  time step.
\item Once we arrive at $t_n$ with the adjoint simulation, we save the
  last state of the adjoint to RAM, making sure we have continuity in
  the adjoint variables across checkpoints. We then clear the memory
  and begin with running the forward problem from $t_{n-1}$ to $t_n$,
  once again saving flow fields to RAM.
\item We repeat steps 2-4, moving successively backwards in the
  checkpoints until we reach the initial starting time $t_0=0$. At
  this point, we evaluate the final time conditions for the quantities
  we wish to optimize and supply these to our optimization routine to
  generate the new (and improved) mixing strategy for our system.
\item The direct-adjoint looping strategy is continued until a
  user-specified criterion is reached; at this time, the simulations
  terminate.
\end{enumerate}

\subsection{Examples of mixing optimization}

We will induce mixing in our geometries by embedding moving stirrers
of elliptical cross-section. These stirrers will move on a
pre-determined path. The entire configuration is contained in a
circular vessel. Even though our formalism allows for a point-by-point
definition and manipulation of the stirrer geometry, we will instead
use the parametric form introduced in equation~(\ref{Mask:ChiEq}) with
$f$ taking the form

\begin{eqnarray}
  f &=& \biggl[ \left(\frac{(x-x_{0,i})\cos {\alpha_i} -
      (y-y_{0,i})\sin {\alpha_i}}{a_i} \right)^2 \nonumber \\ && +
      \left( \frac{(x-x_{0,i})\sin {\alpha_i} +
        (y-y_{0,i})\cos {\alpha_i}}{b_i} \right)^2
      \biggr]^{\frac{1}{2}},
\end{eqnarray}
where $x_{0,i}$ and $ y_{0,i}$ denote the center of the elliptical
solid $i$, $a_i$ and $b_i$ are the two perpendicular axes,
respectively, and $\alpha_i$ is the angle of attack with respect to
the horizontal coordinate direction. This parameterization will yield
a more low-dimensional version of general shape optimization. It is
possible to optimize the axes $a_i$ and $b_i$ independently, however,
as we seek to enforce a constant cross-sectional area of our stirrers,
we will enforce $a_i b_i = 1.$

\subsubsection{Shape optimization}

The above simplifications for the optimization of the stirrer shape
carries through to the optimality condition that furnishes, together
with the optimization routine, a new and improved geometry after every
iteration. With the area constraint in place, we only control the axis
$a_i$ and reformulate the cost functional gradient with respect to our
(restricted) control variables, expressed in terms of the adjoint
variable $a_i^{\dag},$ as follows:

\begin{eqnarray}
  a_i^{\dag} & = & \int_0^{T^{F}} \frac{\partial
    \mathcal{L}}{\partial a_i} \ \text{d} t, \nonumber \\
  &= & \int_0^{T^{F}} \frac{\partial \mathcal{L}}{\partial
    \chi_i}\frac{\partial \chi_i}{\partial f}\frac{\partial
    f}{\partial a_i} \ \text{d} t.
\end{eqnarray}
In the above expression, the gradient $\displaystyle{\frac{\partial
    \mathcal{L}}{\partial \chi_i}}$ follows directly from
equation~(\ref{eq:augL}). From equation~(\ref{Mask:ChiEq}) we can see
that
\begin{equation}
  \frac{\partial \chi_i}{\partial f} =
  \begin{cases}
    0, & \qquad \vert f(x,y,a_i,b_i) \vert < r_i, \\
    -\displaystyle{\frac{\pi}{4h}}\sin
    \left(\displaystyle{\frac{\pi (f-r_i)}{2h}}\right),
    & \qquad r_i < \vert f(x,y,a_i,b_i) \vert < r_i + 2h, \\
    0, & \qquad \text{otherwise.}\\
  \end{cases}
\end{equation}
The remaining derivative $\displaystyle{\frac{\partial f}{\partial
    a_i}}$ is straightforward. The above expressions can then be
combined to obtain $a_i^{\dag}$, which is then used to optimize the
axis for optimal mixing.

\subsubsection{Speed optimization}

One may note that our definition of the solid velocity,
equation~(\ref{FLUSI:US}), includes a rotational term, linked to
$\omega_i,$ as well as the general velocity term $\bm{u}_{C_i}$. It is
certainly possible to optimize both terms independently; but this will
be left as a future effort. As this article focuses primarily on the
algorithm and its validation, we choose to draw our attention on the
optimization of the rotational velocity $\omega_i$, as a
representative test of the direct-adjoint optimization framework.

\section{\label{sec:validation} Validation of gradient direction and optimization results}

\subsection{A simple gradient check}

Before embarking on various test cases for the optimization of mixing
strategies, we perform a consistency check of our adjoint
framework. To this end, we consider a scalar control variable $q$
together with its adjoint equivalent $q^{\dag}.$ The optimality
condition

\begin{eqnarray}
  \frac{\delta\mathcal{L}}{\delta q} = 0
\end{eqnarray}
then establishes a link between the adjoint control variable
$q^{\dag}$ and the cost functional gradient $\delta
{\mathcal{J}}/\delta q.$ We have
\begin{eqnarray}
  q^{\dag}  = -\frac{\delta\mathcal{J}}{\delta q}
\end{eqnarray}
which we will use to test our direct-adjoint system. More
specifically, we evaluate the right-hand side by a finite-difference
approximation according to

\begin{eqnarray}
  \frac{\delta\mathcal{J}}{\delta q} \approx
  \frac{\mathcal{J}(q+\epsilon) - \mathcal{J}(q)}{\epsilon},
\end{eqnarray}
for a small value of $\epsilon.$ This expression is based on the
forward problem only. This expression is then contrasted to the value
$q^{\dag}$ which stems from the adjoint system. While we will not be
able to make a quantitative comparison between the two expressions, we
can match the sign-distributions of the various gradients. In other
words, we evaluate the consistency relation

\begin{eqnarray}
  \text{sgn}\left( q^{\dag}\cdot \epsilon \right) =
  \text{sgn}(\mathcal{J}(q) - \mathcal{J}(q+\epsilon)).
\end{eqnarray}
The results of this check is listed in table~(\ref{VerifTab}), where
we show that the sign combinations across all scenarios (using the
axis $a$ and the rotational speed $\omega$ as control variables in our
case) match accordingly. This test verifies the correct directionality
of the adjoint-based gradient which, in turn, will ensure an improved
mixing strategy from iteration to iteration.

\begin{table}
  \centering
  \begin{tabular}{ |r|c|c| }
    \hline
    & $a + \epsilon$ & $a - \epsilon$ \\
    \hline
    $\omega + \epsilon $ & $+$ & $+$ \\
    \hline
    $\omega - \epsilon $ & $-$ & $-$ \\
    \hline
  \end{tabular}
  \qquad \qquad
  \begin{tabular}{ |r|c|c| }
    \hline
    & $\epsilon a^{\dag}$ & $-\epsilon a^{\dag}$ \\
    \hline
    $\epsilon \omega^{\dag}$ & $+$ & $+$ \\
    \hline
    $-\epsilon \omega^{\dag}$ & $-$ & $-$ \\
    \hline
  \end{tabular}
  \caption{\label{VerifTab} Consistency check based on the sign
    distribution of the control variable gradient: (left) based on a
    finite-difference approximation of the forward problem, (right)
    based on the adjoint system.}
\end{table}

\subsection{Definition of test configurations}

To further validate the direct-adjoint optimization framework for
enhancing mixing, we choose a suite of test problems that
progressively challenge the computational procedure but still comply
with our intuition for an optimal solution. The purpose of this
exercise is less in treating a physically interesting case, but rather
in demonstrating the efficiency and effectiveness of the adjoint-based
optimization scheme.

\begin{figure}
  \centering
  \includegraphics[width=0.67\textwidth]{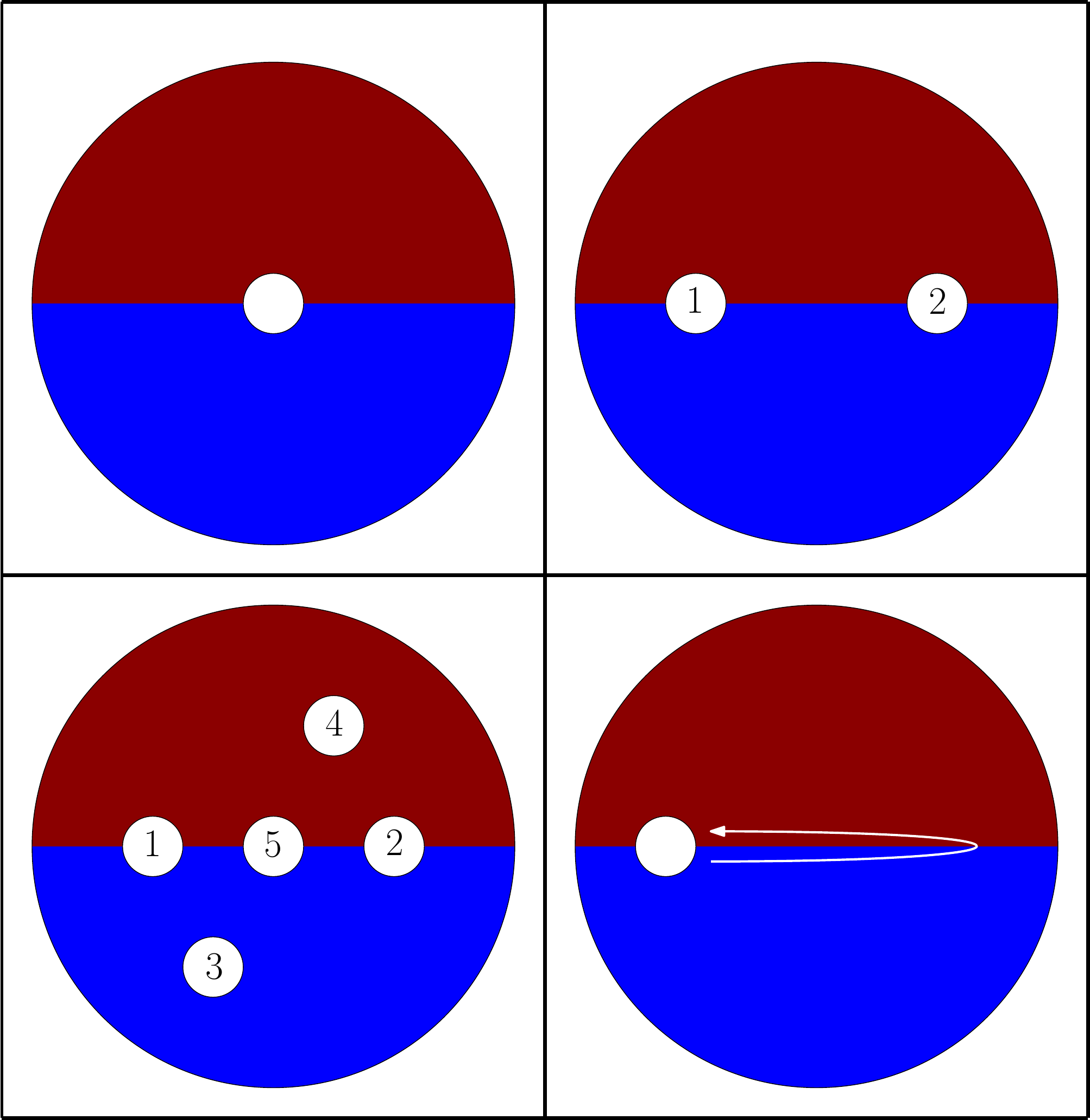}
  \caption{\label{GeometriesPic} Sketch of the initial configurations
    for the four test cases. (top left) Case 1: one centered cylinder
    rotating about its axis. (top right) Case 2: two cylinders on the
    horizontal axis, rotating in opposite directions. (bottom left)
    Case 3: five rotating cylinders, placed such that three cylinders
    are aligned along the horizontal axis, while the remaining two are
    vertically offset. (bottom right) Case 4: one rotating cylinder
    moving from left to right with the velocity of the horizontal
    movement dictated by a $\cos$-function.}
\end{figure}

In each chosen scenario all involved stirrers begin as circular
cylinders with unit radius and an initial rotational speed of $0.25$
(either in the clockwise or anti-clockwise direction). The various
configurations, ranging from a stationary, rotating cylinder to five
rotating cylinders, to a moving and rotating cylinder, are sketched in
figure~\ref{GeometriesPic}. For all simulations, we chose a Reynolds
number of $Re=1000$ and a P\'{e}clet number of $Pe=1000.$ The time
horizon for the optimization is $T^{F}=32;$ and the penalization term
$C_{\eta}$ has been set to $C_{\eta}=0.001,$ where the value was
chosen in accordance with similar scenarios as present in {\tt{FLuSI}}
as well as being consistent with the value reliant on the grid
distance. The passive scalar $\theta$ is initially stratified with
$\theta=1$ in the upper half of the cylindrical domain $\Omega$ and
$\theta=0$ in the lower half. In all cases, we will present the
results for a weakly and highly penalized optimization setting
(choosing the parameter $\lambda$).

\subsection{Case 1: one stationary, rotating stirrer}

We commence by considering the case of a single, rotating (initially
cylindrical) stirrer located in the center of a circular vessel. We
optimize the rotational speed as well as the stirrer shape which we
assume generally elliptical. By lengthening or shortening the the
stirrer's axis and increasing the speed at which the stirrer spins, we
seek to enhance the mixing efficiency (measured by the variance of the
passive scalar) over a given time interval. We place limitations on
the shape of the stirrer by holding constant its cross-sectional area
and on the speed of the rotation by capping the maximum energy
injected into the mixture. We note that the shape of the stirrer
directly affects the energy transferred into the fluid, as highly
eccentric shapes require a larger input effort, but simultaneous may
yield improved mixing.

This case serves as a first benchmark for the direct-adjoint
optimization framework; in particular, we wish to gauge the
convergence behavior, probe the influence of the penalization
parameter, and assess the physical fidelity of the obtained solution.

\subsubsection{Highly penalized system}

\begin{figure}
  \centering
  \begin{tabular}{cc}
    \includegraphics[width=0.5\textwidth]{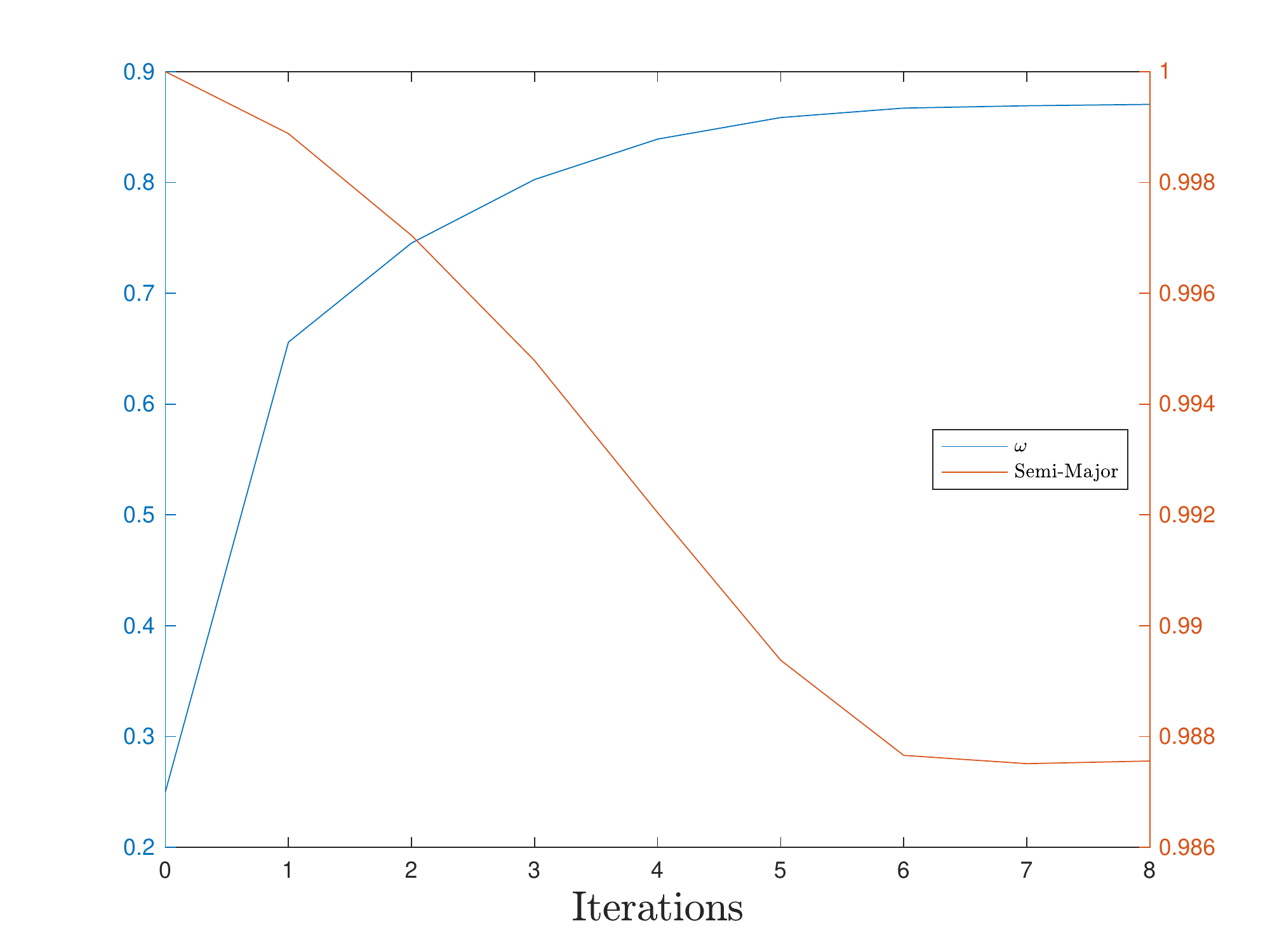} &
    \includegraphics[width=0.5\textwidth]{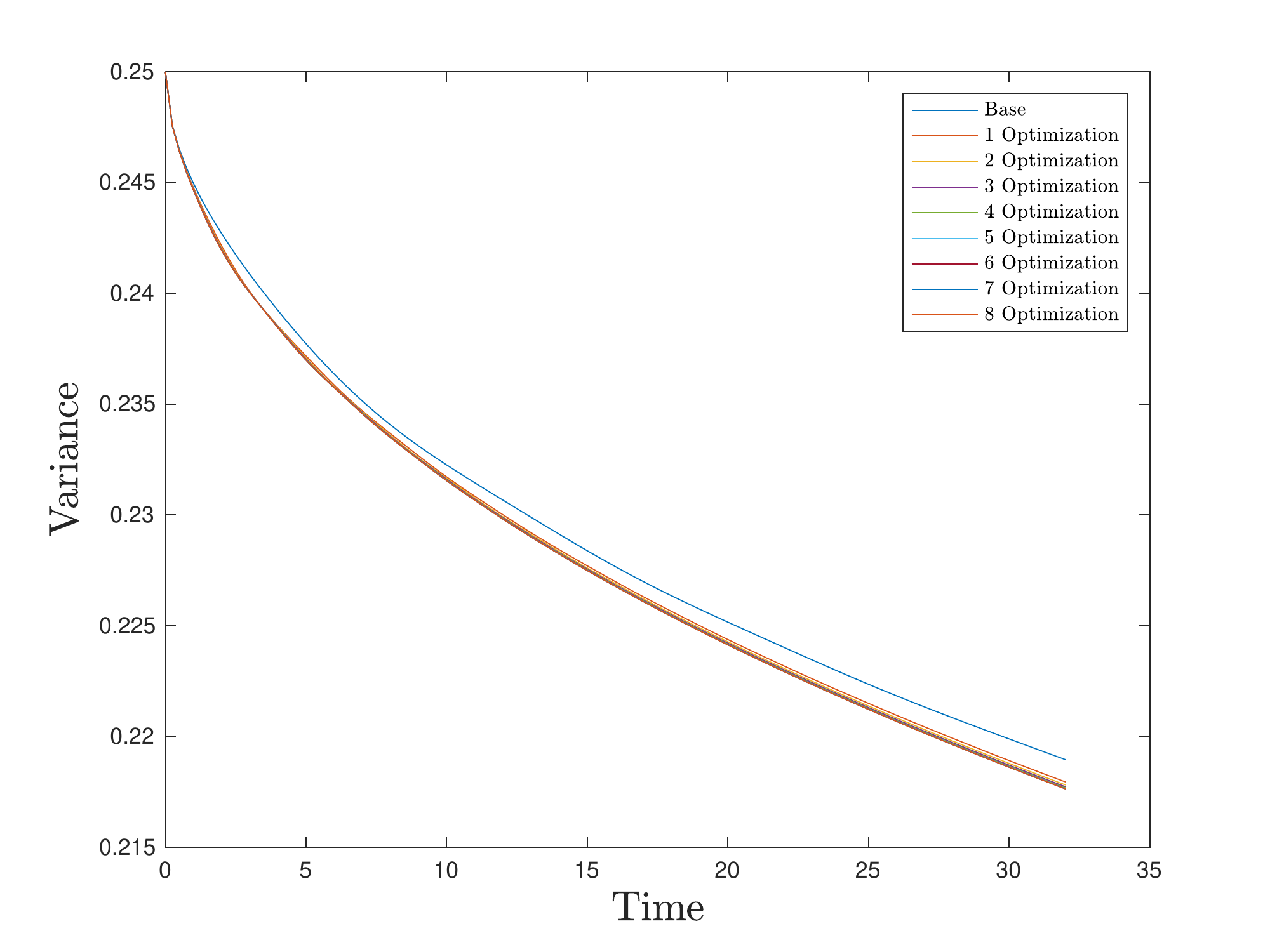}
  \end{tabular}
  \caption{\label{1CCHPResults} Case 1: mixing optimization using one
    stationary, rotating stirrer. A highly penalized optimization
    setting has been used. (a) Rotational speed $\omega$ and axis $a$
    versus the number of direct-adjoint iterations. (b) Variance of
    the passive scalar versus time $t \in [0,\ T^{F}].$}
\end{figure}

As a proof of concept, we present the results of our optimizations
with a rather high penalization parameter of $\lambda_1 = 10^{-3}$. As
expected, the optimization increases the rotational velocity and the
eccentricity of the stirrer. The progression in these control
parameters versus the number of iterations is shown in
figure~\ref{1CCHPResults}a where we observe a monotone shift towards
more eccentric shapes (in red) and a gradual increase in the rotation
speed $\omega$ (in blue). After six iterations, convergence is
achieved, caused by the input energy penalization.

Considering the evolution of the variance over the chosen time
interval $t \in [0,\ T^{F}]$ we observe a clustering of
the various iterations; it appears that an over-penalization has
allowed only marginal improvement in mixing efficiency. This matches
the convergence of the shape and rotational speed variables in
figure~\ref{1CCHPResults}a.

From a physical point of view, the marginal mixing improvement can be
attributed to the inability of the (penalized) stirrer to produce
small-scale structures which could be dissipated or to induce
significant advective mixing processes. Instead, after nine iterations
we remain within the solid-body rotation regime, an example of which
is shown in the left column of figure~\ref{1CC}. The lack of
small-scale structures renders diffusion ineffective, and explains the
disappointing decrease in variance; the final mixing is solely due to
the diffusion of an extended fluid interface created by the faster
spinning (near-)cylinder.

\subsubsection{Weakly penalized system}

To induce more effective mixing, the system must be able to use
advective processes to create small-scale structures and filaments
which then give rise to significant mixing by diffusion. The highly
penalized system of the previous section does not venture into the
proper parameter regime to encourage this behavior. For this reason,
we drastically lower the penalization value to $\lambda=10^{-4}.$

\begin{figure}
  \centering
  \begin{tabular}{cc}
    \includegraphics[width=0.5\textwidth]{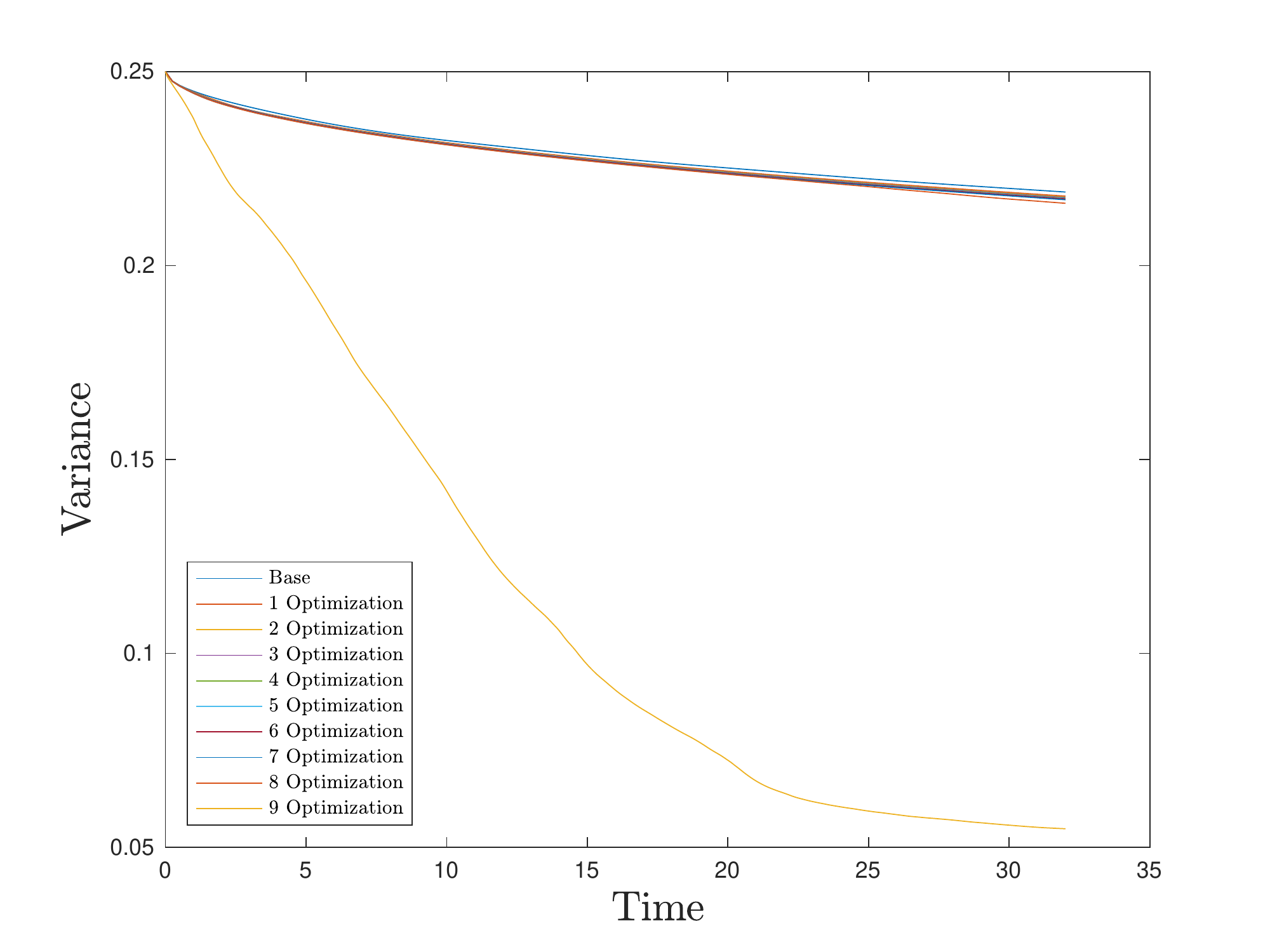} &
    \includegraphics[width=0.5\textwidth]{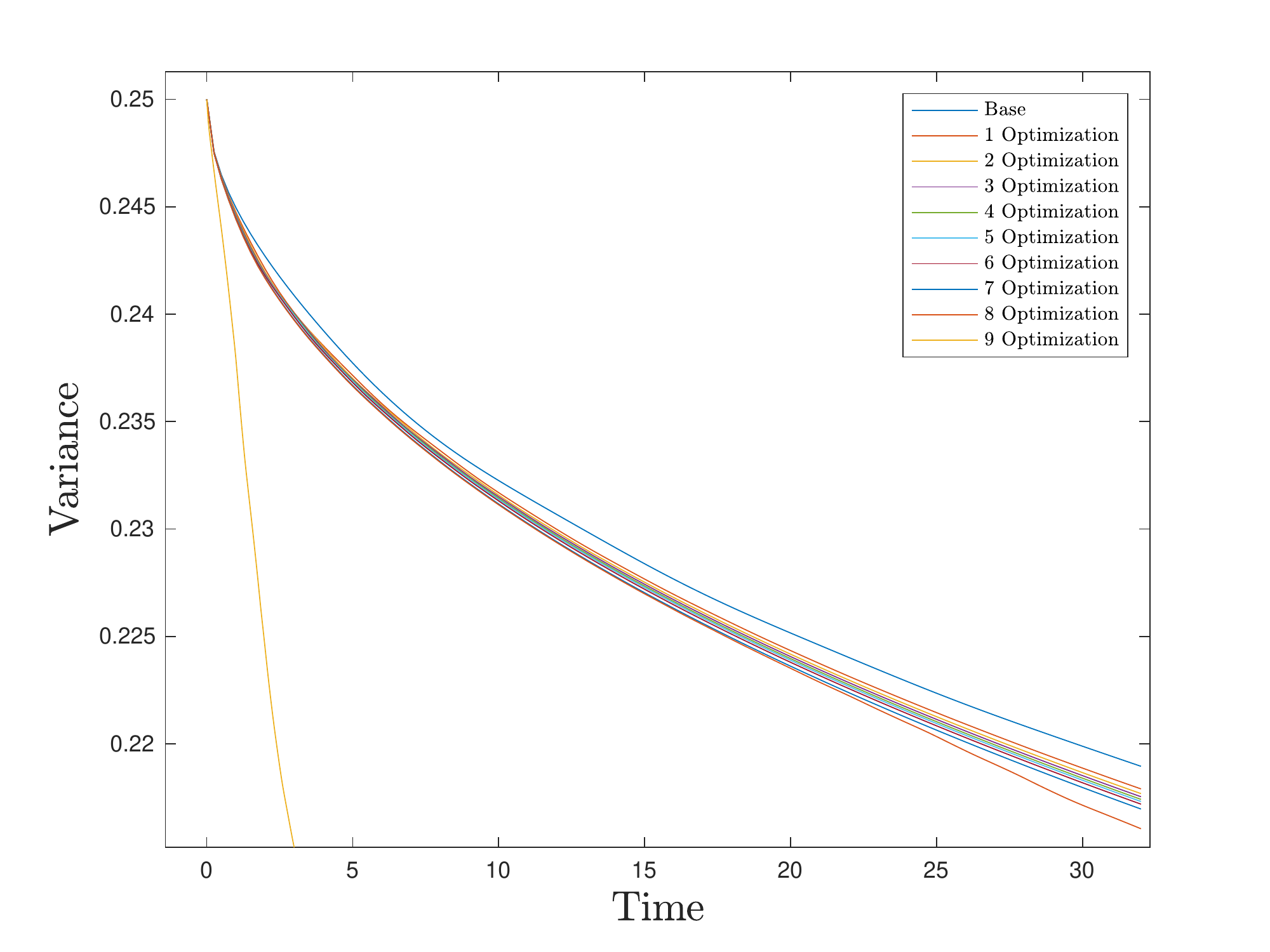}
  \end{tabular}
  \includegraphics[width=0.75\textwidth]{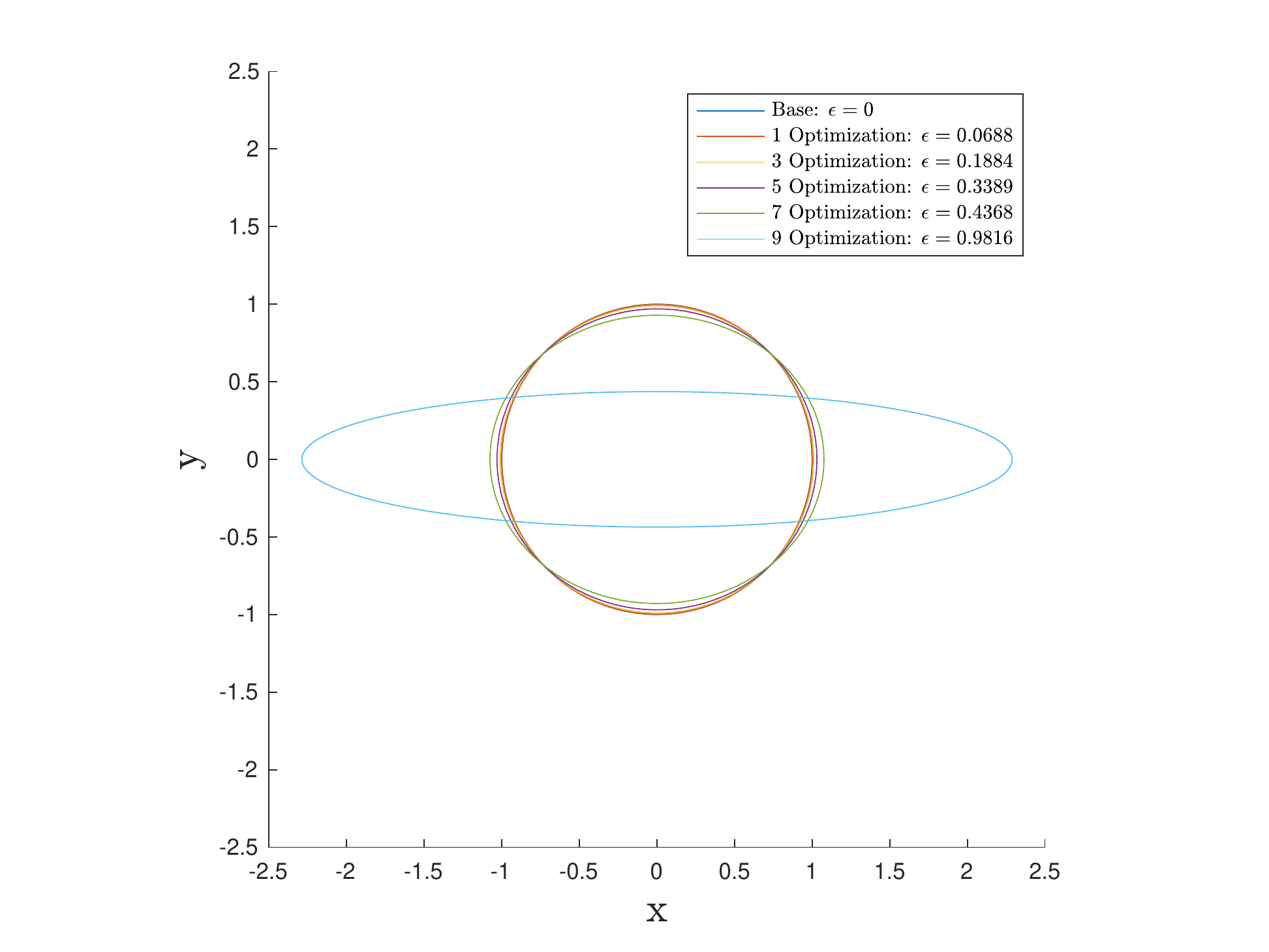}
  \caption{\label{1CCVarPic} Case 1: mixing optimization using one
    stationary, rotating stirrer. (a) Variance, as defined in equation
    (\ref{eq:Variance}), of the scalar field $\theta$ versus time $t
    \in [0,\ T^{F}].$ (b) Zoomed-in view to illustrate the decrease in
    variance for the first eight iterations. (c) Contour of cylinder
    shapes as result of successive iterations.}
\end{figure}

When considering the variances that follow from each direct-adjoint
iteration in figure~\ref{1CCVarPic}a, we note a clustering over the
first optimizations, similar to the highly penalized approach. In this
range, we still remain in the solid-body regime and thus solely rely
on diffusion for our mixing. However, we note that the seventh and
eighth iteration brings about a marked change and leads to a
significant decrease in the variance when compared to the preceding
steps (see figure~\ref{1CCVarPic}b for a closer view). The associated
stronger mixing is created by the availability of an advective process
caused by the elliptic stirrer that is now sufficiently elongated (and
spinning sufficiently fast) to shed vortices off its tips. These
vortices form the sought-after small-scale structures that intensify
the diffusion process by increasing the length of the interface
between the two fluids. These increased small scale dynamics enhance
the adjoint's ability to create further mixing significantly and
therefore result in the drastic decrease in variance as can be
observed in figure~\ref{1CCVarPic}a.

During the optimization, we notice a substantial increase in the
adjoint variables -- to a degree that requires the marked reduction of
the step-size in the optimization routine. This increase is expected,
as a more efficient process that accomplished mixing is available
after the stirrer has been modified to induce vortex shedding. In
other words, after six iterations on the cost functional surface, we
have reached the edge of the diffusion-dominated plateau and
progressed towards lower variances by exploiting
advection-diffusion-dominated mixing. The high penalization parameter
$\lambda$ in the previous section inhibited the exploration of this
regime.

\begin{figure}
  \centering
  \begin{tabular}{lcr}
    \includegraphics[width=0.35\textwidth]{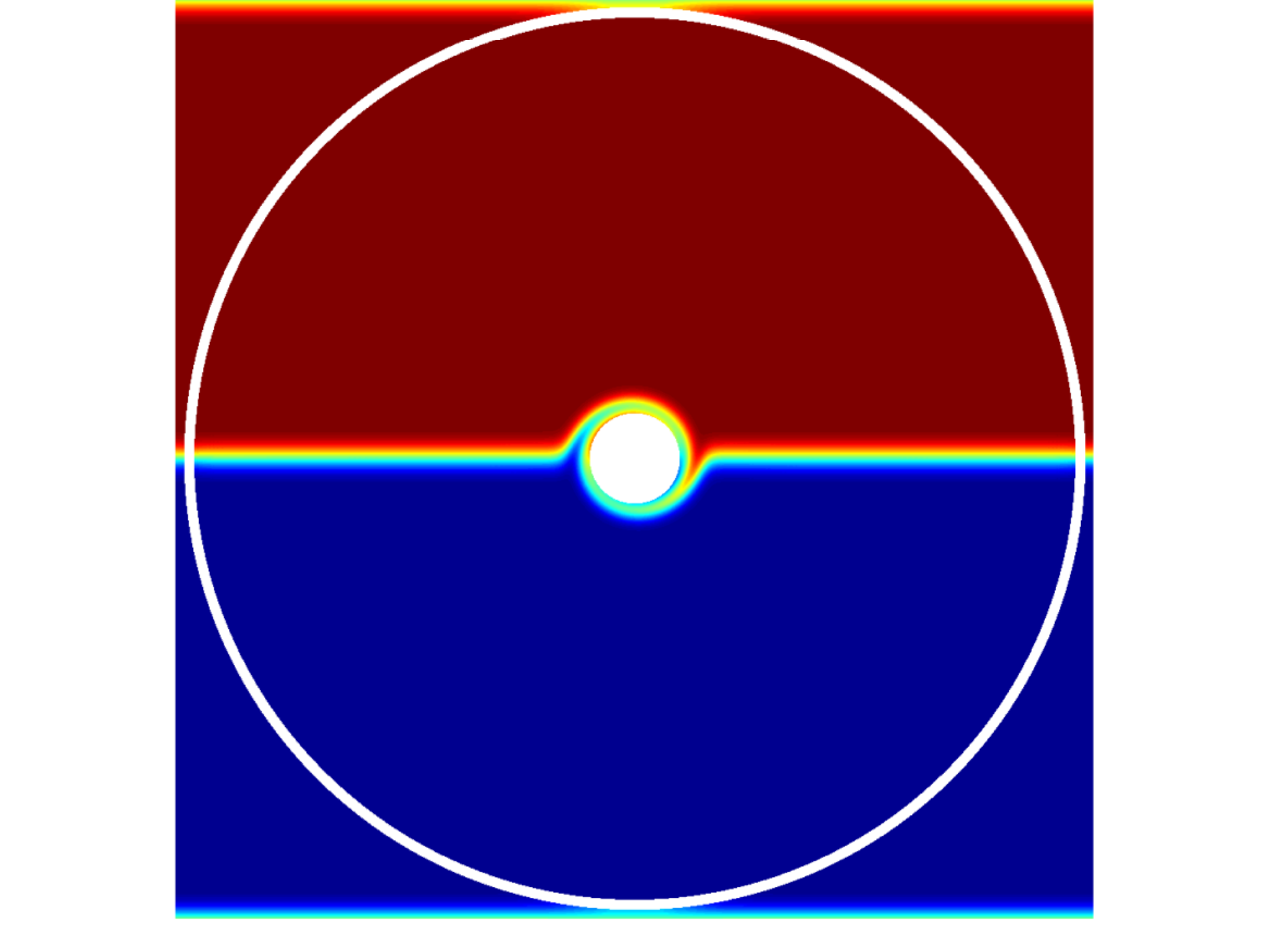} &
    \includegraphics[width=0.35\textwidth]{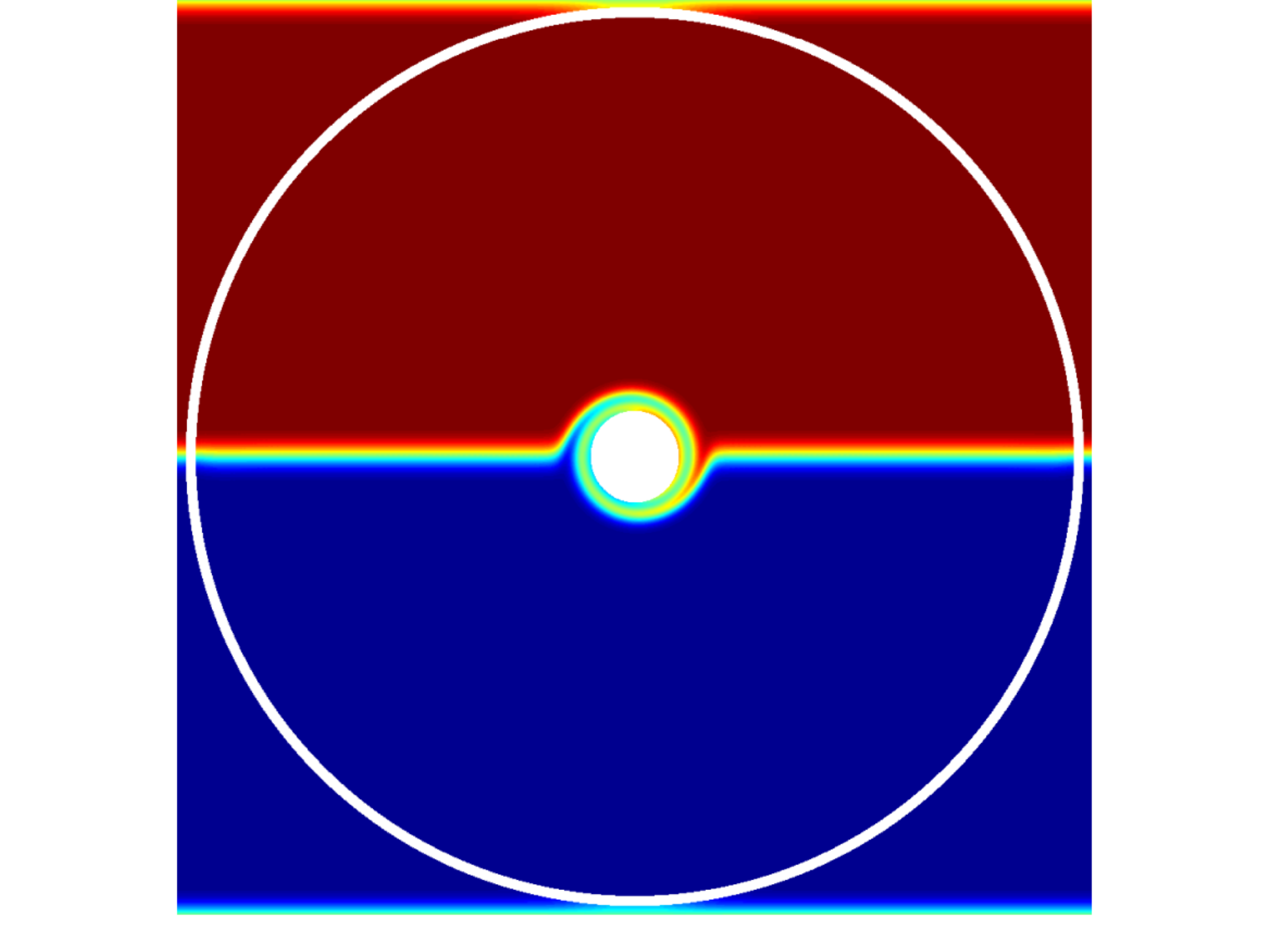} &
    \includegraphics[width=0.35\textwidth]{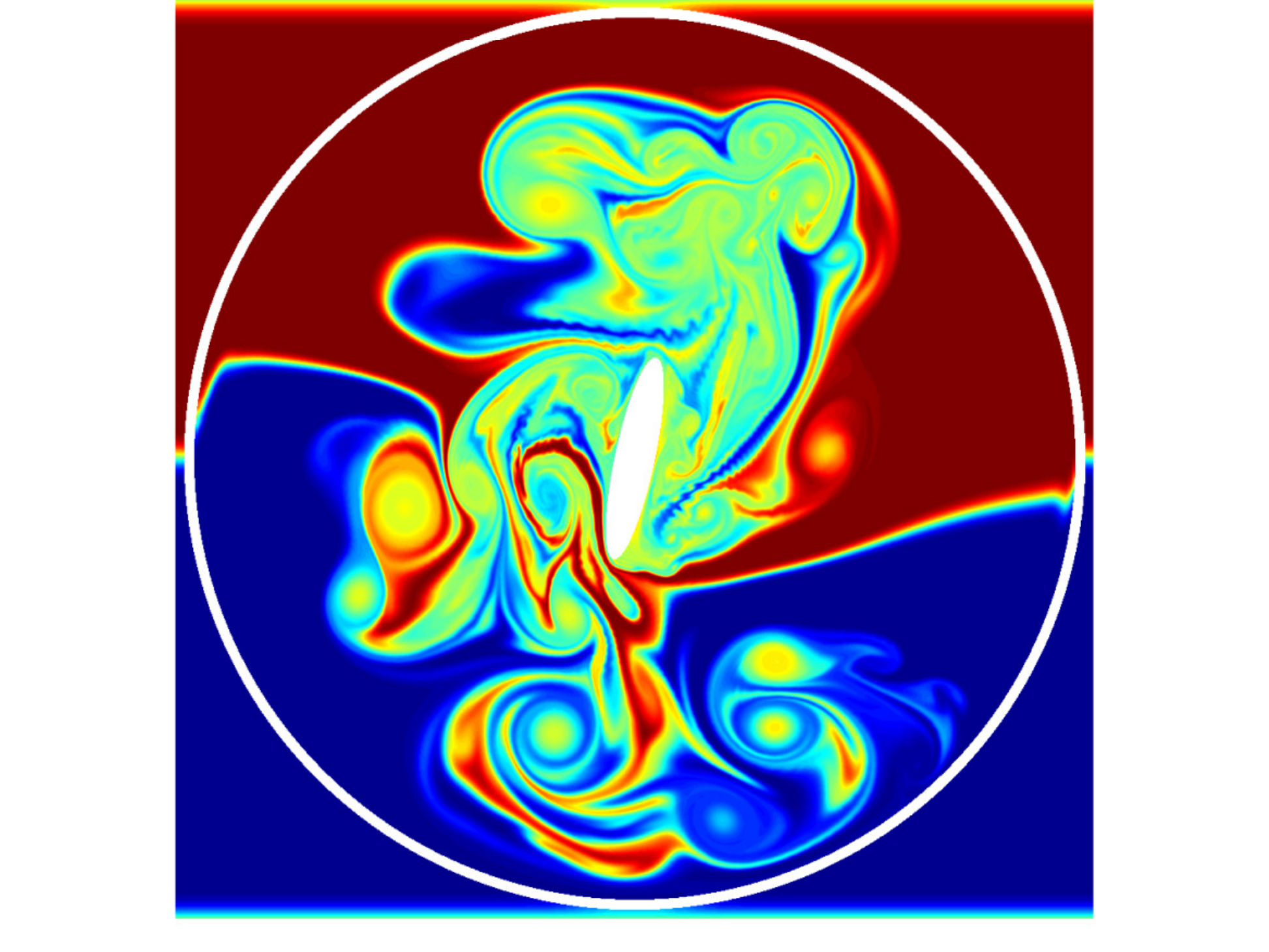} \\
    \includegraphics[width=0.35\textwidth]{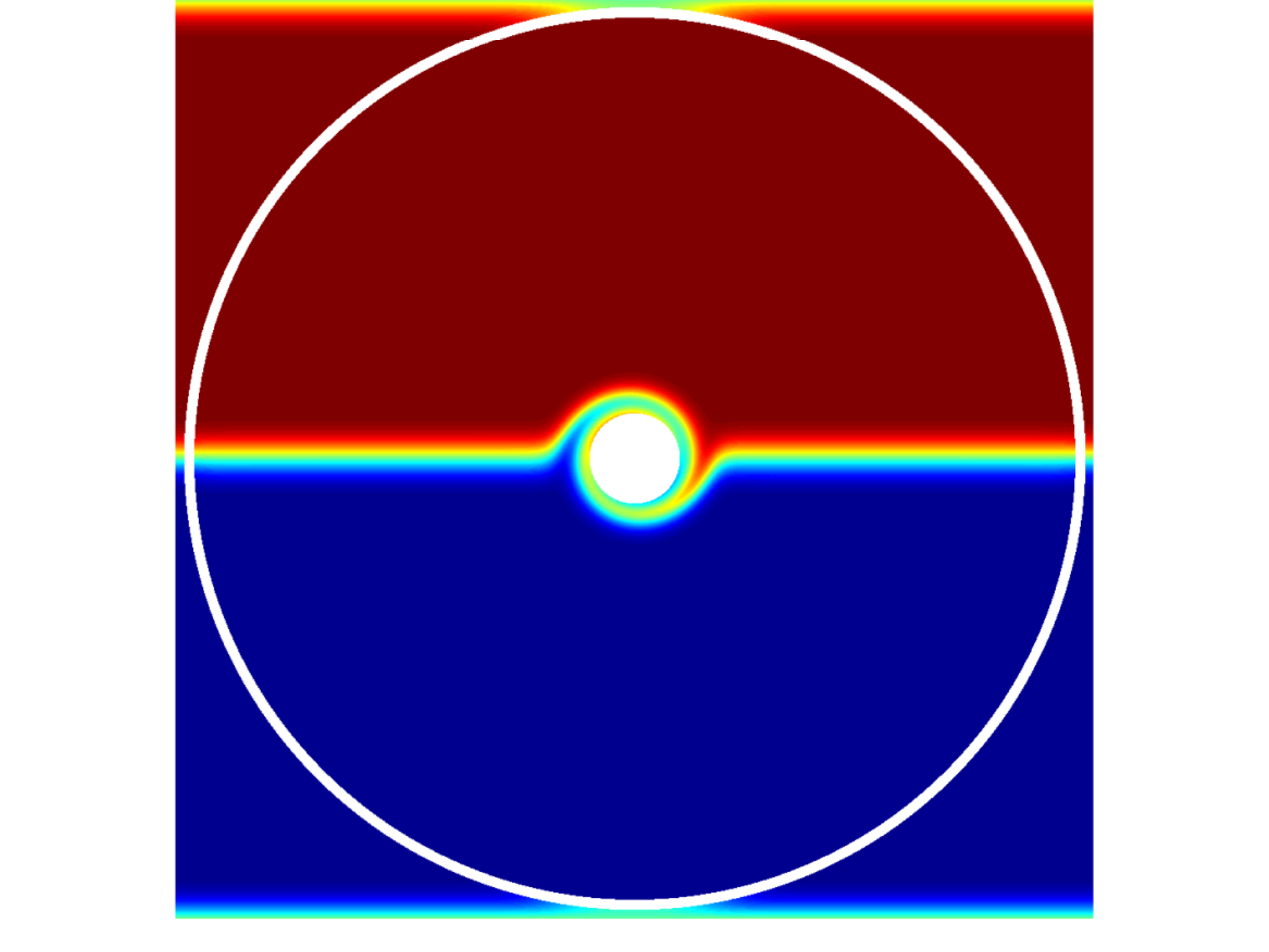} &
    \includegraphics[width=0.35\textwidth]{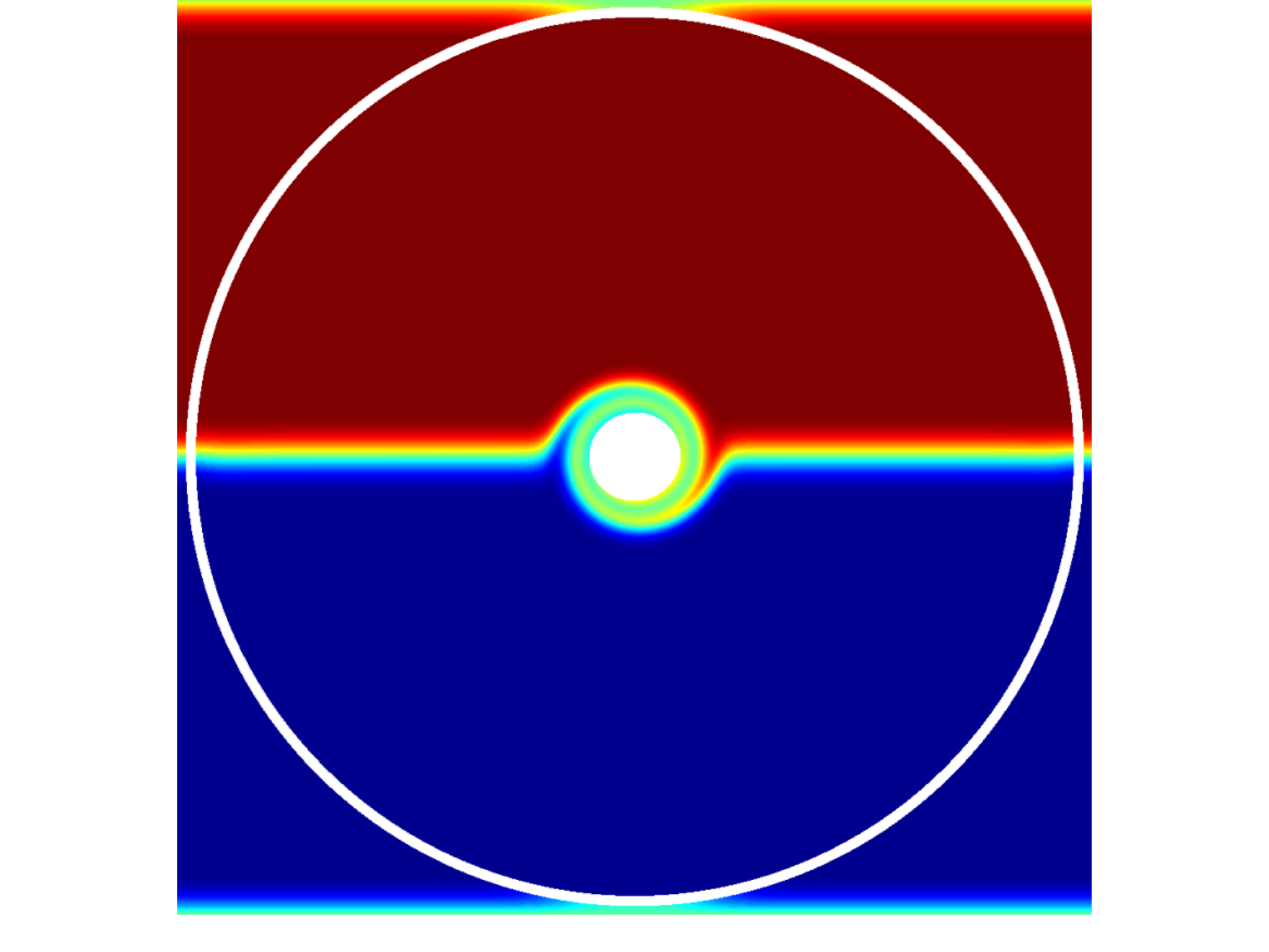} &
    \includegraphics[width=0.35\textwidth]{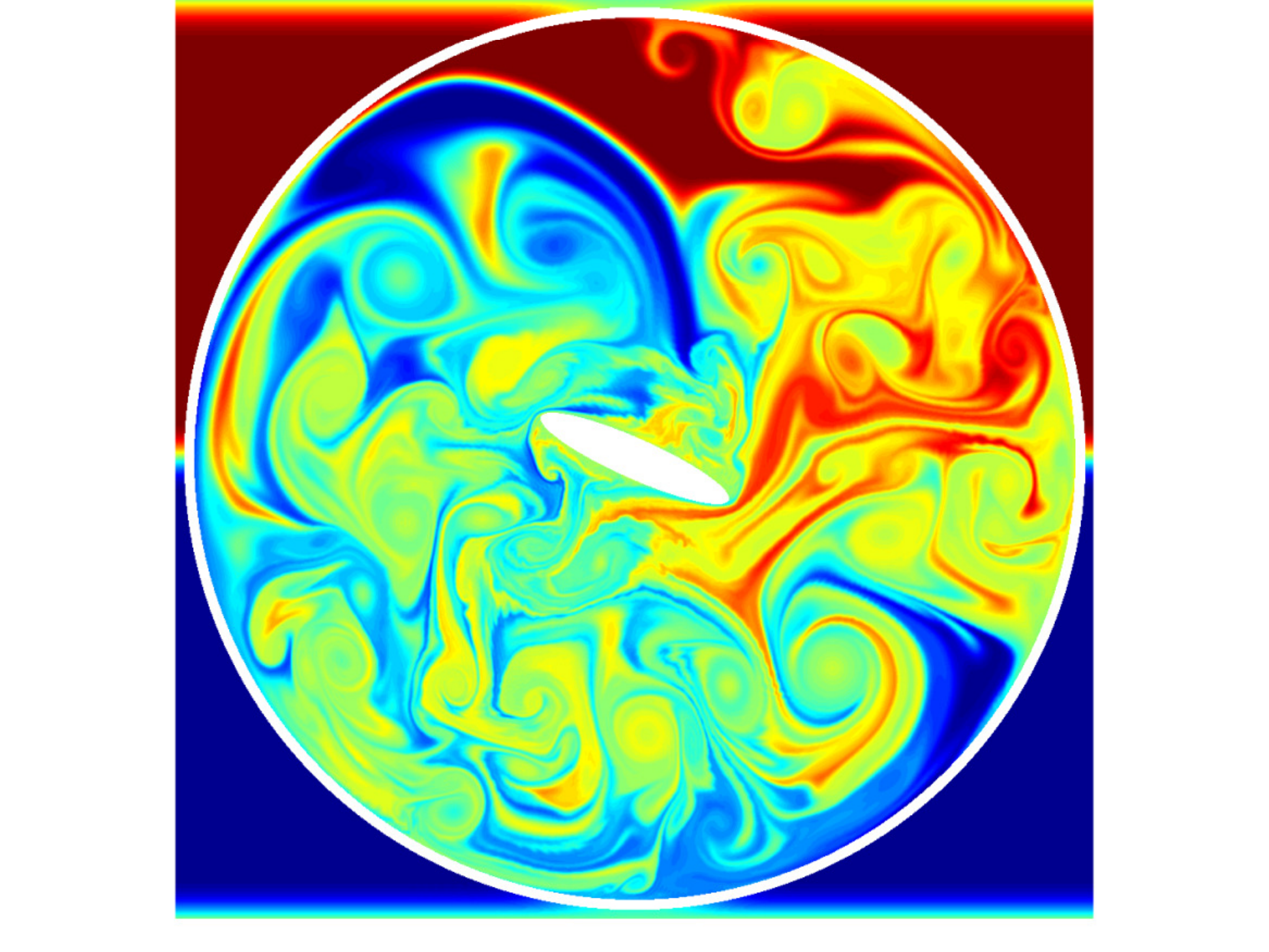} \\
    \includegraphics[width=0.35\textwidth]{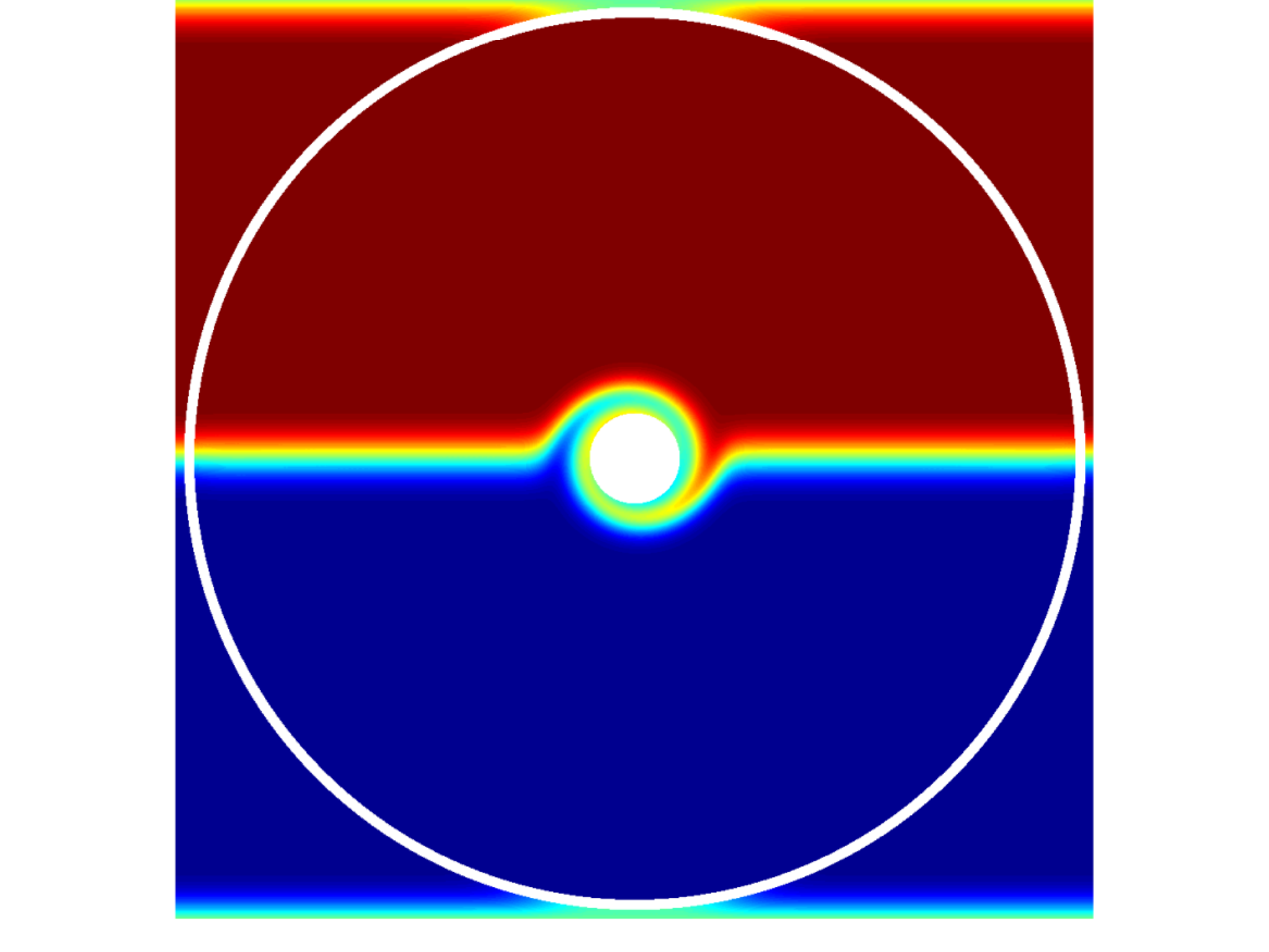} &
    \includegraphics[width=0.35\textwidth]{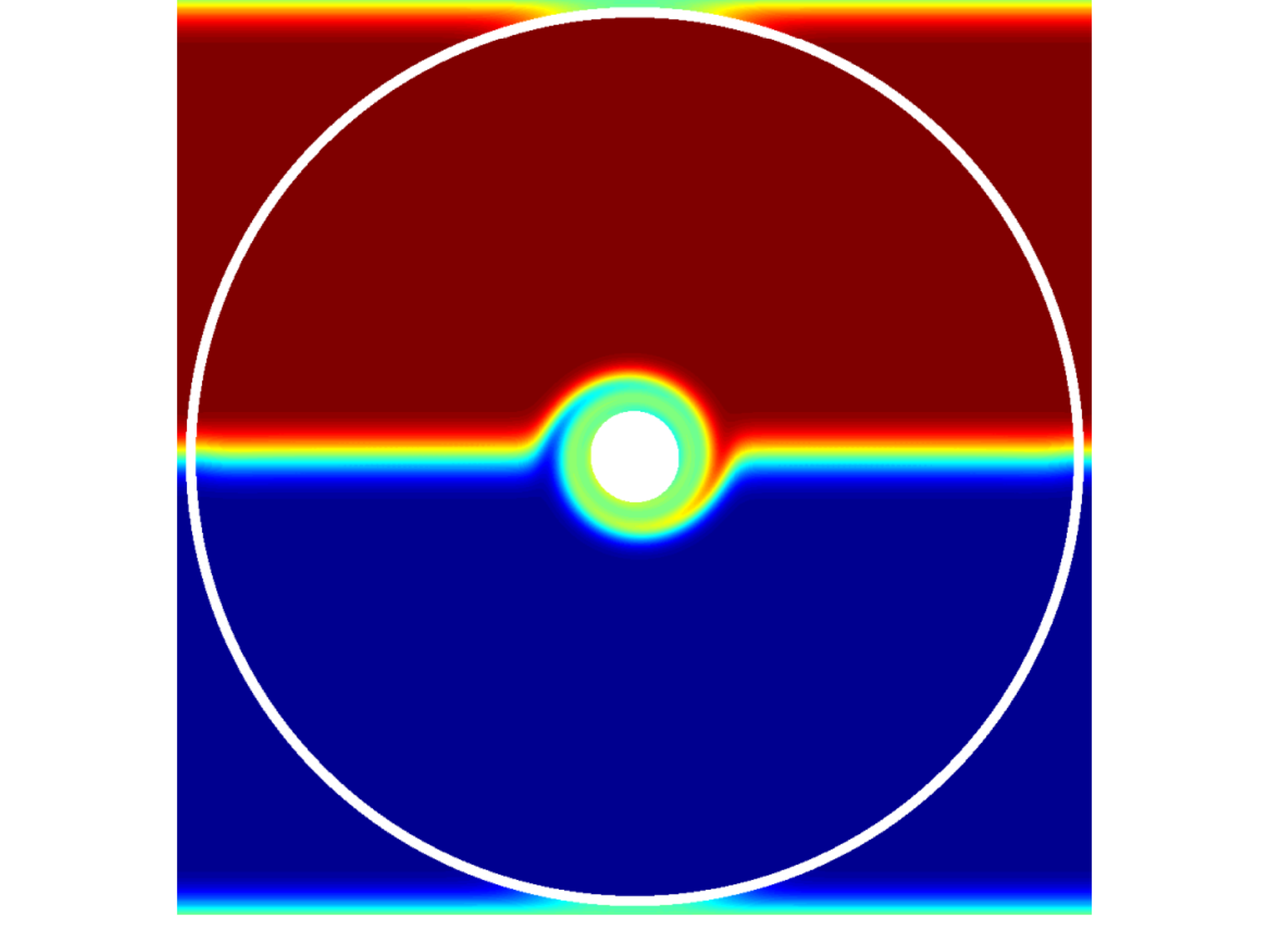} &
    \includegraphics[width=0.35\textwidth]{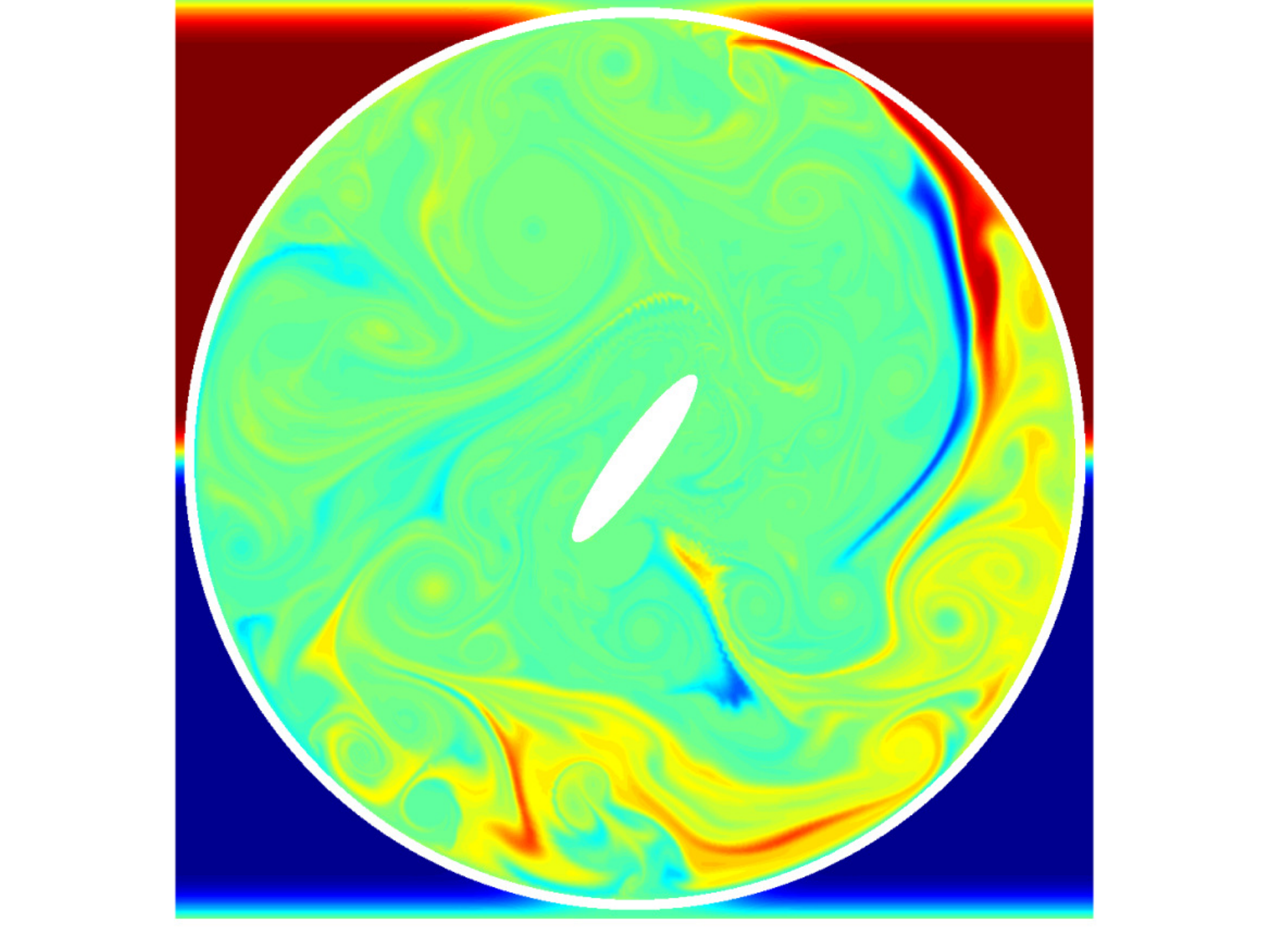} \\
    \includegraphics[width=0.35\textwidth]{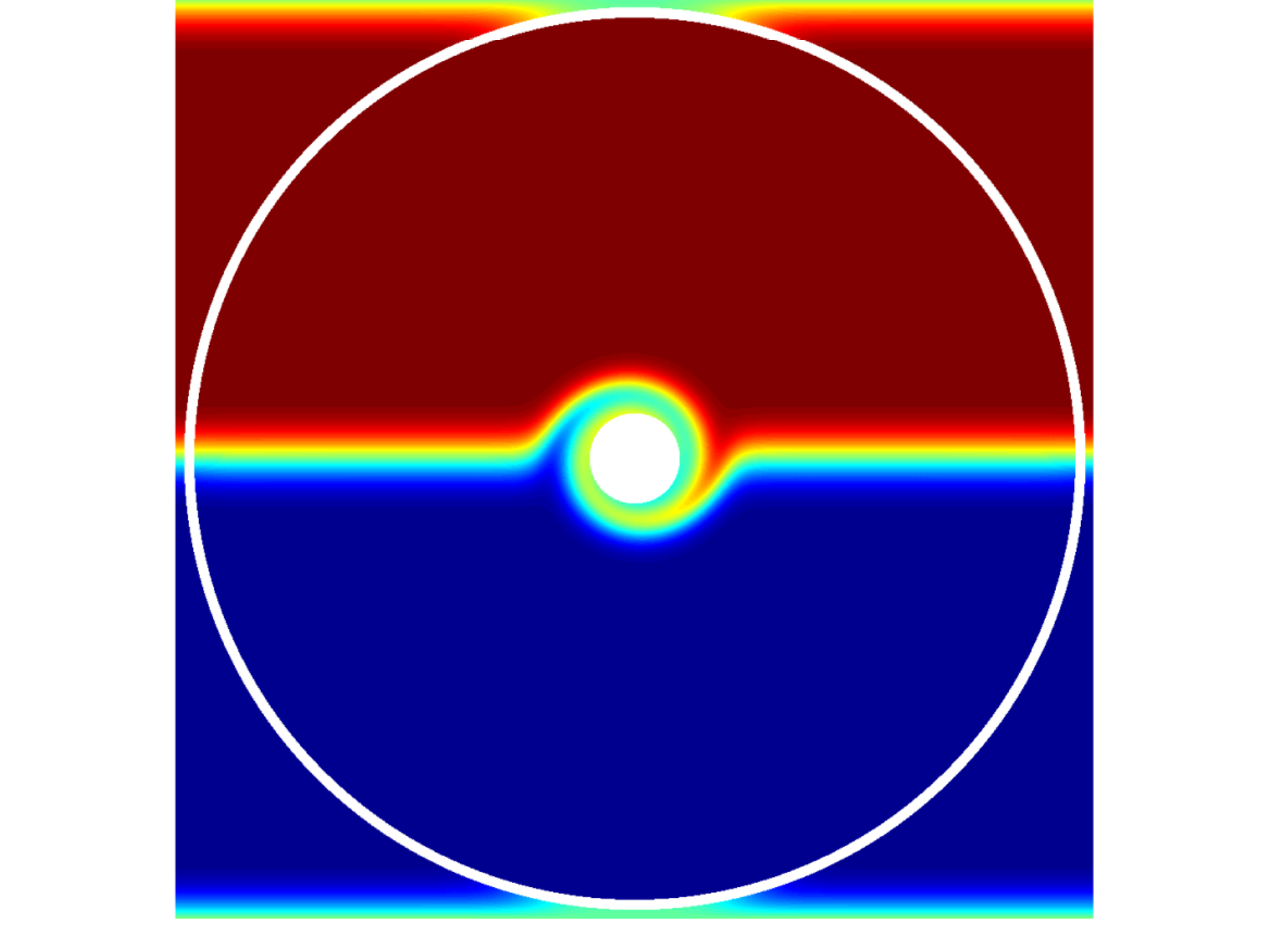} &
    \includegraphics[width=0.35\textwidth]{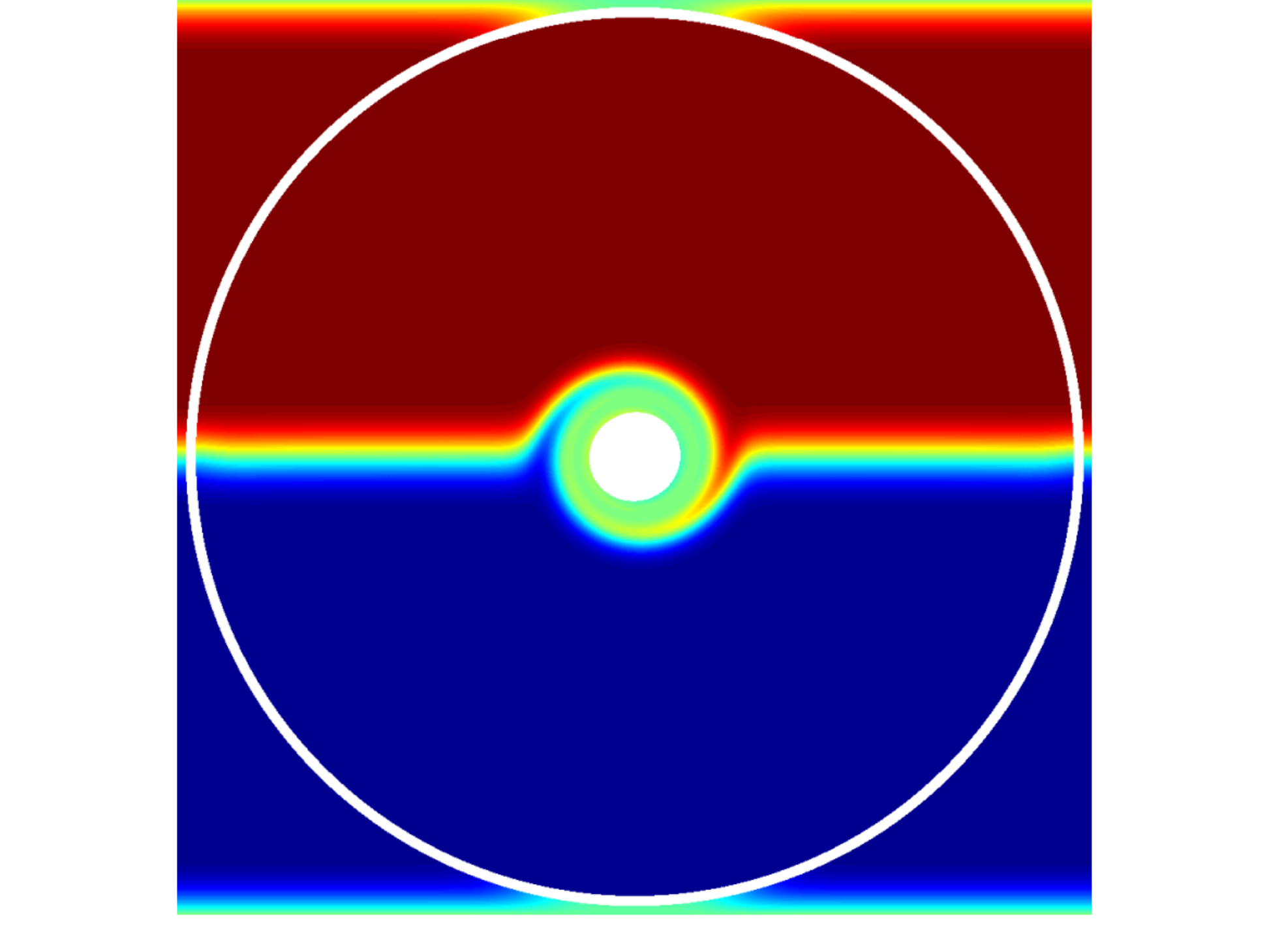} &
    \includegraphics[width=0.35\textwidth]{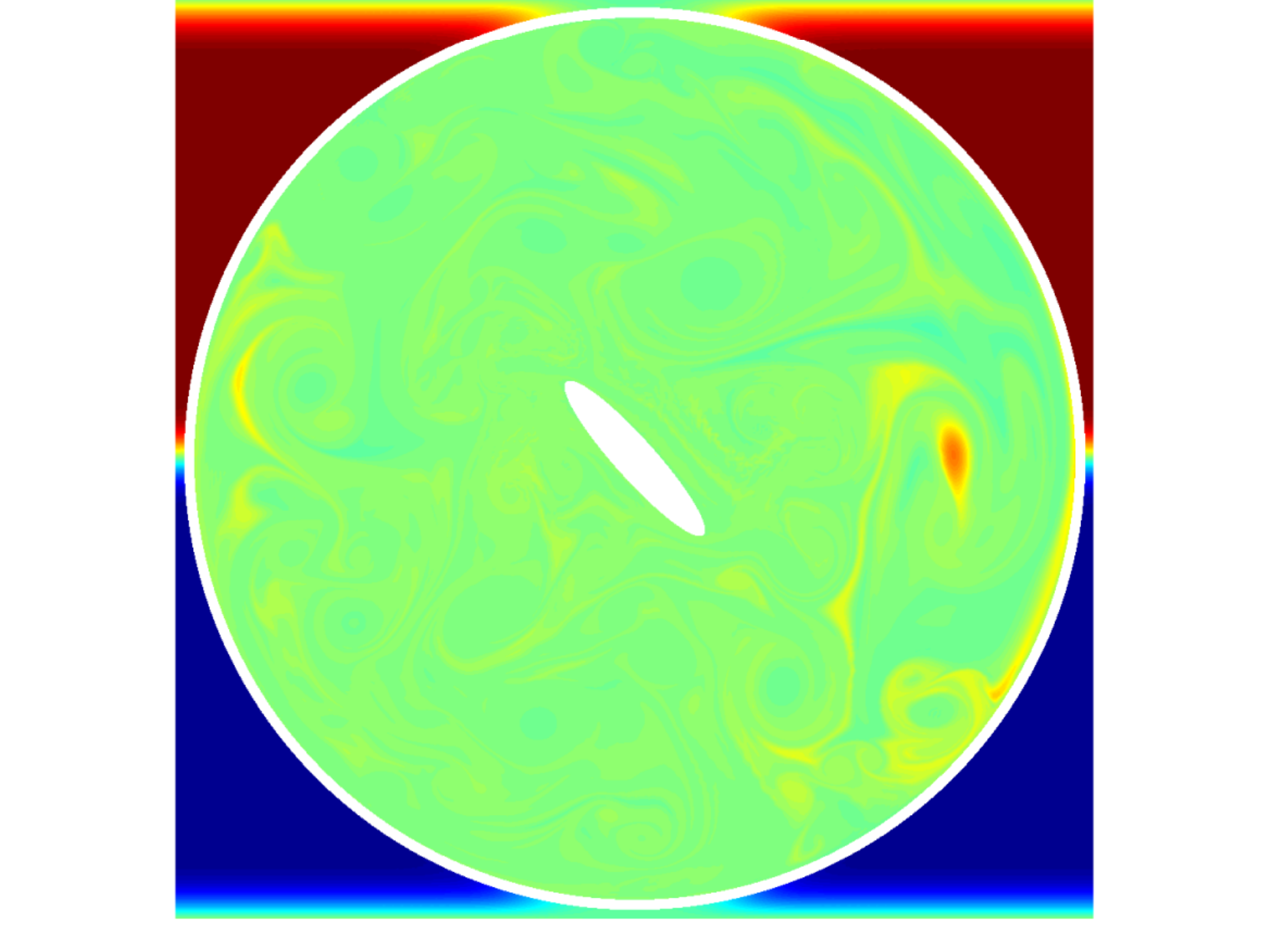} \\
  \end{tabular}
  \caption{\label{1CC} Case 1: mixing optimization using one
    stationary, rotating stirrer. Left column: unoptimized
    configuration, with snapshots at $t = 8, 16, 24, 32$ (top to
    bottom). Middle Column: after four direct-adjoint optimizations,
    with snapshots at $t = 8, 16, 24, 32$ (top to bottom). Right
    column: after nine direct-adjoint optimizations, with snapshots at
    $t = 8, 16, 24, 32$ (top to bottom). For videos of these scenarios
    please refer to {\tt{1Before.mp4}}, {\tt{1Intermediate.mp4}} and
    {\tt{1After.mp4}} for the left, middle and right column,
    respectively.}
\end{figure}

The shape of the variance for the ninth iteration (see
figure~\ref{1CCVarPic}a) exhibits a leveling off during
the later stages of the direct simulation (between $t
\approx 25$ and $t =32$). This suggests that at this
point the mixing process driven by the elliptical stirrer is nearly
complete, and any further mixing is primarily due to diffusion. In
fact, when comparing the gradient of the variance to previous
iterations, we observe a corresponding similarity and thus can
conclude that advection no longer plays an important role. This
conclusion is further corroborated by regarding the right-hand column
of figure~\ref{1CC}, which shows snapshots from the final iteration;
the uniformity of the passive scalar field $\theta$ is evident towards
the end of the temporal optimization horizon.

While this simple example has supplied information about the
convergence behavior and the role of the penalization parameter in
including or excluding physical mixing strategies to accomplish
optimal mixing results, we now proceed to more complex cases and
further probe the direct-adjoint optimization framework.

\newgeometry{left=1cm}

\restoregeometry
\subsection{Case 2: two stationary, rotating stirrers}

We complicate the geometry and optimization scheme by introducing a
second stirrer which we place along the horizontal axis of the mixing
dish; as before, the initial shape of both stirrers is taken as
circular. Again, we also consider the influence of the penalization
parameter (controlling the maximum amount of energy added to the
system) on the convergence behavior and the chosen physical
optimization strategy.

\subsubsection{Highly penalized system}

We observe, similar to the previous one-cylinder case, that a high
energy penalization parameter prevents the optimal mixing strategy to
explore options other than solid-body rotation and diffusion of the
spinning boundary layer around the (mostly) circular stirrers. With
the existence of a second stirrer, there is an additional possibility
for the optimization scheme of exiting this regime: by placing the two
stirrers close to each other, we can have their respective rotating
boundary layers interact and exchange sufficient adjoint (gradient)
information to induce advection-dominated strategies and a
corresponding drop in variance. In our case, the two stirrers appear
unaware of each other; no cooperative strategy is pursued by the
optimization scheme and, as a result, little progress is made in
improving mixing efficiency.

\begin{figure}
  \centering
  \begin{tabular}{cc}
    \includegraphics[width=0.5\textwidth]{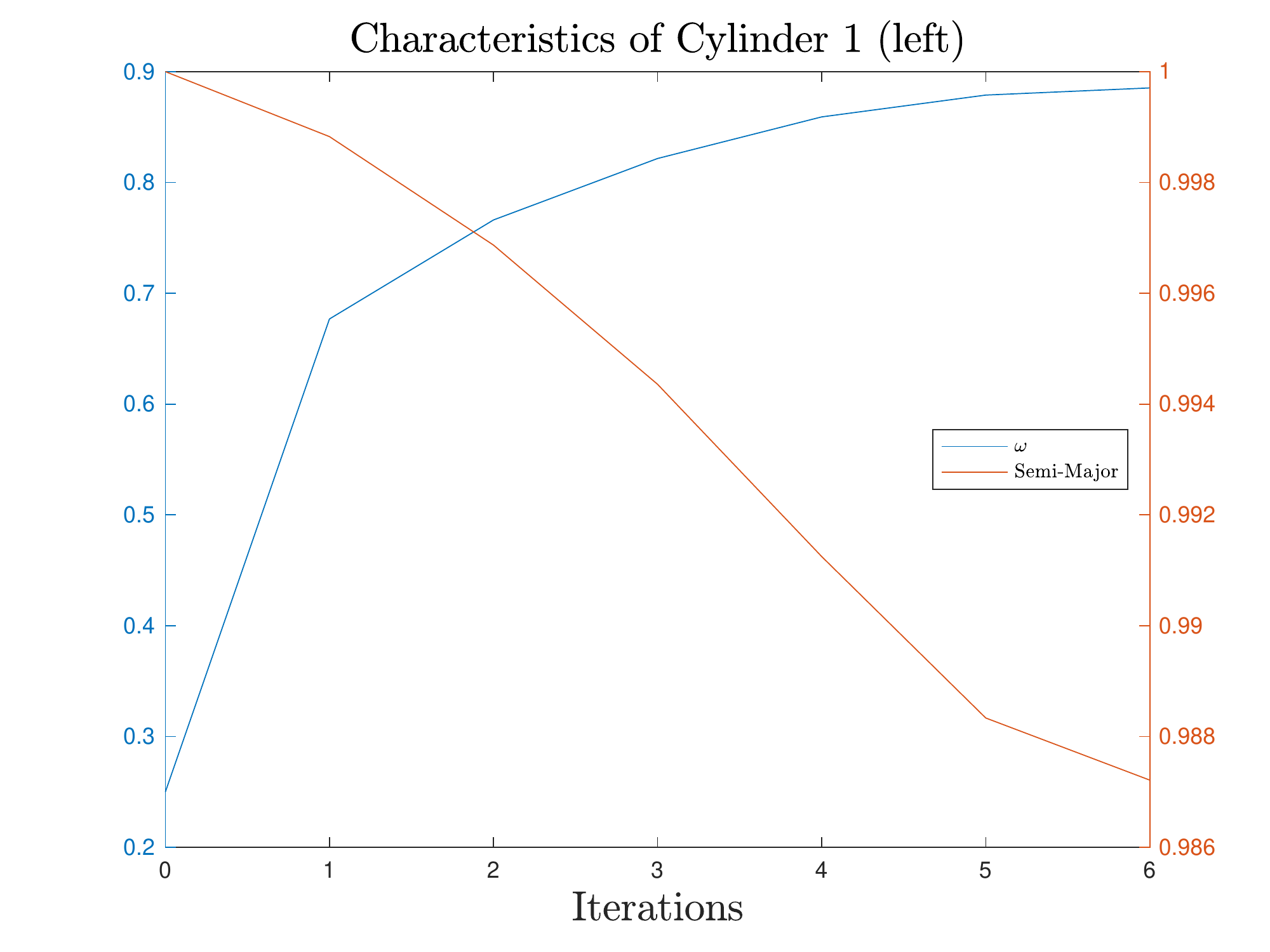} &
    \includegraphics[width=0.5\textwidth]{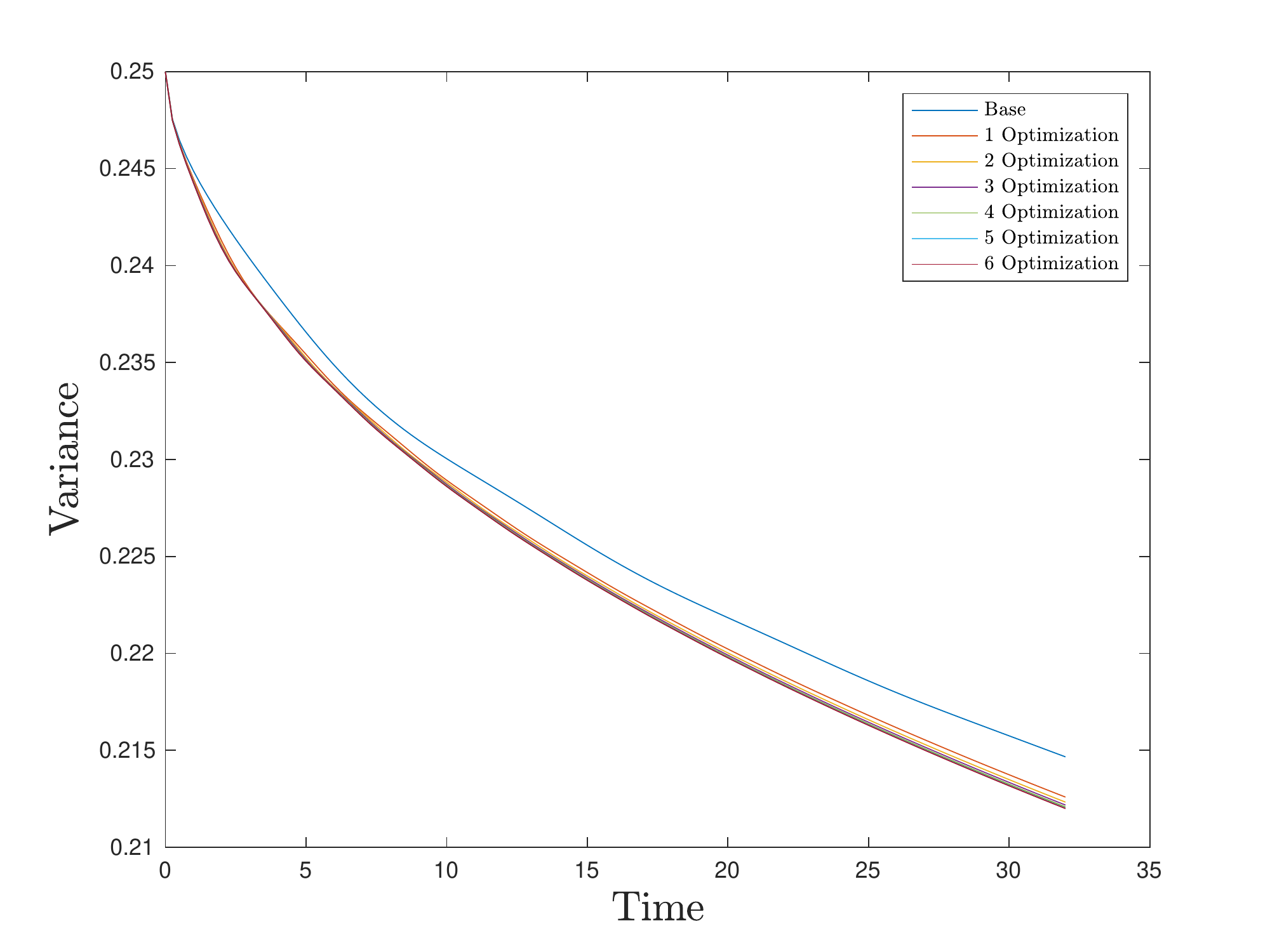}
  \end{tabular}
  \caption{\label{2CCHPResults} Case 2: mixing optimization using two
    stationary, rotating stirrers. A highly penalized optimization
    setting has been used. (a) Rotational speed $\omega$ and axis
    length $a$ versus the number of direct-adjoint iterations for the
    first (left) cylinder. (b) Variance of the passive scalar versus
    time $t \in [0,\ T^{F}].$}
\end{figure}

Figure~\ref{2CCHPResults}a shows the evolution of the axis length
(while maintaining the cross-sectional area) and the rotational speed
of the first (left) stirrer as we progress through six iterations of
the direct-adjoint looping. We see an increase in the rotational
speed, but only an insignificant change in the stirrer's
eccentricity. The control variables for the second (right) stirrer are
identical to the ones shown in figure~\ref{2CCHPResults}a. It is not
surprising that the temporal evolution of the variance over these six
iterations appears rather clustered (see figure~\ref{2CCHPResults}b).

\subsubsection{Weakly penalized system}

Applying a lower value of $\lambda$ to this two-stirrer configuration
is expected to yield similar results as observed before: by allowing
from energy expenditure, advective processes will become a feasible
option, vortex shedding from elliptical stirrers will commence and
substantially more efficient mixing will ensue.

\begin{figure}
  \centering
  \includegraphics[width=0.67\textwidth]{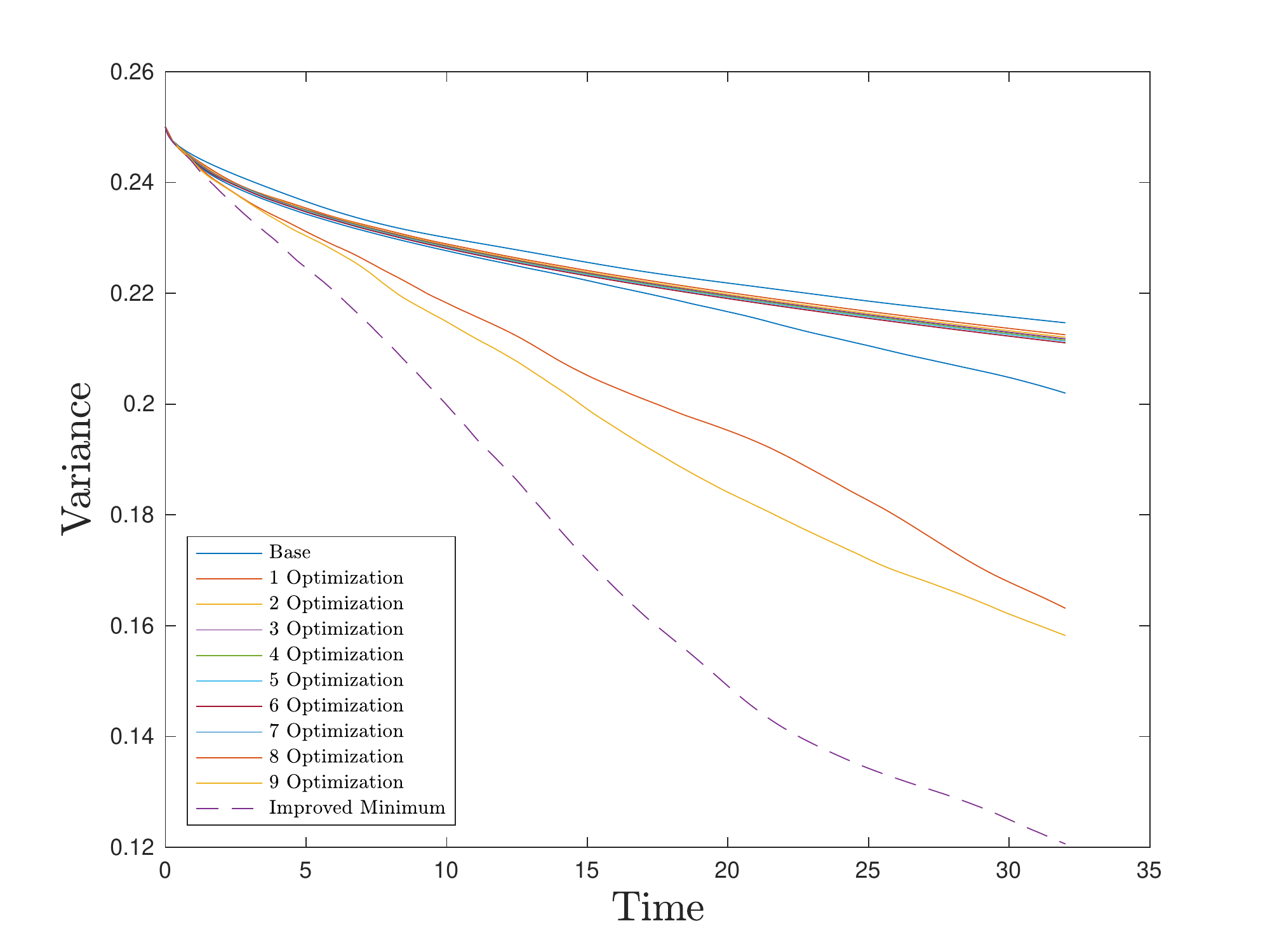}
  \caption{\label{2CCVarPic} Case 2: mixing optimization using two
    stationary, rotating stirrers. Variance, as defined in equation
    (\ref{eq:Variance}), of the scalar field $\theta$ versus time $t
    \in [0,\ T^{F}].$ The solid lines represent a local
    optimum, where the left stirrer remains rather inactive. The
    dashed line represents an improved optimum, by mirroring the
    control variables onto the left stirrer before continuing the
    gradient-based optimization.}
\end{figure}

\begin{figure}
\begin{center}
  \begin{tabular}{lcr}
    \includegraphics[width=0.35\textwidth]{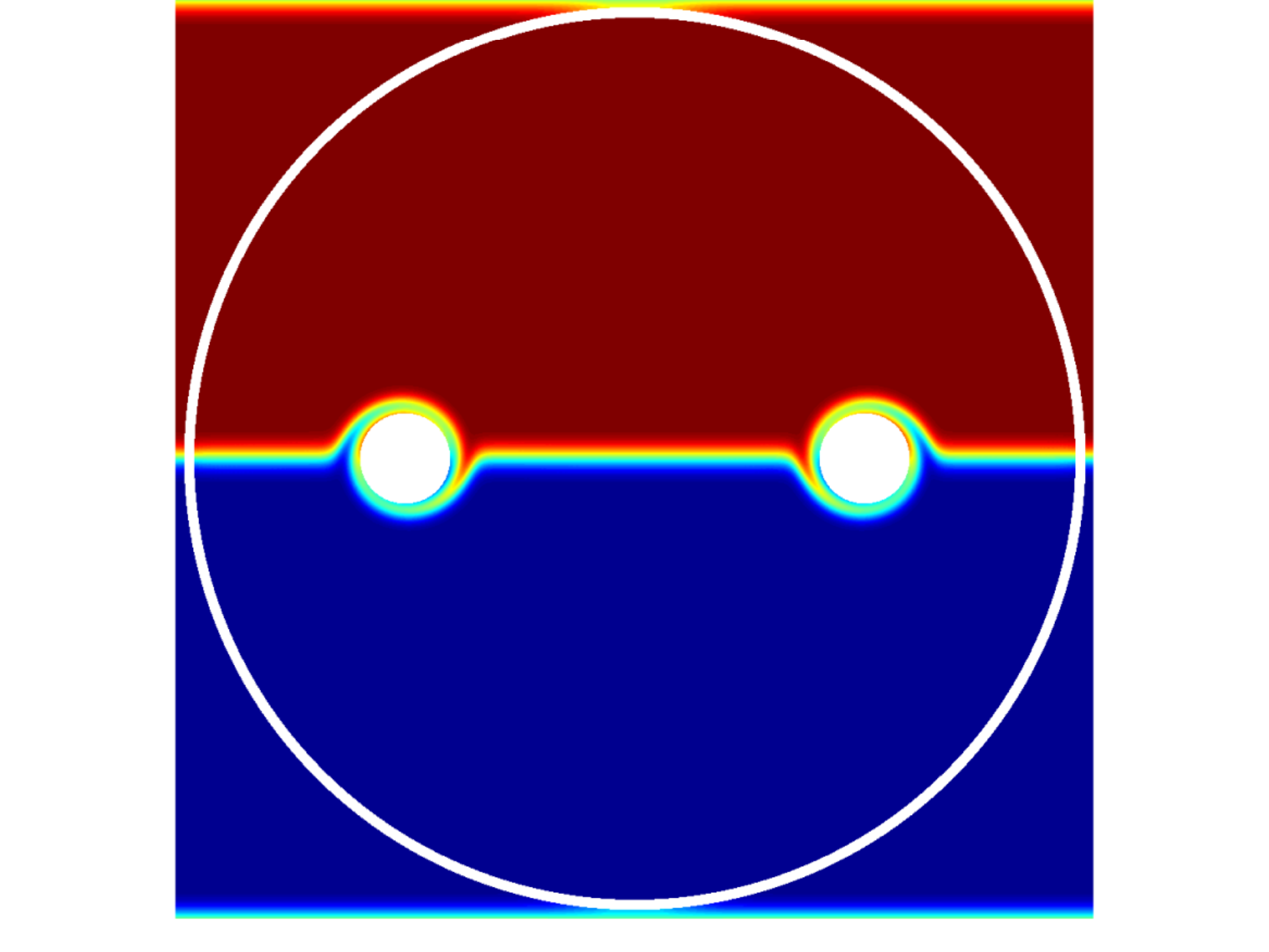} &
    \includegraphics[width=0.35\textwidth]{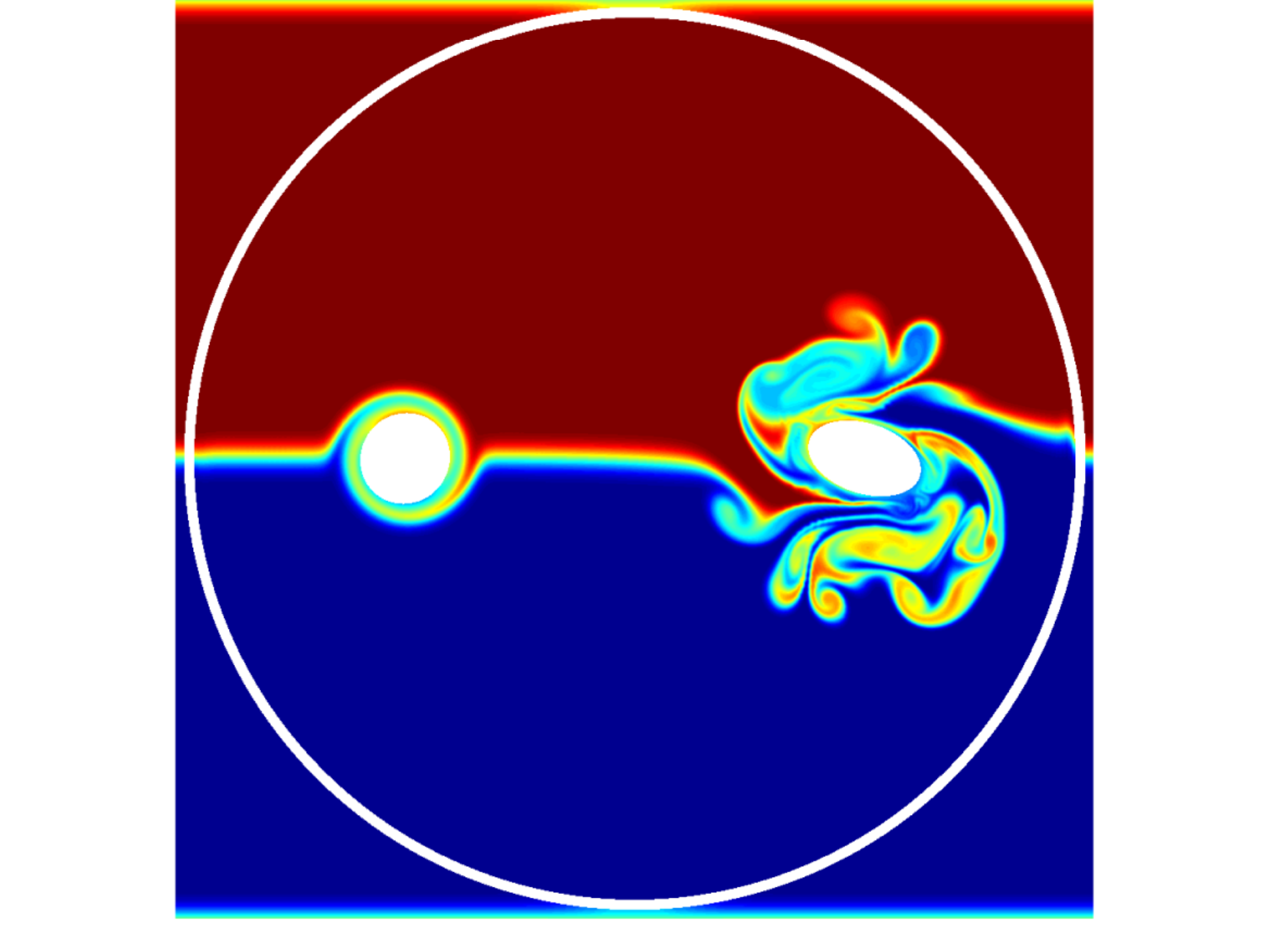} &
    \includegraphics[width=0.35\textwidth]{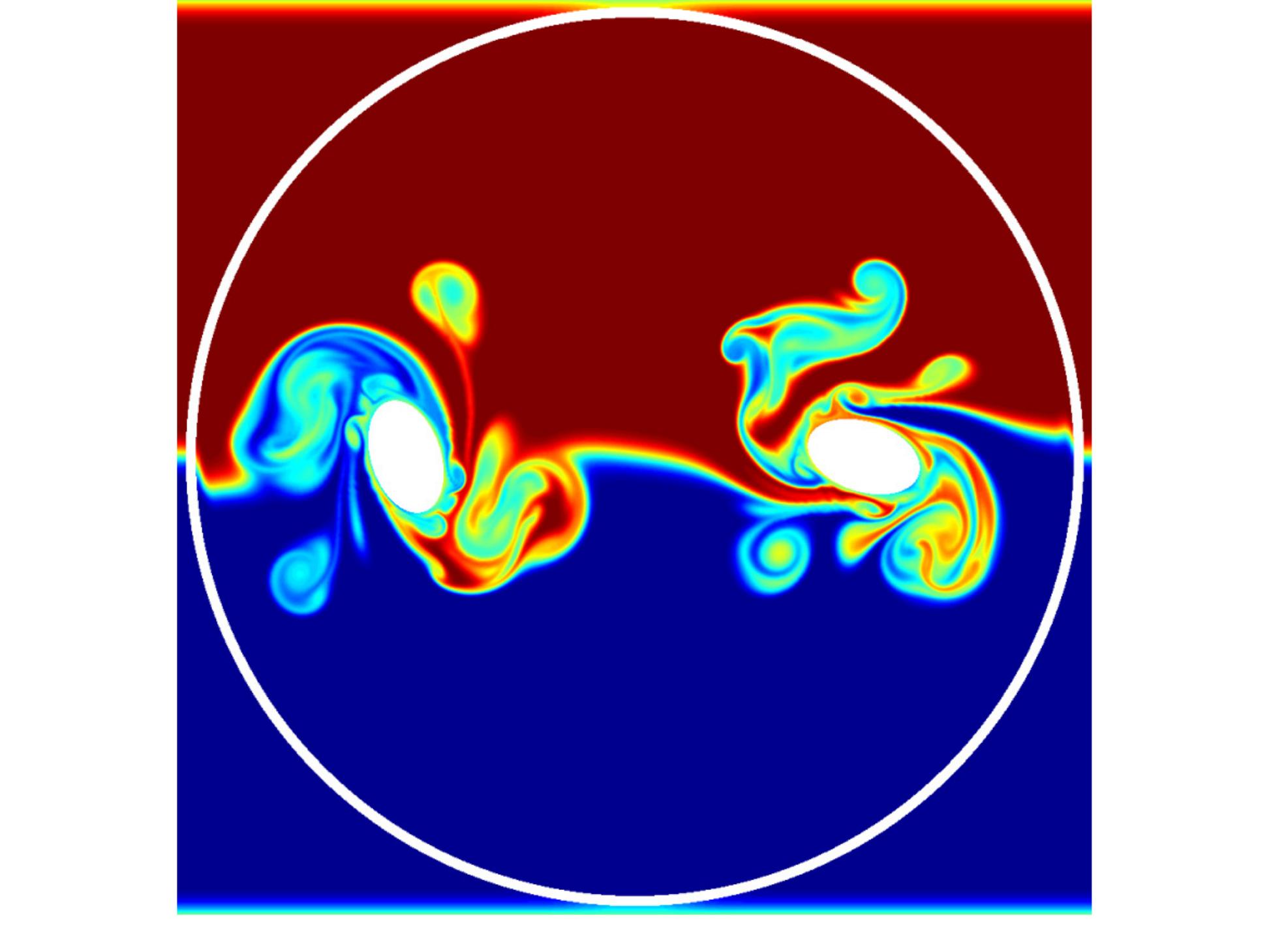} \\
    \includegraphics[width=0.35\textwidth]{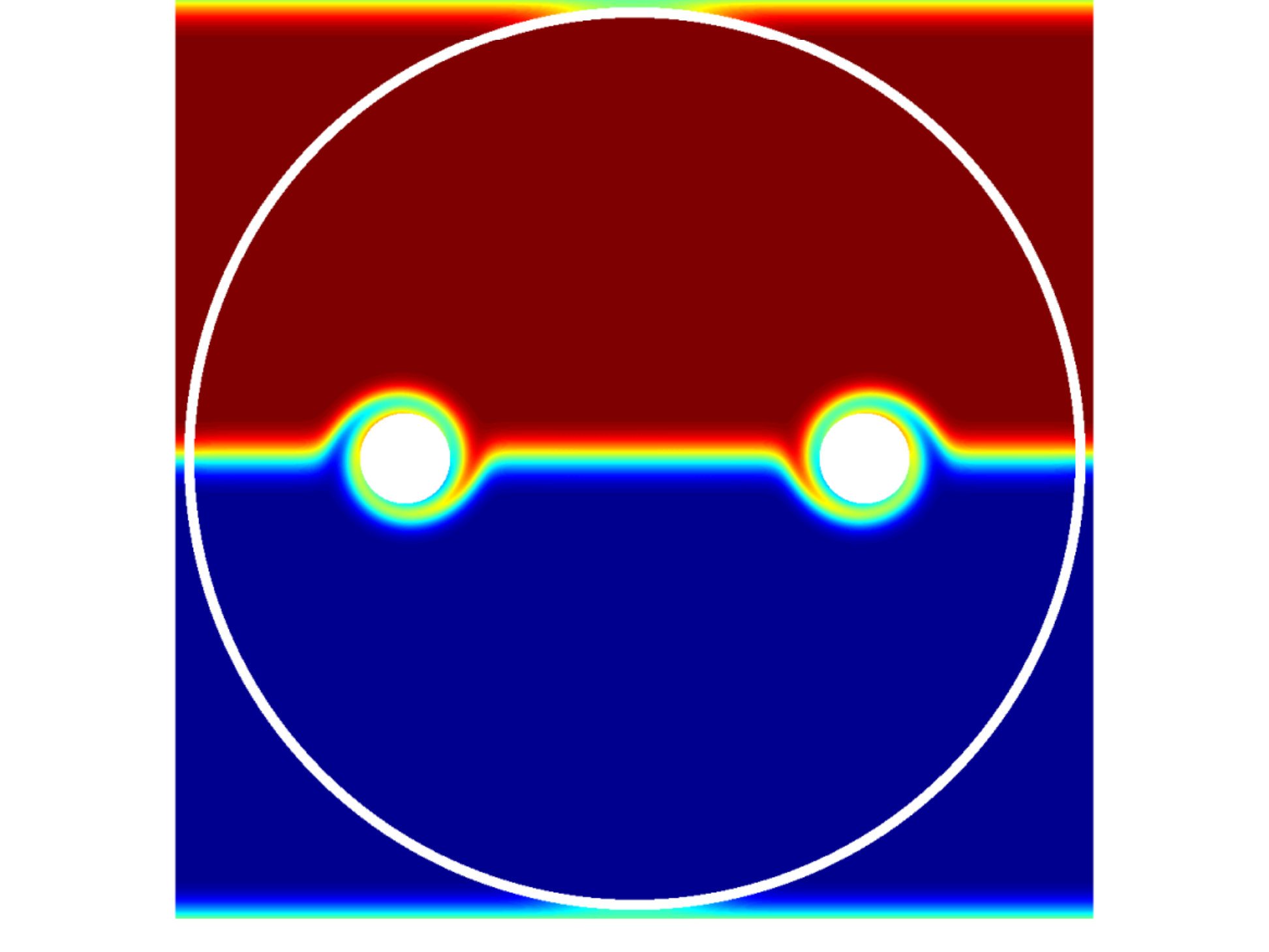} &
    \includegraphics[width=0.35\textwidth]{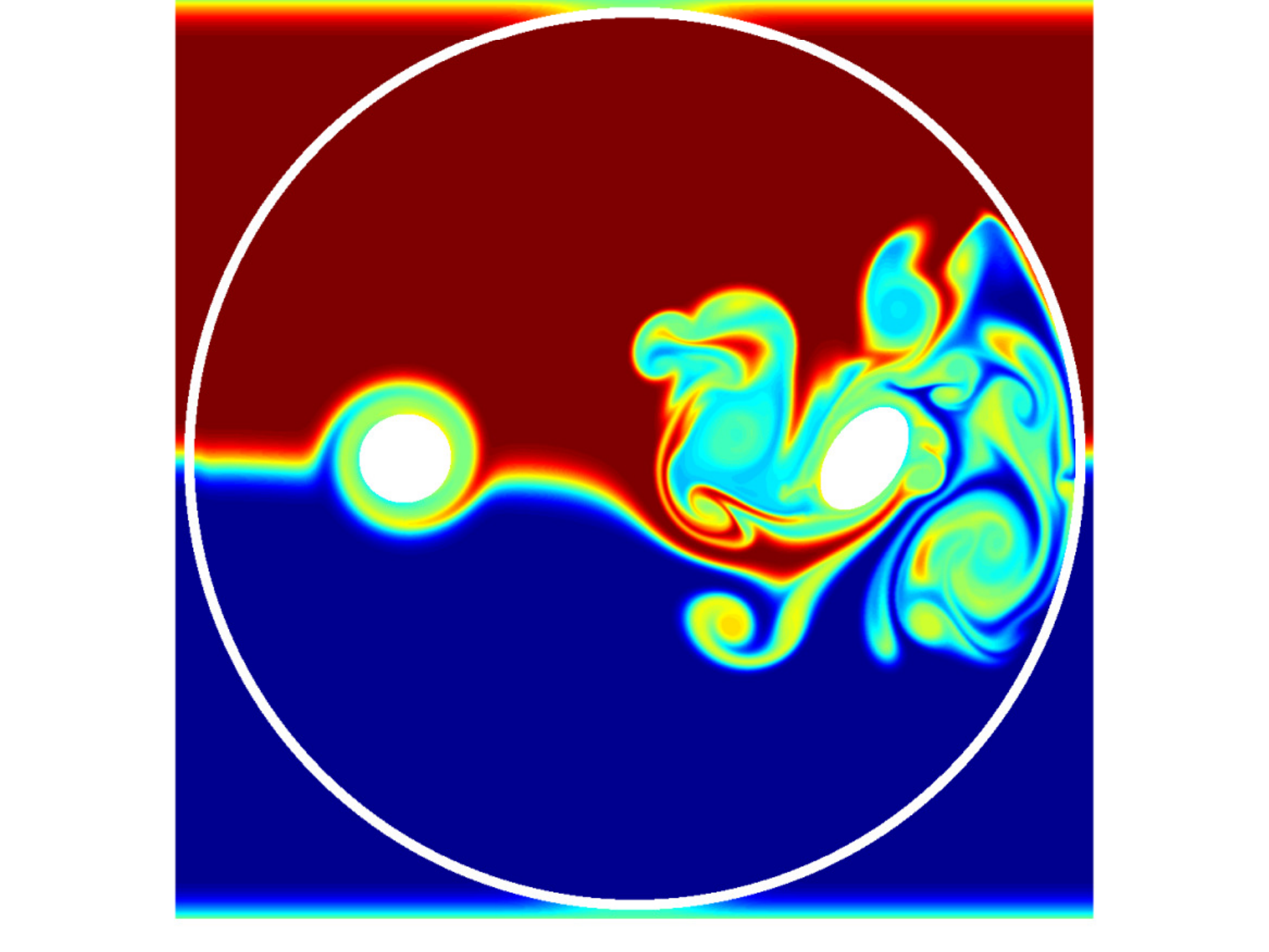} &
    \includegraphics[width=0.35\textwidth]{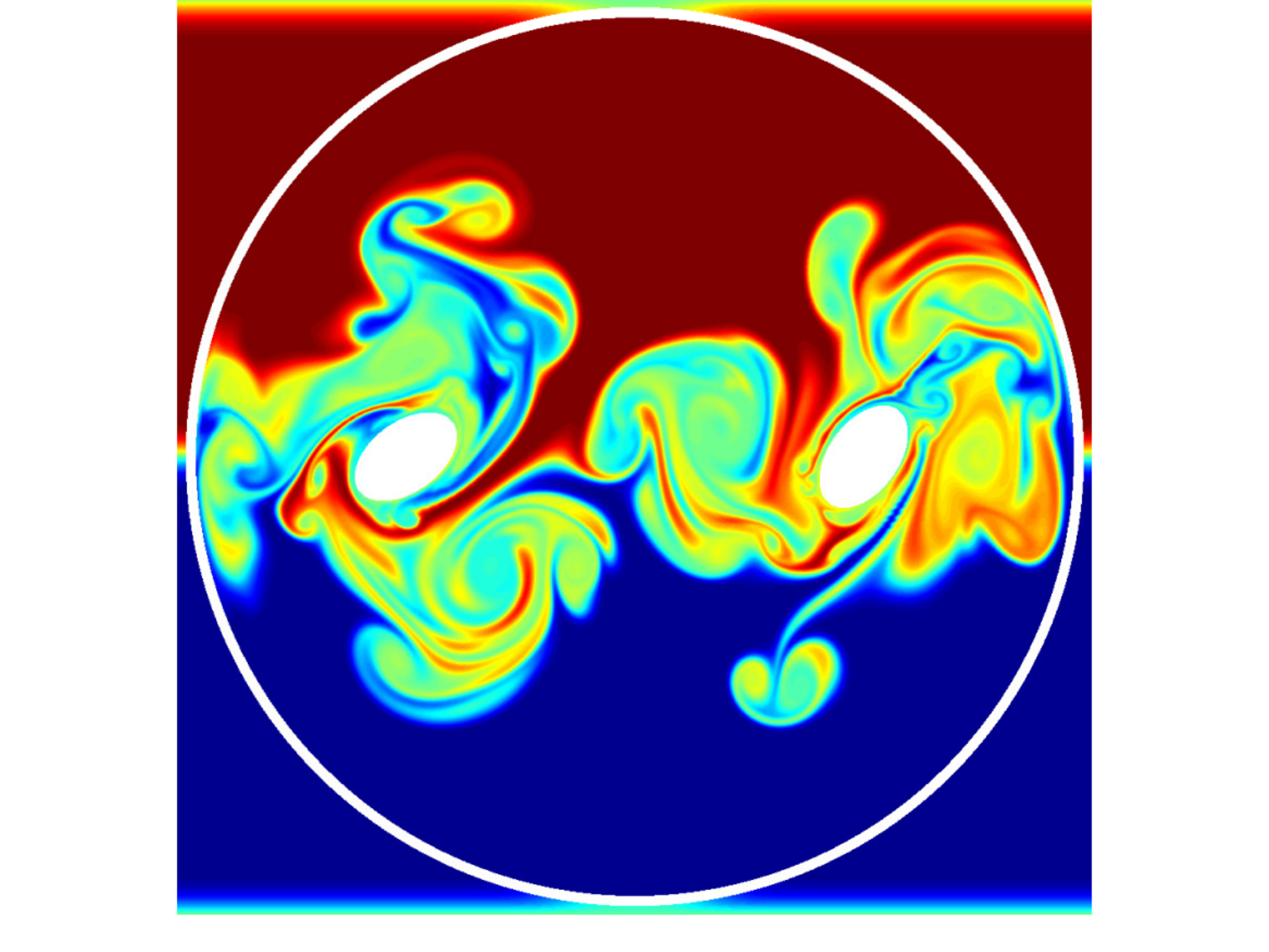} \\
    \includegraphics[width=0.35\textwidth]{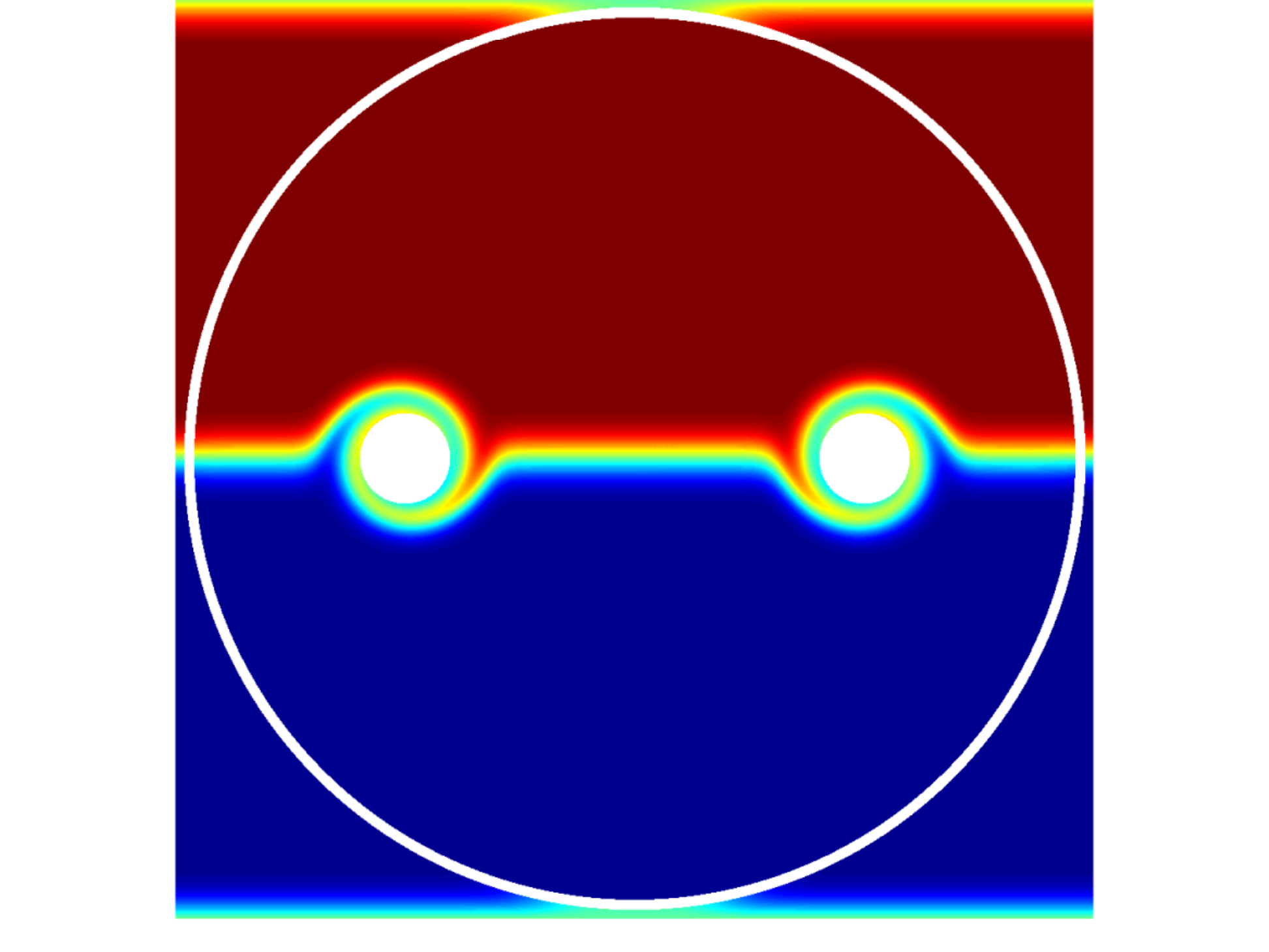} &
    \includegraphics[width=0.35\textwidth]{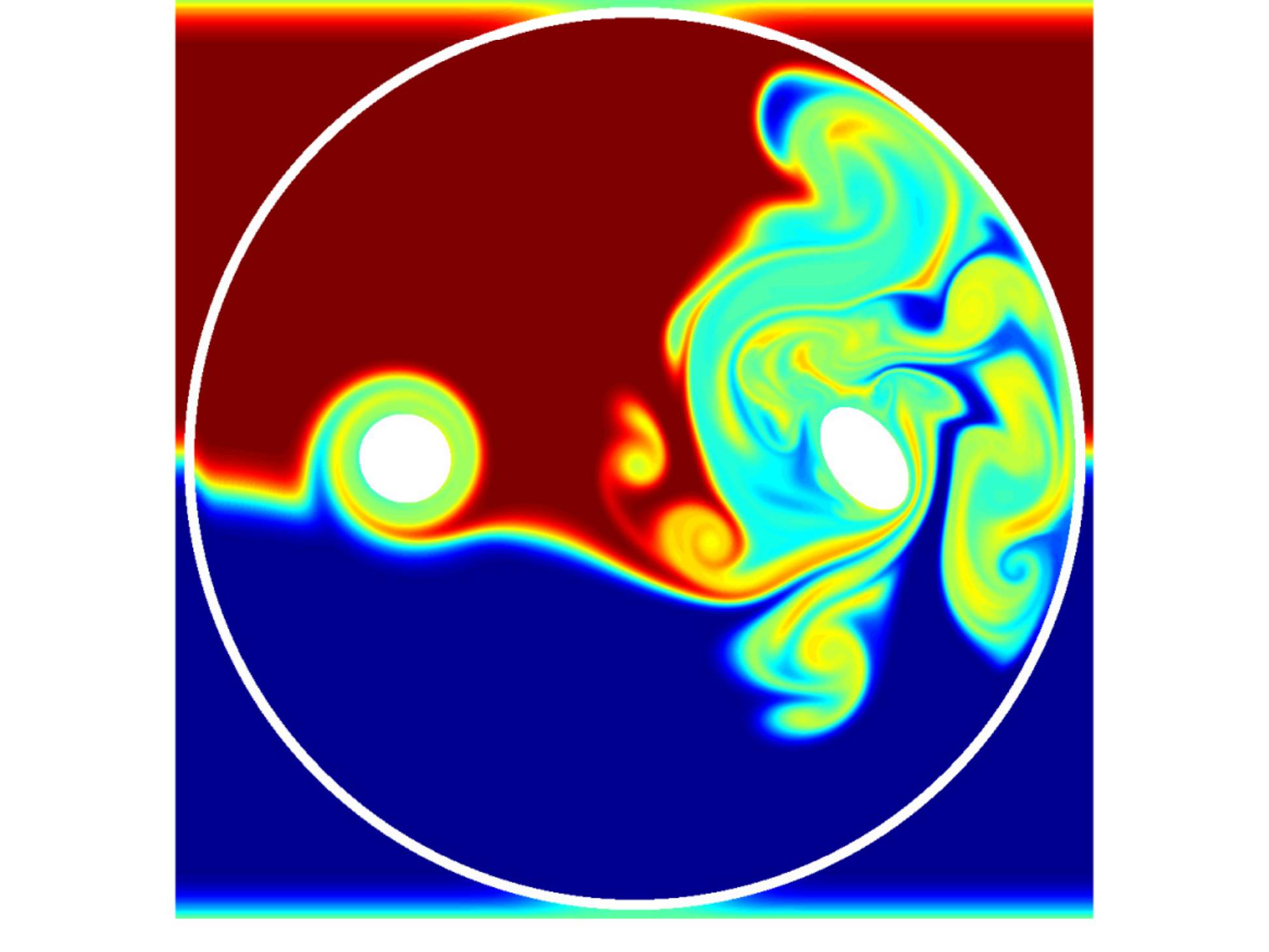} &
    \includegraphics[width=0.35\textwidth]{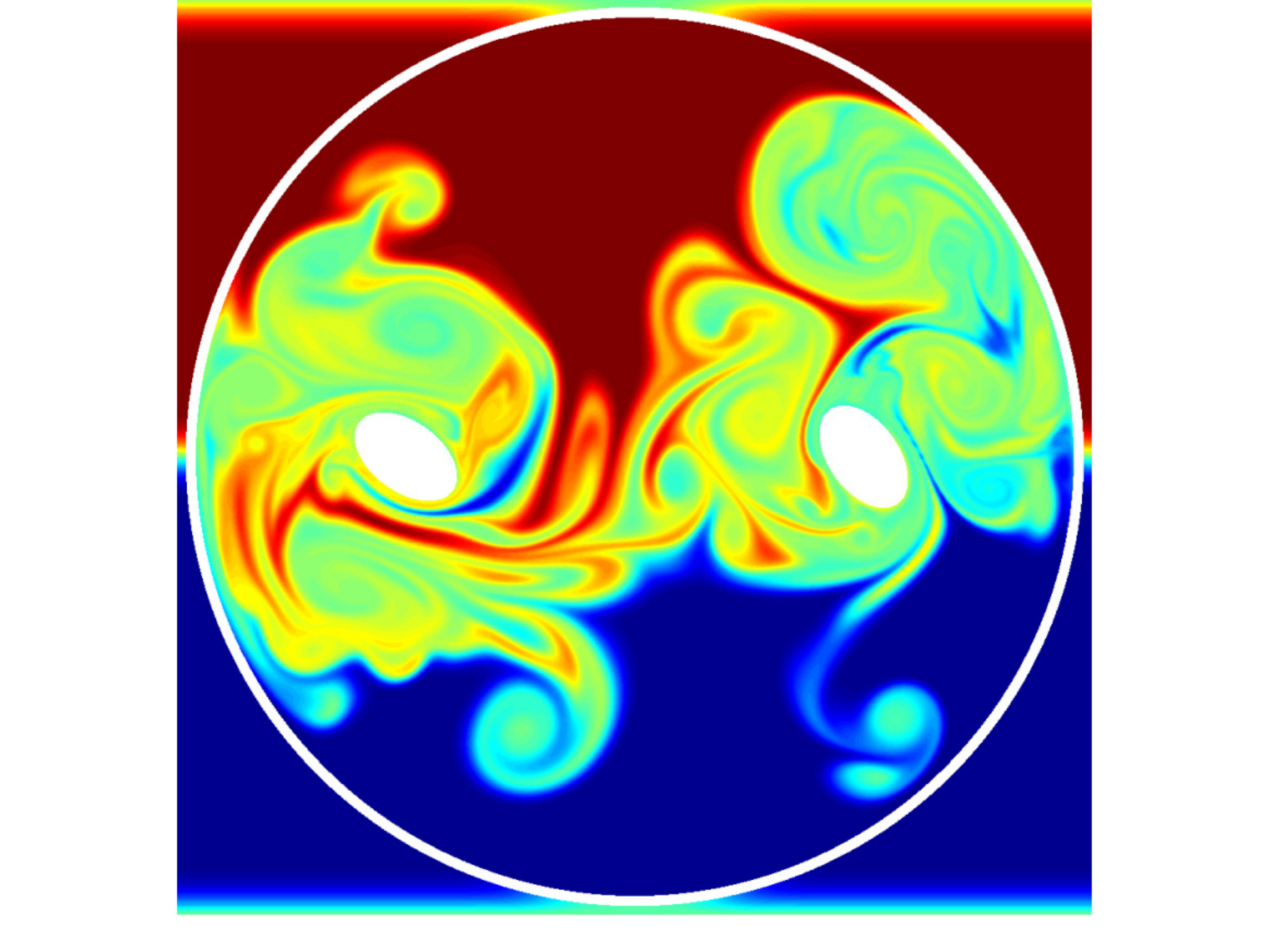} \\
    \includegraphics[width=0.35\textwidth]{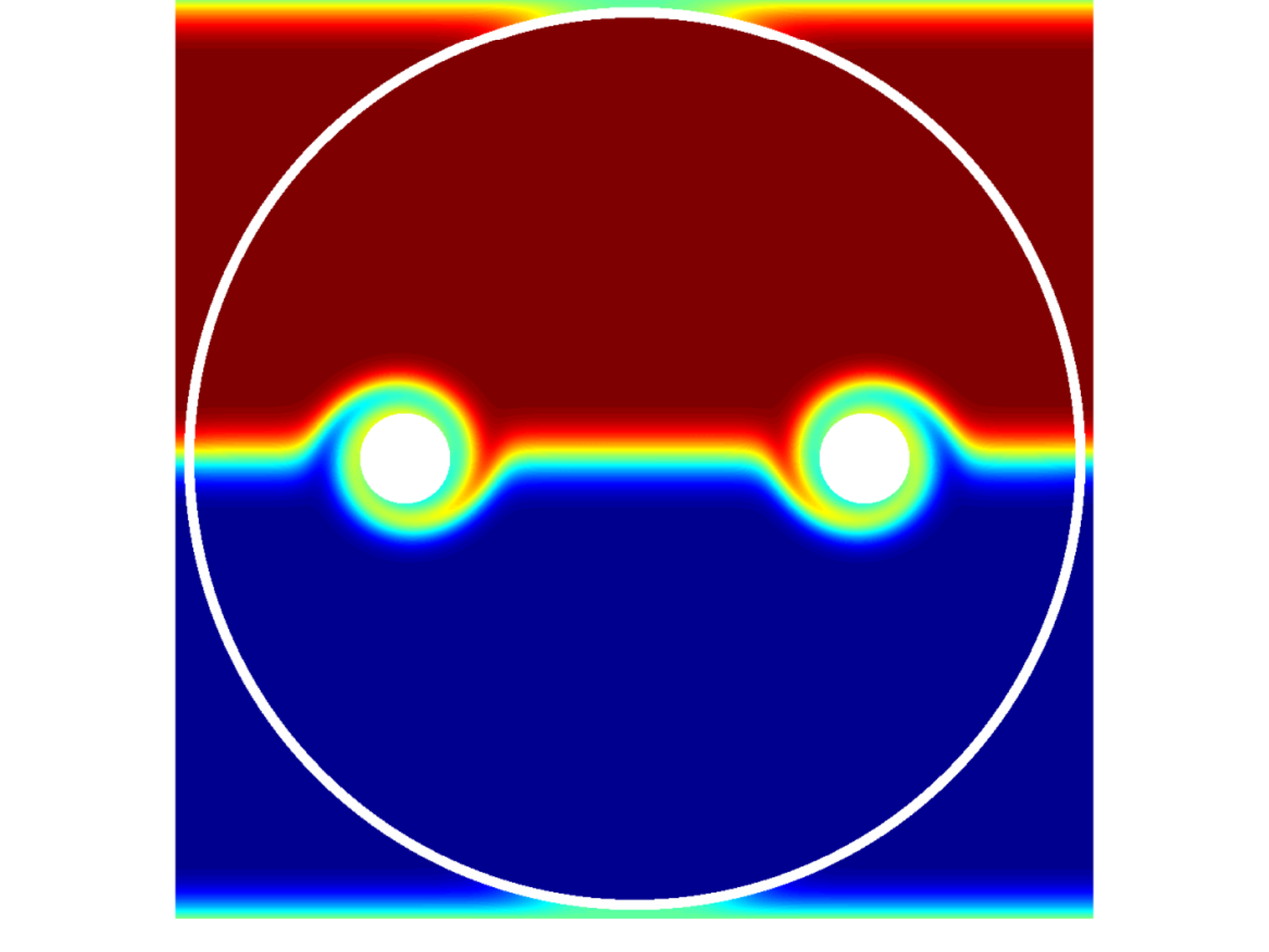} &
    \includegraphics[width=0.35\textwidth]{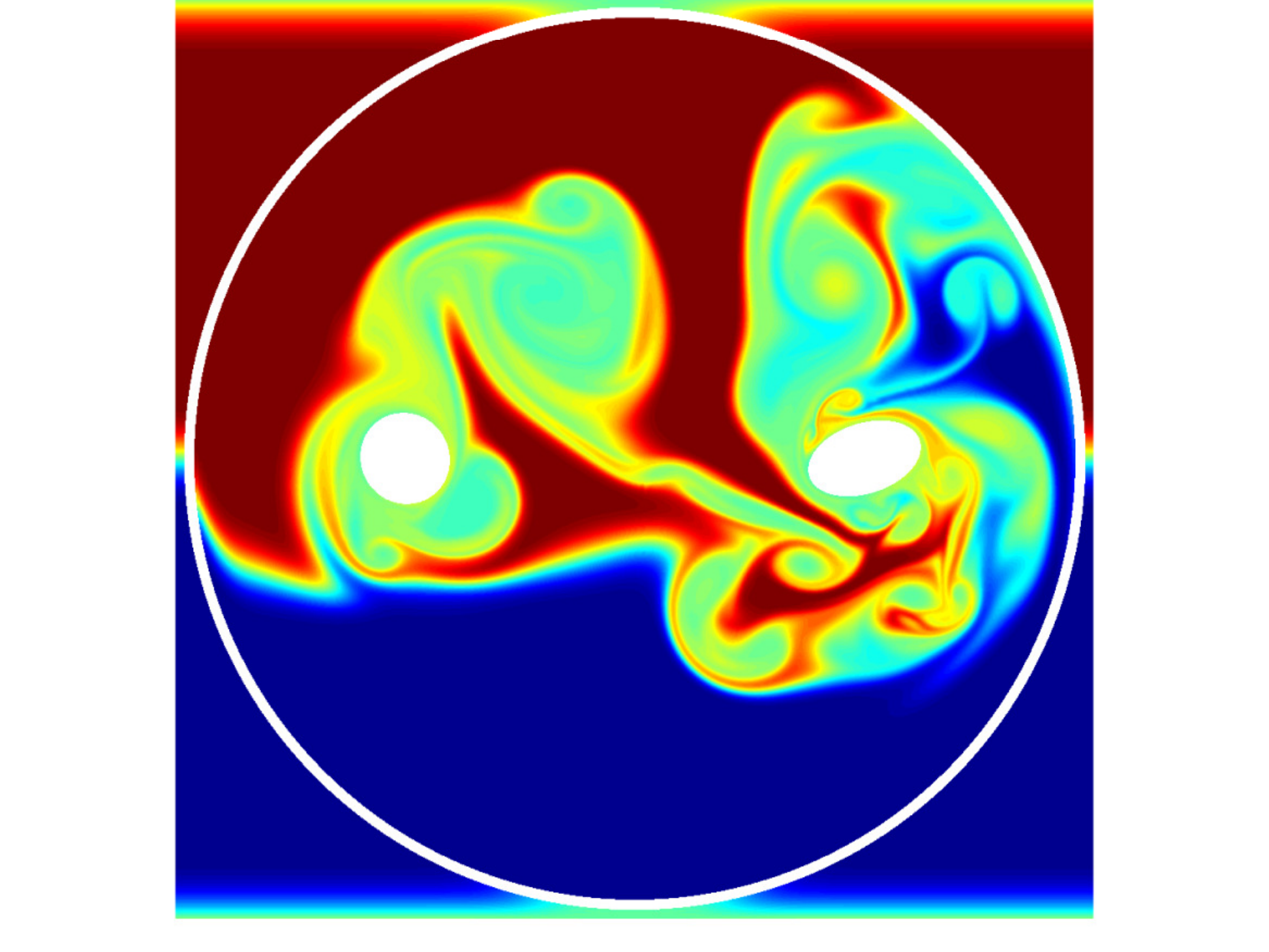}&
    \includegraphics[width=0.35\textwidth]{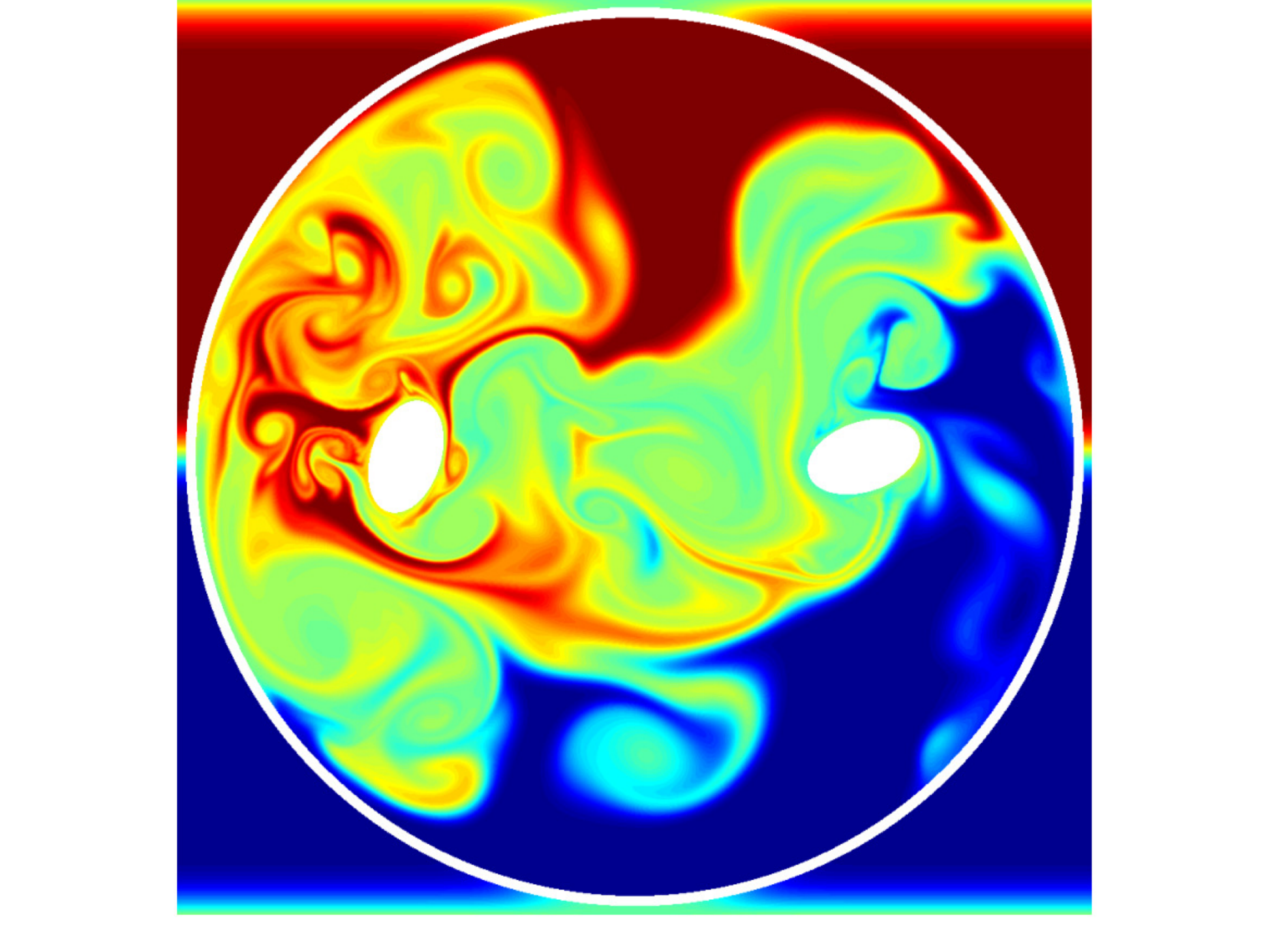} \\
  \end{tabular}
  \end{center}
  \caption{\label{2CC} Case 2: mixing optimization using two
    stationary, rotating stirrers. Left column: unoptimized
    configuration, with snapshots at $t = 8, 16, 24, 32$ (top to
    bottom). Middle column: after seven direct-adjoint optimizations,
    with snapshots at $t = 8, 16, 24, 32$ (top to bottom). Right
    column: Enforced minimum by mirroring the axis length across both
    cylinders at $t = 8, 16, 24, 32$. For videos of these scenarios
    please refer to {\tt{2Before.mp4}}, {\tt{2After.mp4}} and
    {\tt{2ImprovedMin.mp4}} for the left, middle and right column,
    respectively.}
\end{figure}

While this behavior is certainly prevalent (as shown in
figure~\ref{2CCVarPic}), when considering snapshots in time after the
seventh iteration of the direct-adjoint optimization we observe that
while the right stirrer has been optimized into a fast-rotating,
elliptical shape (as anticipated), the left cylinder is still nearly
circular in nature and appears to only mix by diffusing its rotating
boundary layer (see the right column of figure~\ref{2CC}). In fact,
the bulk of the variance drop can be ascribed to the right stirrer.

This observation highlights an issue and shortcoming of gradient-based
optimization. In order for the left stirrer to ``engage'' in the
mixing process, it has to pass through a (locally) less optimal
configuration. In other words, we have to first allow an increase in
variance contribution from the left stirrer, before we can
substantially lower the global variance by having the left stirrer
contribute to the overall mixing process. Within our optimization
framework, where after every iteration we proceed along the local
gradient, this required procedure is excluded. However, this is a
well-known and acknowledged issue of gradient-based optimization
schemes: we are able to find local minima, but have no guarantee (or
strategy) to find a global optimum.

In our case, we can improve the situation by mirroring the control
variables of the second (right) stirrer onto the first (left)
stirrer. In this manner, we induce sufficient gradients in the cost
functional for {\it{both}} cylinders to encourage further progress in
the reduction of the global variance. In this manner, we achieve lower
variance levels than before (see the dashed line in
figure~\ref{2CCVarPic}). The final shape of the stirrers for the
improved strategy consists of two ellipses of marked eccentricity,
with their respective semi-major axes positioned $90^\circ$ to each
other (the behavior of this improved minimum can be seen in the video
{\tt{2ImprovedMin.mp4}}).

For a more objective manner of reaching a global optimum,
sophisticated optimization strategies have to be employed in addition
to the gradient-based framework; these strategies require the
(commonly stochastic) evaluation of various parameter settings and
quickly become prohibitively expensive for large-scale, PDE-based
optimizations.

\newgeometry{left=1cm}

\restoregeometry

\subsection{Case 3: five stationary, rotating stirrers}

Motivated by the previous configuration, we proceed by adding more
stirrers and further explore the behavior of the optimization scheme
when multiple optima and strategies compete for the best mixing
efficiency.

\subsubsection{Highly penalized system}

\begin{figure}
  \centering
  \begin{tabular}{cc}
    \includegraphics[width=0.5\textwidth]{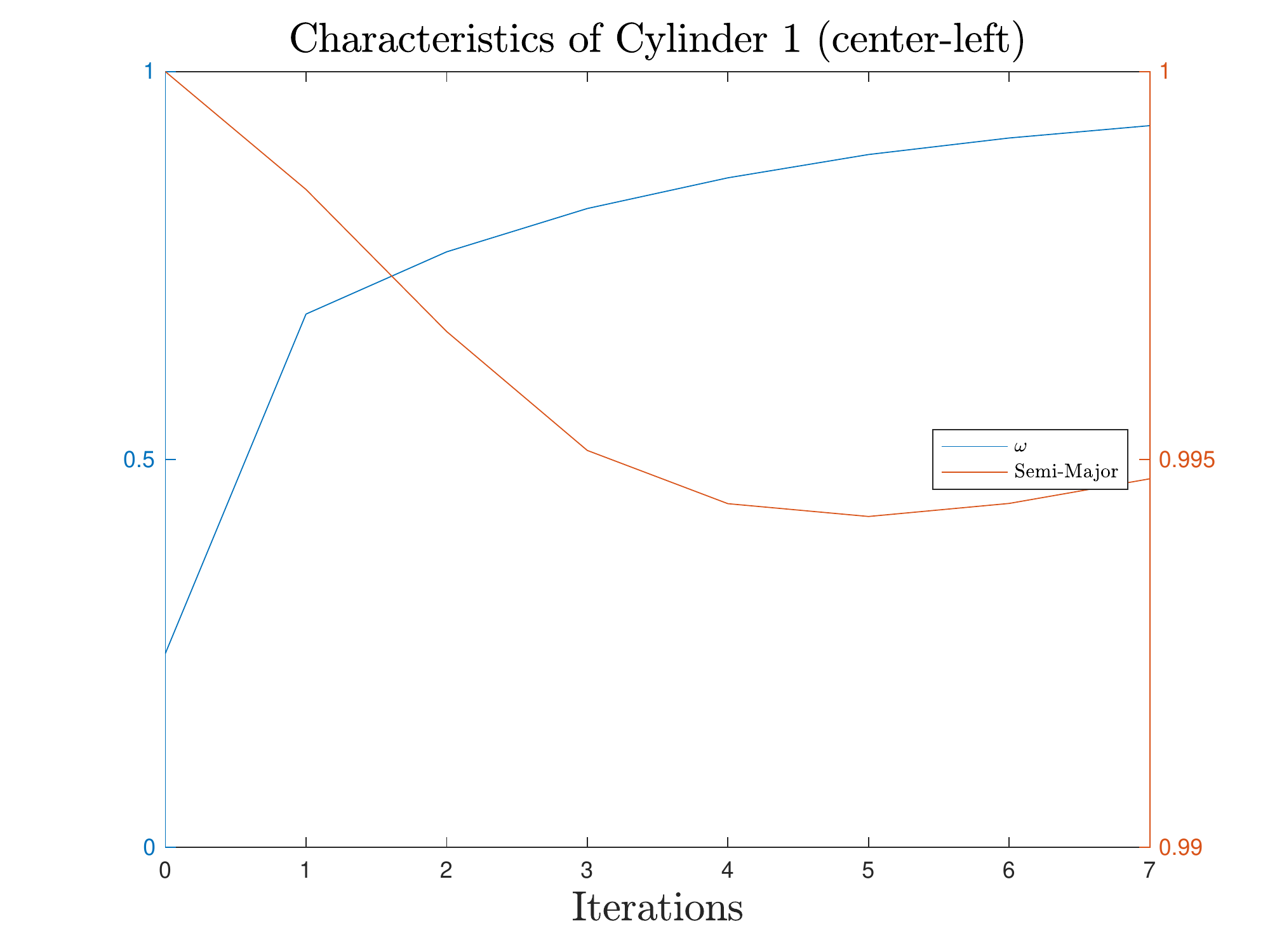} &
    \includegraphics[width=0.5\textwidth]{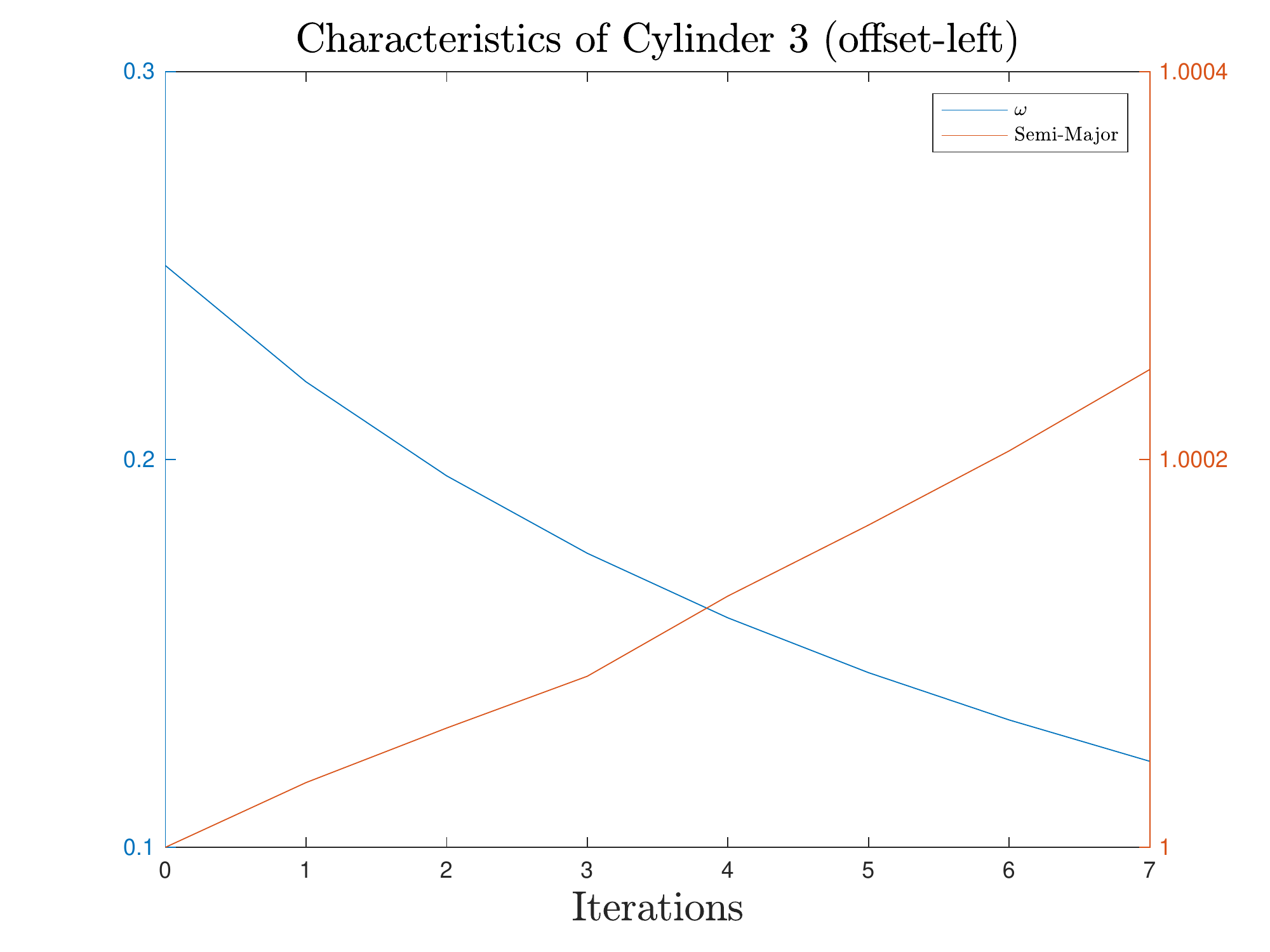} \\
    \includegraphics[width=0.5\textwidth]{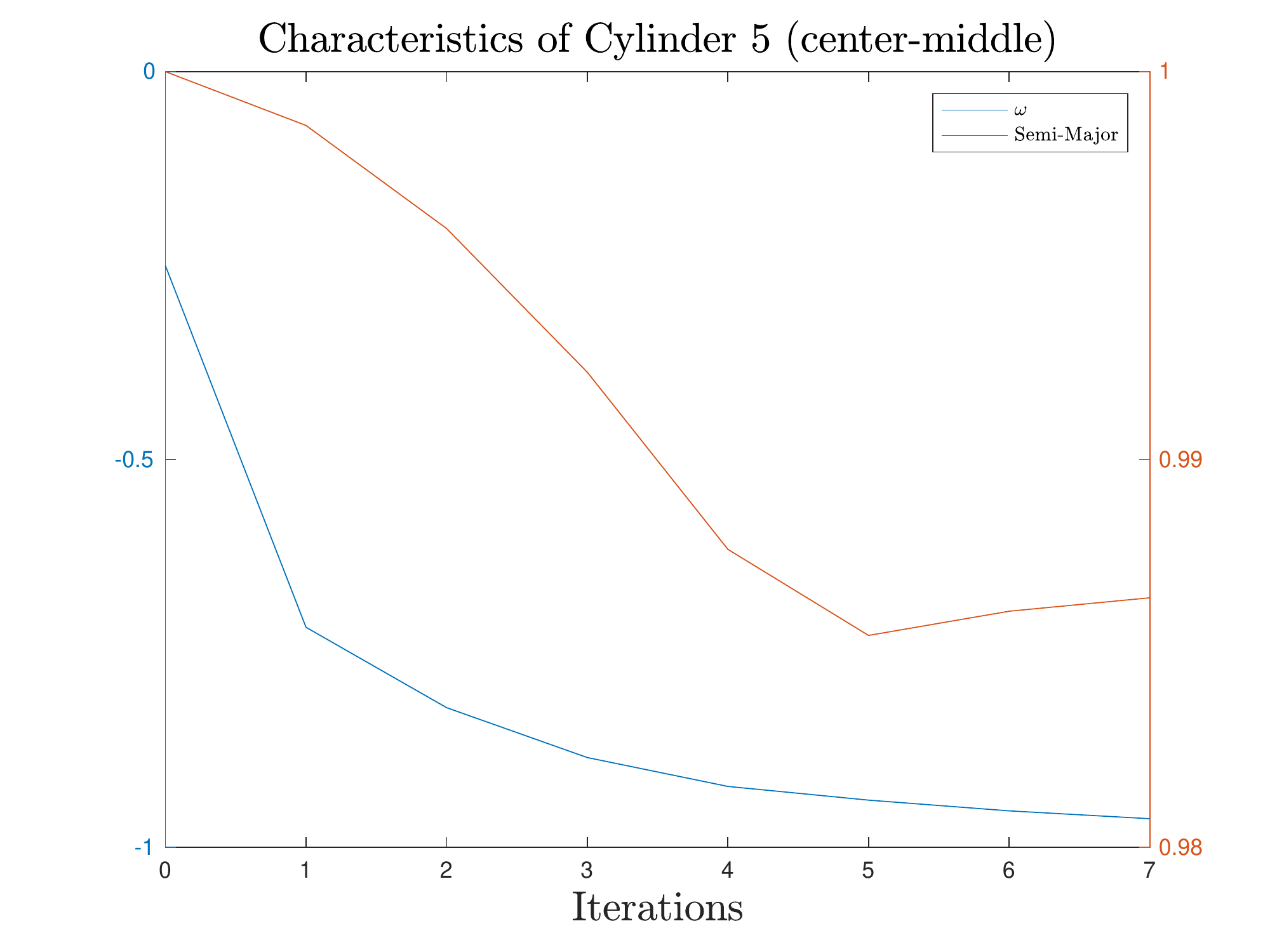} &
    \includegraphics[width=0.5\textwidth]{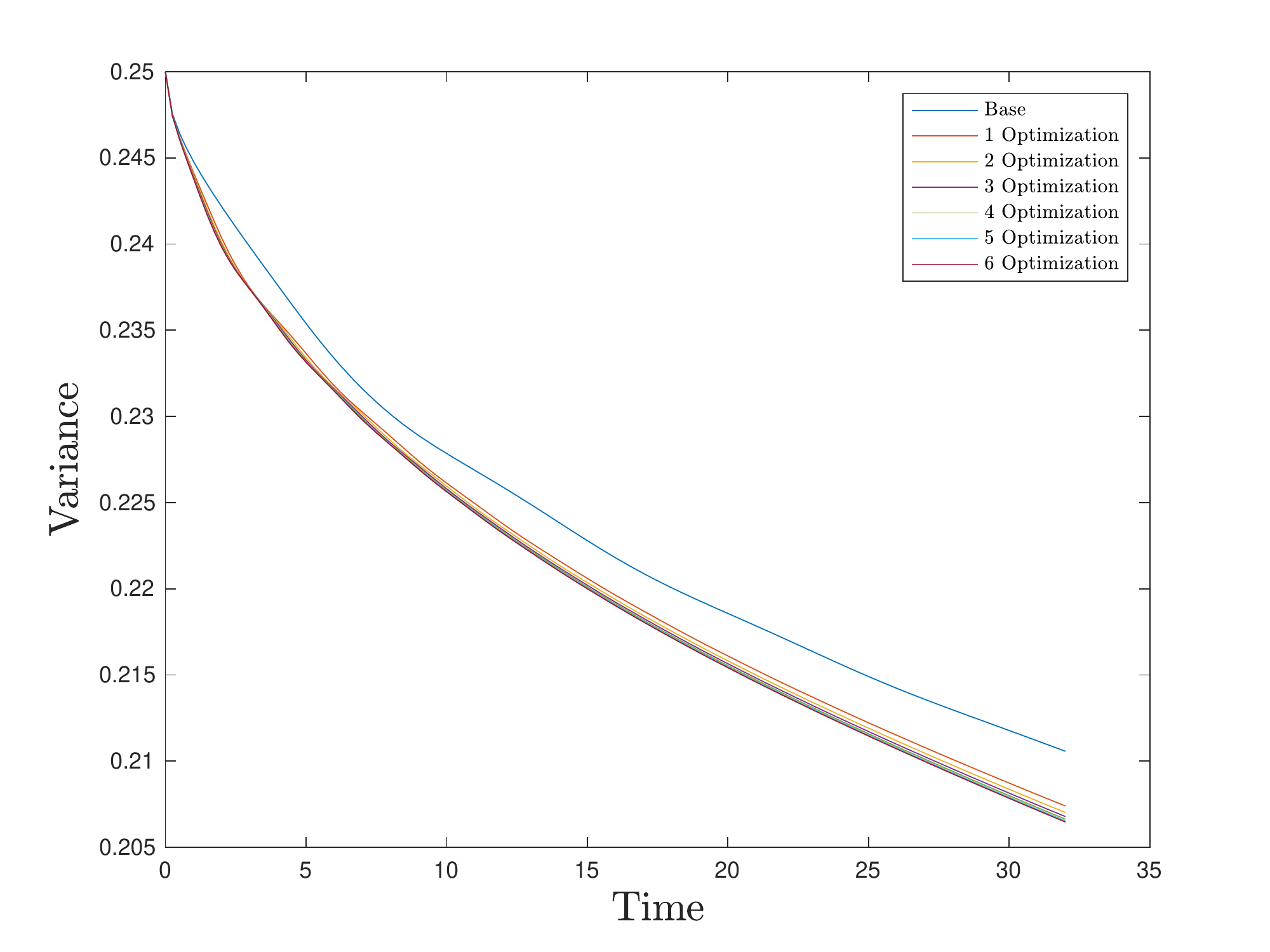}
  \end{tabular}
  \caption{\label{5CCHPResults} Case 3: mixing optimization using five
    stationary, rotating stirrers. A highly penalized optimization
    setting has been used. (a) Rotational speed $\omega$ and axis $a$
    versus the number of direct-adjoint iterations for the first
    (left-most) cylinder. (b) Rotational speed $\omega$ and axis $a$
    versus the number of direct-adjoint iterations for the third
    (bottom) cylinder. (c) Rotational speed $\omega$ and axis $a$
    versus the number of direct-adjoint iterations for the fifth
    (center) cylinder. (d) Variance of the passive scalar versus time
    $t \in [0,\ T^{F}].$}
\end{figure}

To study the behavior of the direct-adjoint optimization scheme on
this more complicated geometry, we first impose a high energy
penalization, as before. The results are displayed in
figure~\ref{5CCHPResults}, for the stirrers labelled 1,3 and 5 (see
\ref{GeometriesPic}). While the two stirrers on the horizontal axis
show behavior similar to the previous cases (i.e., a tendency towards
higher rotational speed and an elliptic shape), eccentricity remains
largely unchanged, while the rotational velocity is significantly
dampened.  At first sight, this may run counter to intuition that
suggests that higher speeds result in improved mixing. However, it
appears that -- under the constraints of a limited energy budget -- it
is more advantageous to invest input energy into the aligned stirrers
rather than squandering it on the offset stirrer that is located in a
rather homogeneous tracer field and thus cannot contributed to the
global variance drop to any significant degree. For this reason, the
optimization scheme (more specifically, the adjoint system) directs
focus on the three aligned stirrers that do make a difference.

Regarding the variances of the scalar field in figure~\ref{5CCVarPic}
we conclude, as before, that we do not fully utilize diffusion
{\it{and}} advection, which is required for efficient mixing. To
exhibit a significant decrease in the variance, we have to explore the
weak energy penalization regime.

\subsubsection{Weakly penalized system}

When removing the constraint of minimal energy expenditure, while
still limiting the total amount of energy distributed among the five
stirrers, we see the emergence of advection-based mixing. The five
stirrers take on elliptical shapes of varying eccentricity and tend to
increase in their spin rate. This process induces a complex system of
shed vortices that not only introduces small-scale features and thin
filaments, but also transports information between the five
stirrers. The ensuing optimization scheme then finds an optimal
collaborative mixing strategy between the five stirrers that optimizes
the global variance while remaining within the imposed energy
constraints.

\begin{figure}
  \centering
  \includegraphics[width=0.67\textwidth]{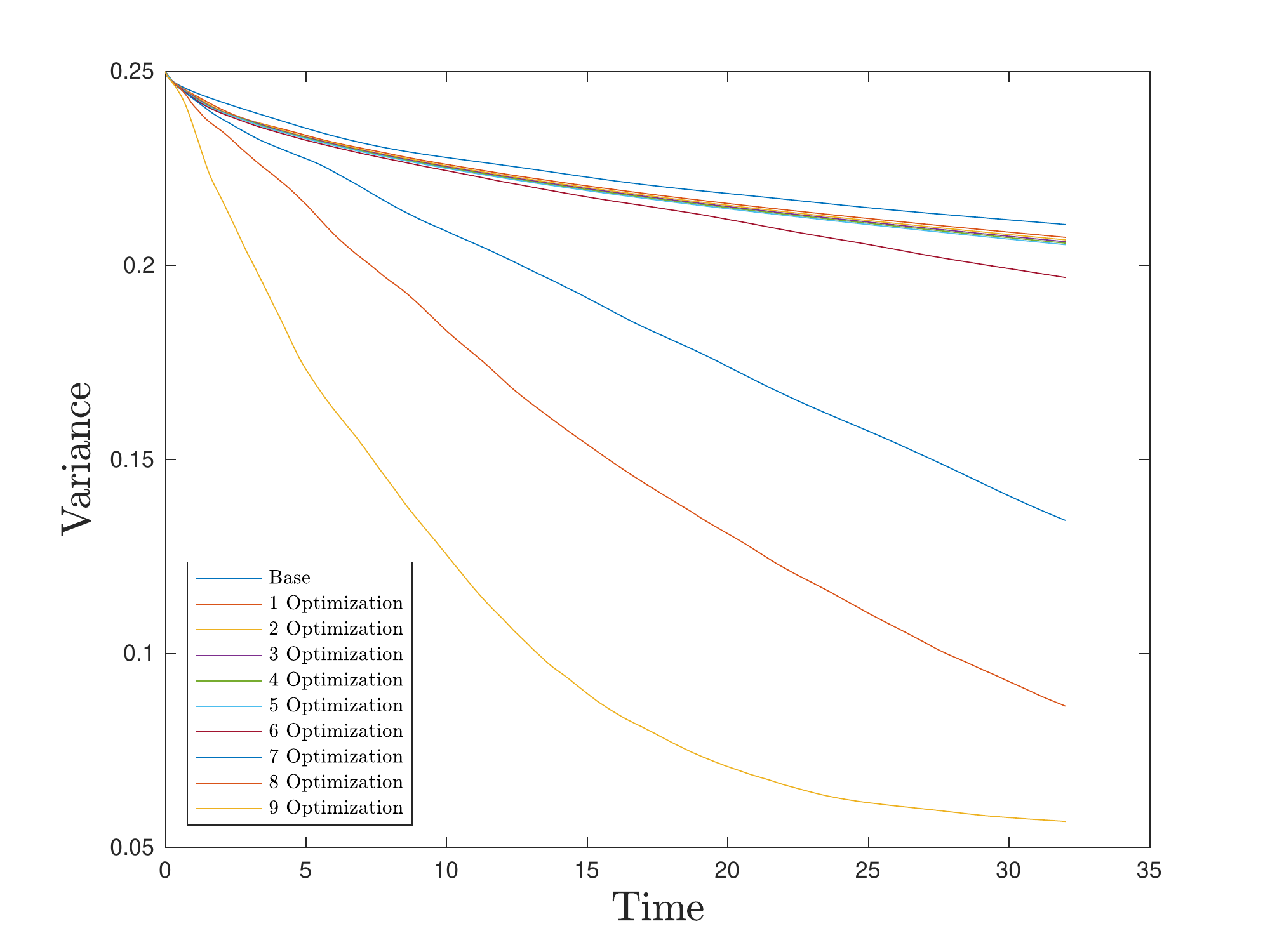}
  \caption{\label{5CCVarPic} Case 3: mixing optimization using five
    stationary, rotating stirrers. Variance, as defined in equation
    (\ref{eq:Variance}), of the scalar field $\theta$ versus time $t
    \in [0,\ T^{F}].$ }
\end{figure}

Throughout the initial iterations, no significant information is
exchanged between the cylinders, and therefore the off-set cylinders
are neglected due to the homogeneity of their surroundings. Even in
the low penalization the optimization regards any energy used by these
cylinders as wasteful.  During subsequent iterations the transport of
information is built up but remains negligible. In particular,
considering the variances in figure~\ref{5CCVarPic}, we note that the
initial five iterations do not lead to significant improvements in
mixedness, as we mainly stay within the previously mentioned
solid-body rotation. Once vortex shedding sets in (starting at the
sixth iteration, mainly with the central stirrer), however, we see a
significantly larger decrease in variance. Continuing further in the
optimization, all remaining stirrers are involved in the mixing
process, as a jet forms between the central cylinders which interacts
with the off-set cylinders implying that energy expended on their
rotation will have an effect on the mixing.  In particular, the
elliptical shape of the stirrers converges towards its final
configuration relative to each other.

Similar to the one-cylinder case, our final optimization reaches a
point where the predominant mixing process is diffusion and the
gradient of the variance is comparable to the purely diffusive limit.

\begin{figure}
  \centering
  \begin{tabular}{cc}
    \includegraphics[width=0.5\textwidth]{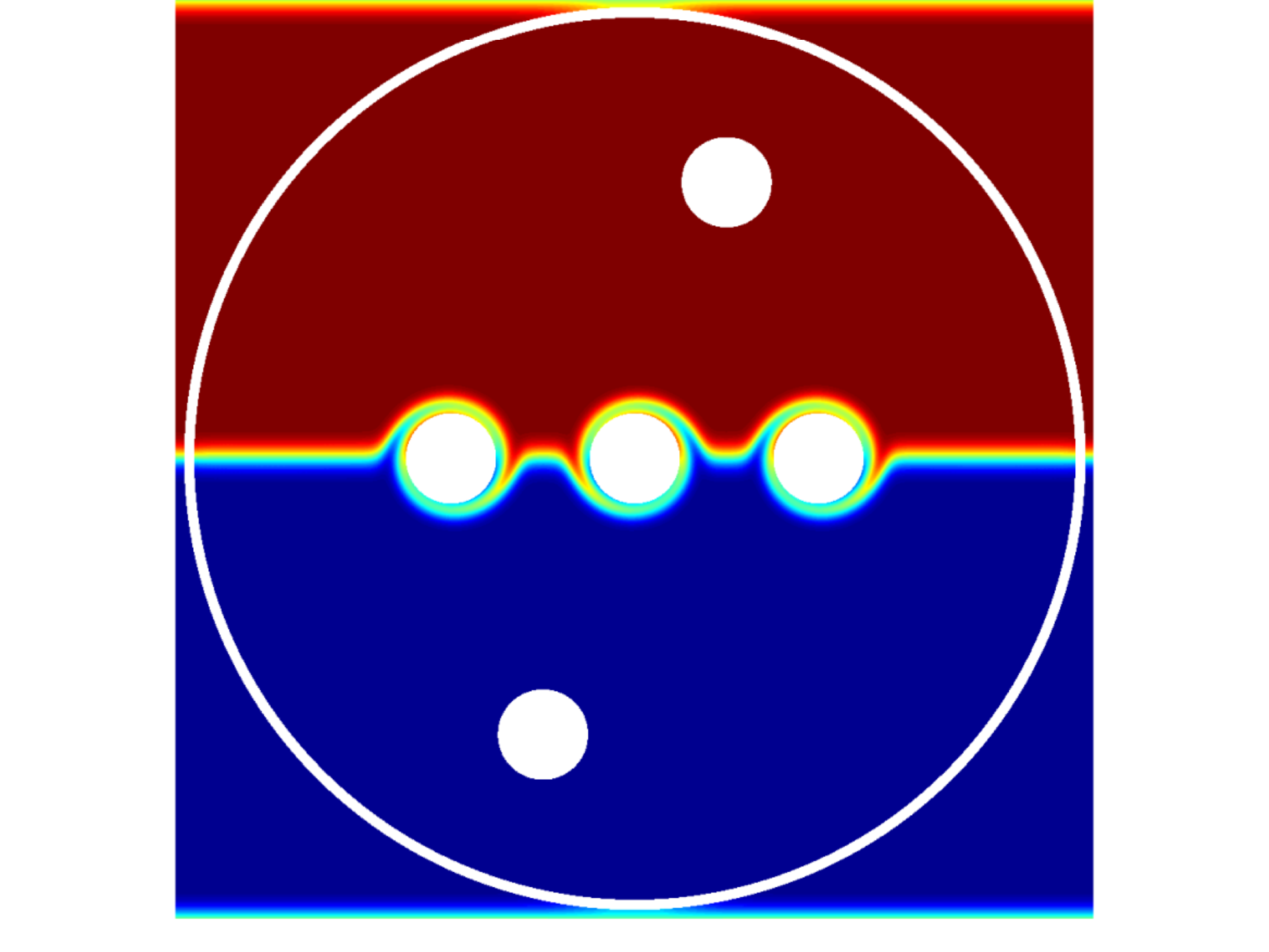} &
    \includegraphics[width=0.5\textwidth]{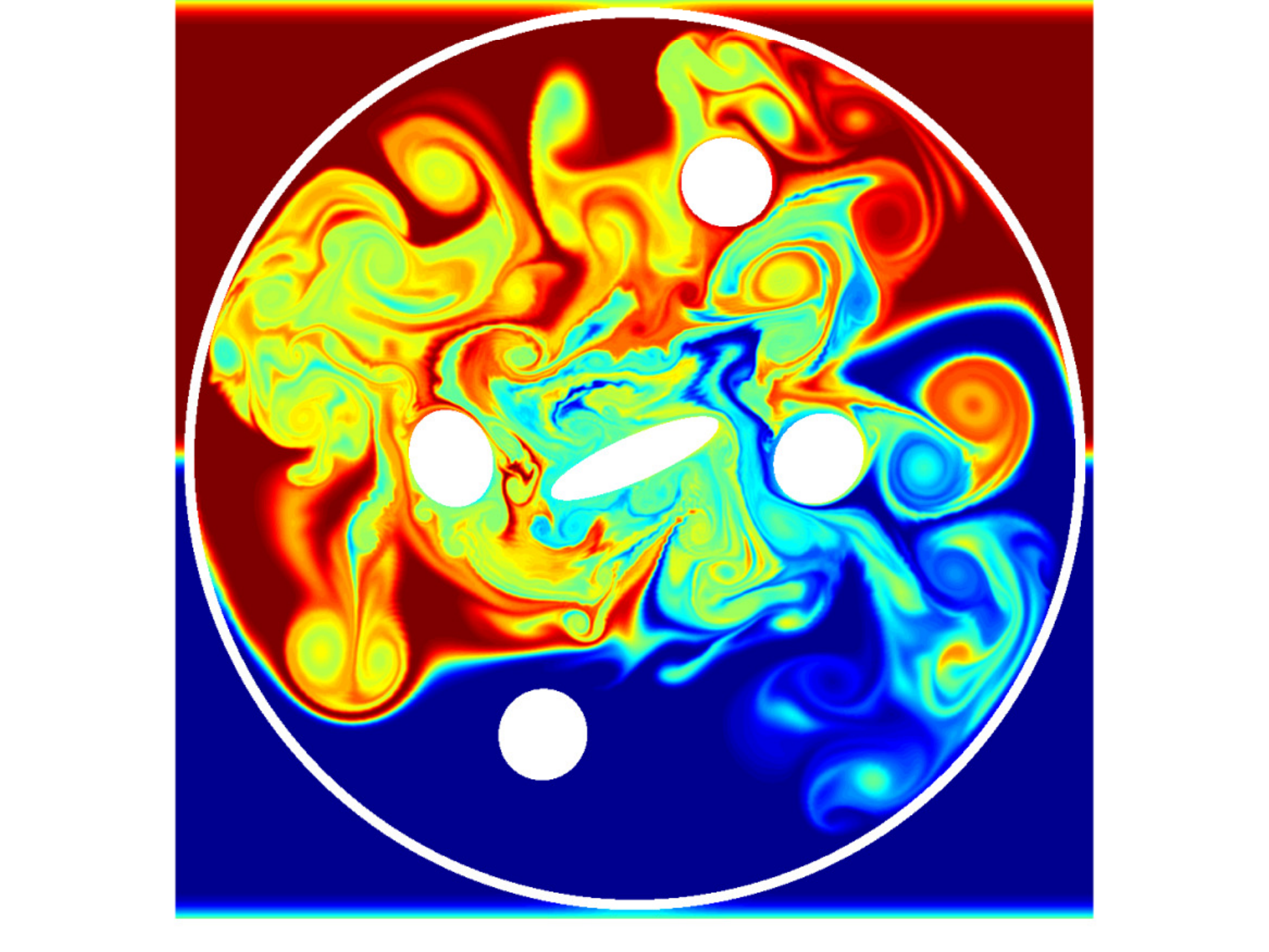} \\
    \includegraphics[width=0.5\textwidth]{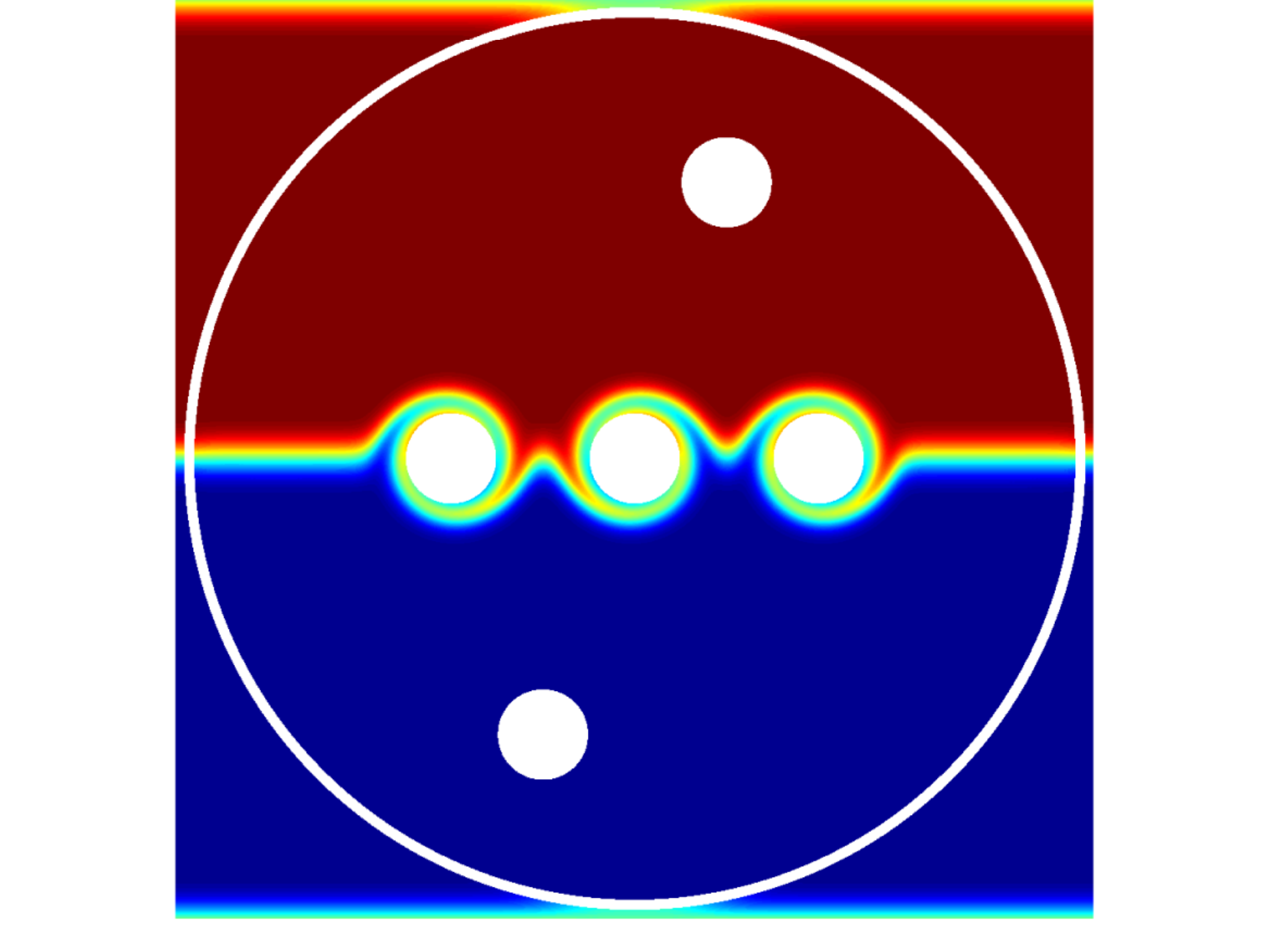} &
    \includegraphics[width=0.5\textwidth]{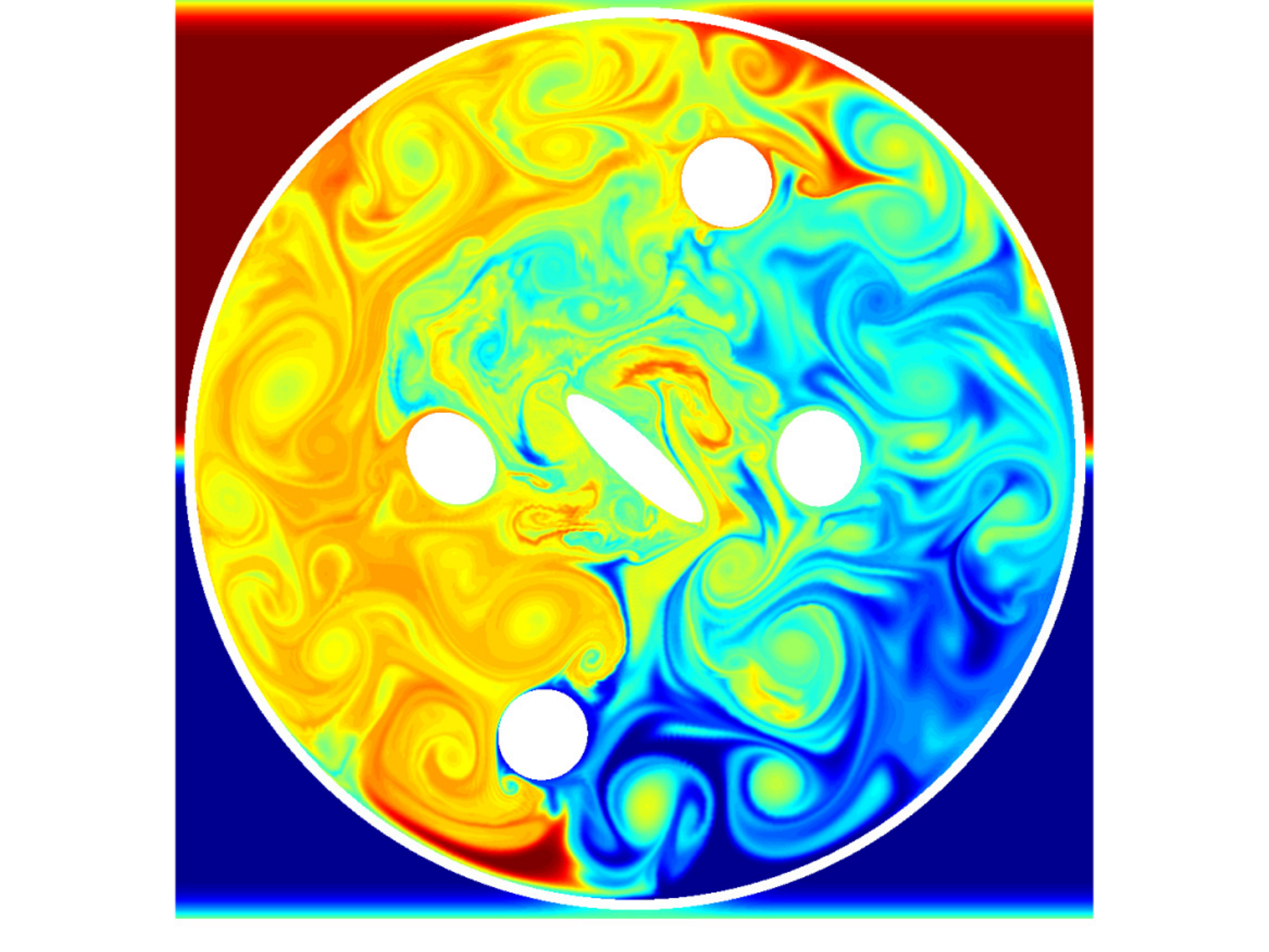} \\
    \includegraphics[width=0.5\textwidth]{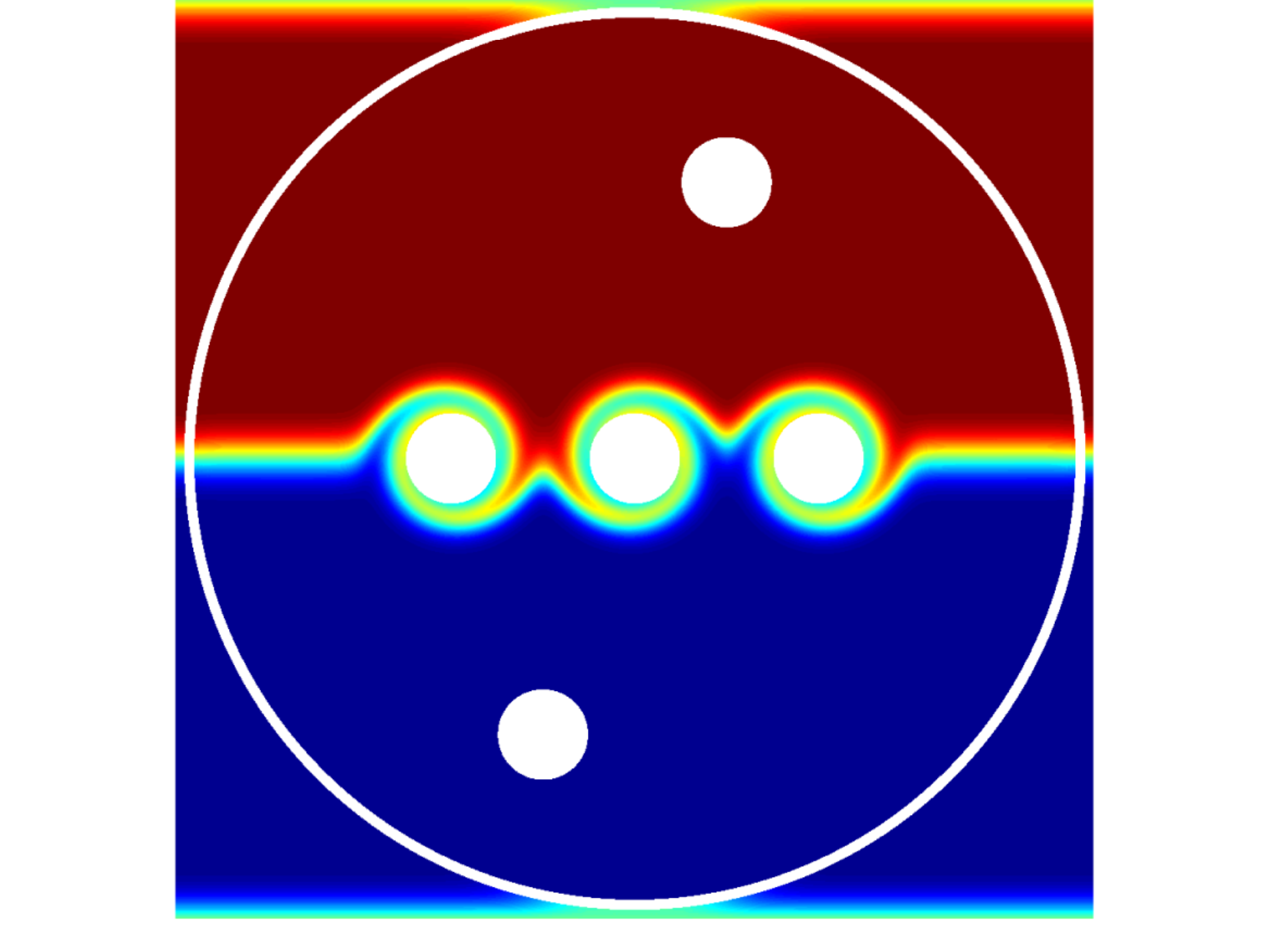} &
    \includegraphics[width=0.5\textwidth]{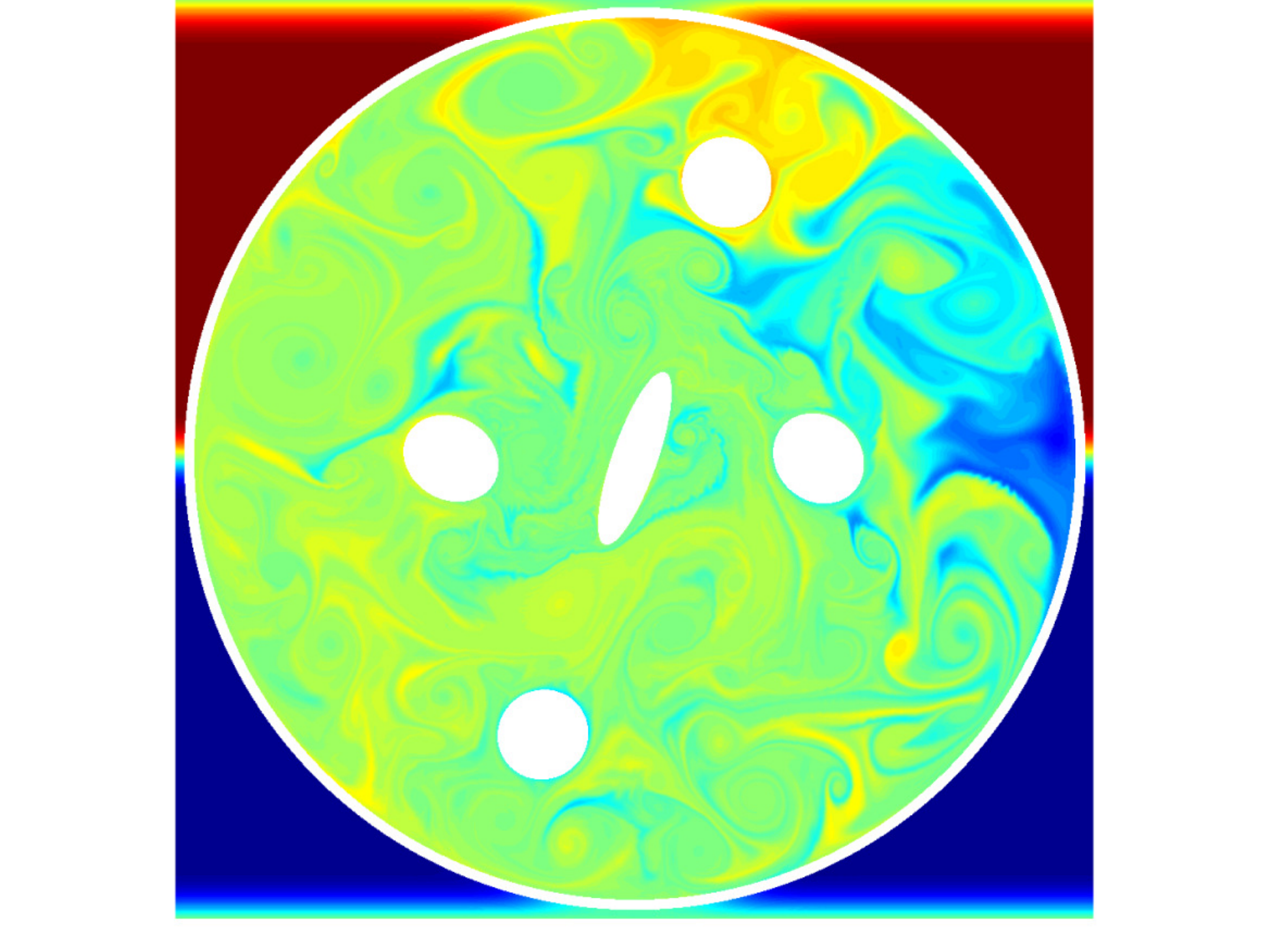} \\
    \includegraphics[width=0.5\textwidth]{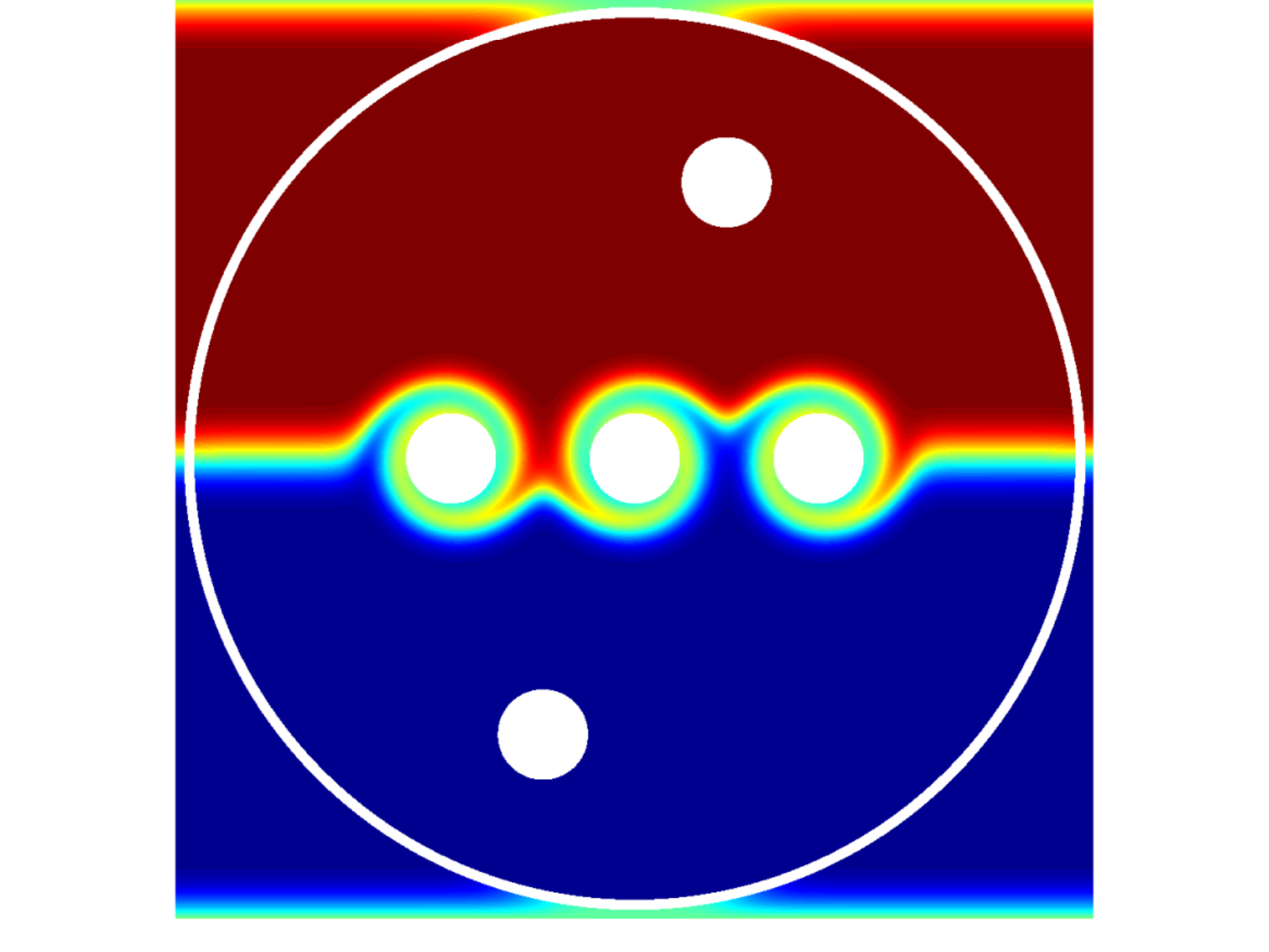} &
    \includegraphics[width=0.5\textwidth]{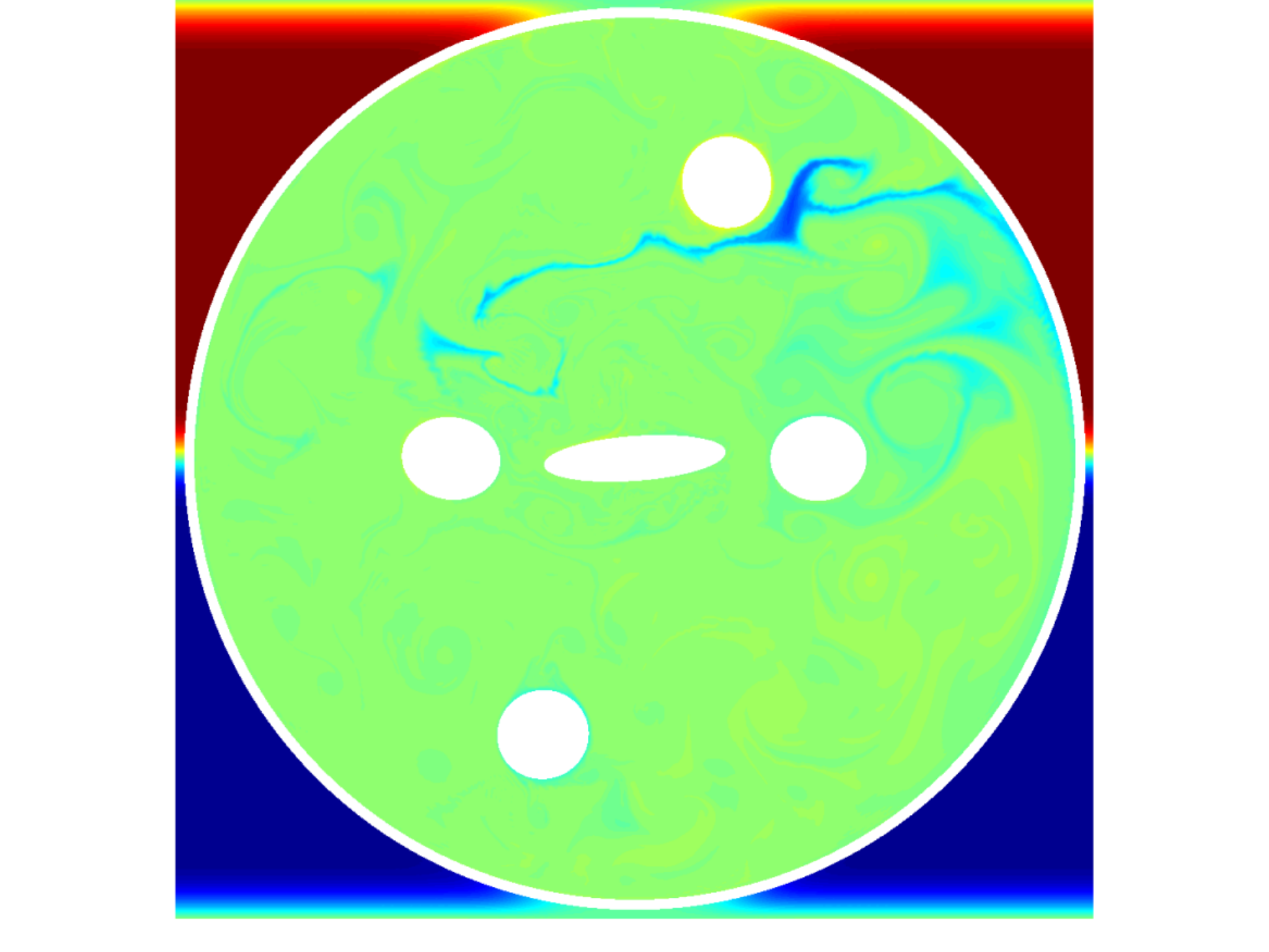}
  \end{tabular}
  \caption{\label{5CC} Case 3: mixing optimization using five
    stationary, rotating stirrers. Left column: unoptimized
    configuration, with snapshots at $t = 8, 16, 24, 32$ (top to
    bottom). Right column: after seven direct-adjoint optimizations,
    with snapshots at $t = 8, 16, 24, 32$ (top to bottom). For videos
    of these scenarios please refer to {\tt{5Before.mp4}} and
    {\tt{5After.mp4}} for the left and right column, respectively.}
\end{figure}

Temporal snapshots of the optimized mixing (after seven iterations)
can be observed in figure~\ref{5CC} in the right column. We remark
that this is the final optimization, since any further lengthening of
the central stirrer (as suggested by the next adjoint step) would
cause a collision with the other horizontally aligned stirrers.

\subsection{Case 4: one horizontally moving, rotating stirrer}

All previous configurations relied on stationary stirrers, and thus
only partially demonstrate the capabilities of the direct-adjoint
method and the associated penalization framework. In this final case,
we present a scenario that considers the optimization of the shape of
a stirrer while being dragged through the binary fluid; the velocity
of the stirrer is defined by a function of the form $\cos(t).$

We neglect energy penalization (i.e. $\lambda=0$) in this case, as
little effect is expected from a pure shape optimization, since the
bulk of the energy expenditure is already contained in the
back-and-forth motion of the stirrer.

\begin{figure}
  \centering
  \includegraphics[width=0.67\textwidth]{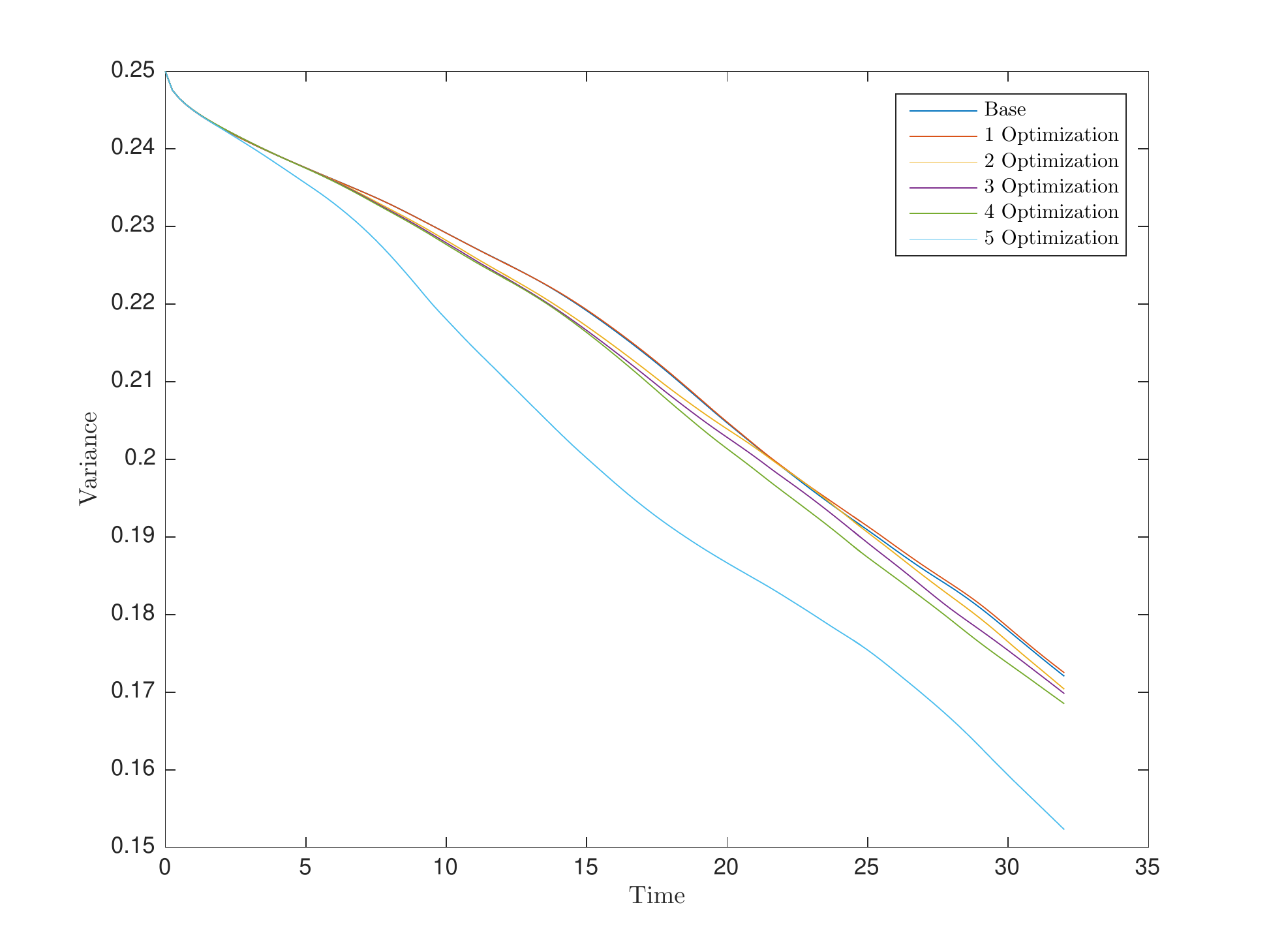}
  \caption{\label{CosVarPic} Case 4: mixing optimization using one
    horizontally moving, rotating stirrer. Variance, as defined in
    equation (\ref{eq:Variance}), of the scalar field $\theta$ versus
    time $t \in [0,\ T^{F}].$}
\end{figure}

We observe in figure~\ref{CosVarPic} a pronounced decrease in the
variance as the initially circular cylinder is lengthened vertically
(for the position at $t=0$). The reason for this optimal configuration
is certainly linked to the fact that dragging an elliptical stirrer
across the interface starting in this vertical position achieves a
great deal of mixing by producing small-scale structures. An
alternative, initially horizontal design would perform significantly
worse. In addition to this obvious observation, we notice that
starting in the vertical position at $t=0$ allows the stirrer to take
on the high-drag vertical position nine times during a full simulation
cycles, while an initially horizontal ellipse would exhibit the same
high-drag position only eight times per cycle. From the early stages
of the cycle (e.g. $t=8$), we create a great many filamented
structures that subsequently get diffused by the flow and help in
ultimately producing a homogeneous mixture. This effect continues
further throughout the simulation and leads to the substantial
decrease in variance, as shown in figure~\ref{CosVarPic}.

Finally, we note that the direct-adjoint system, as introduced, is
ignorant of the physical restrictions by the geometry; specifically,
geometrically overlapping or otherwise colliding structures are not
explicitly accounts for. For this reason, we have to manually
terminate the optimization scheme in the case of such an event.

\begin{figure}
  \centering
  \begin{tabular}{cc}
    \includegraphics[width=0.5\textwidth]{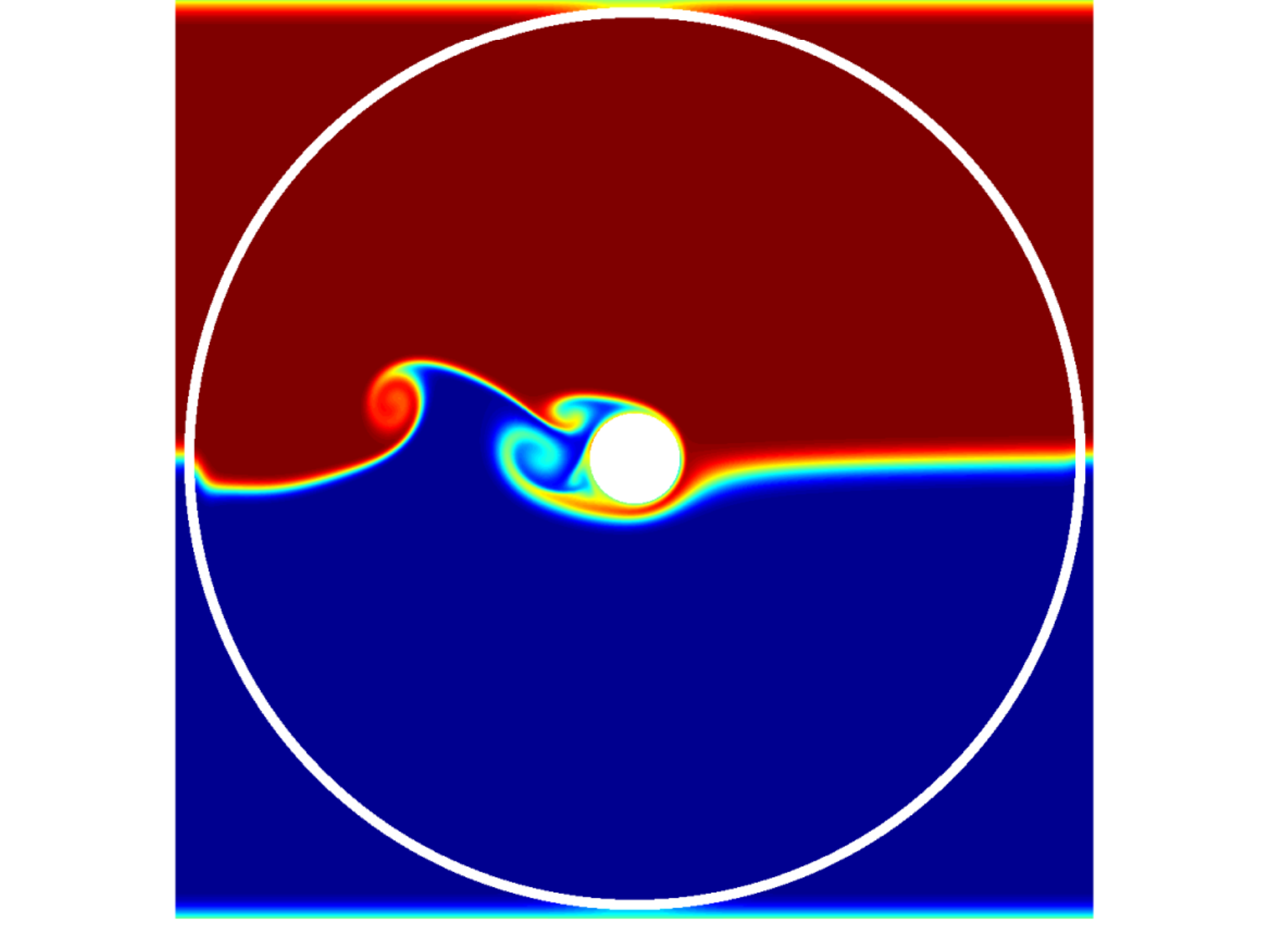} &
    \includegraphics[width=0.5\textwidth]{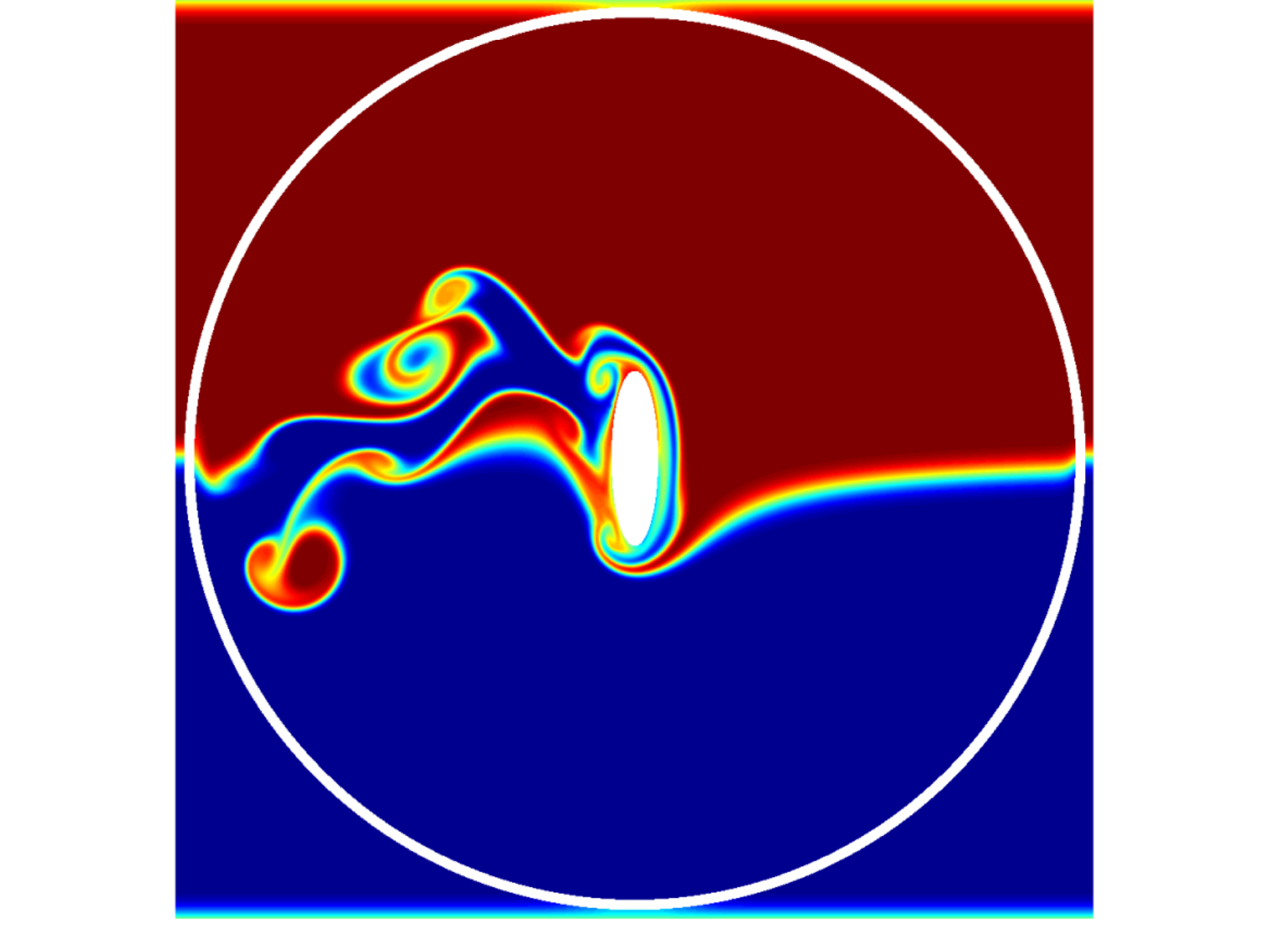} \\
    \includegraphics[width=0.5\textwidth]{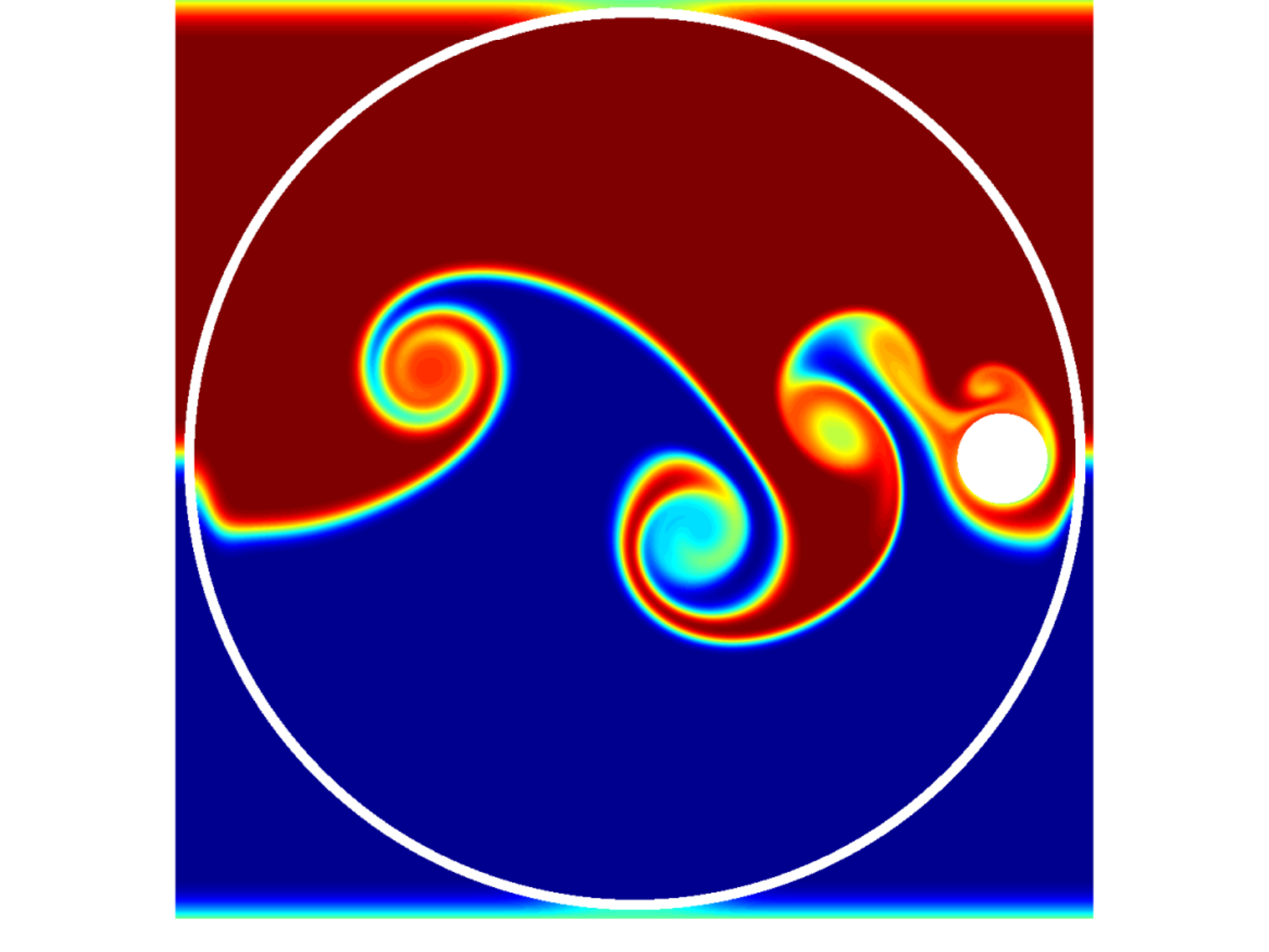} &
    \includegraphics[width=0.5\textwidth]{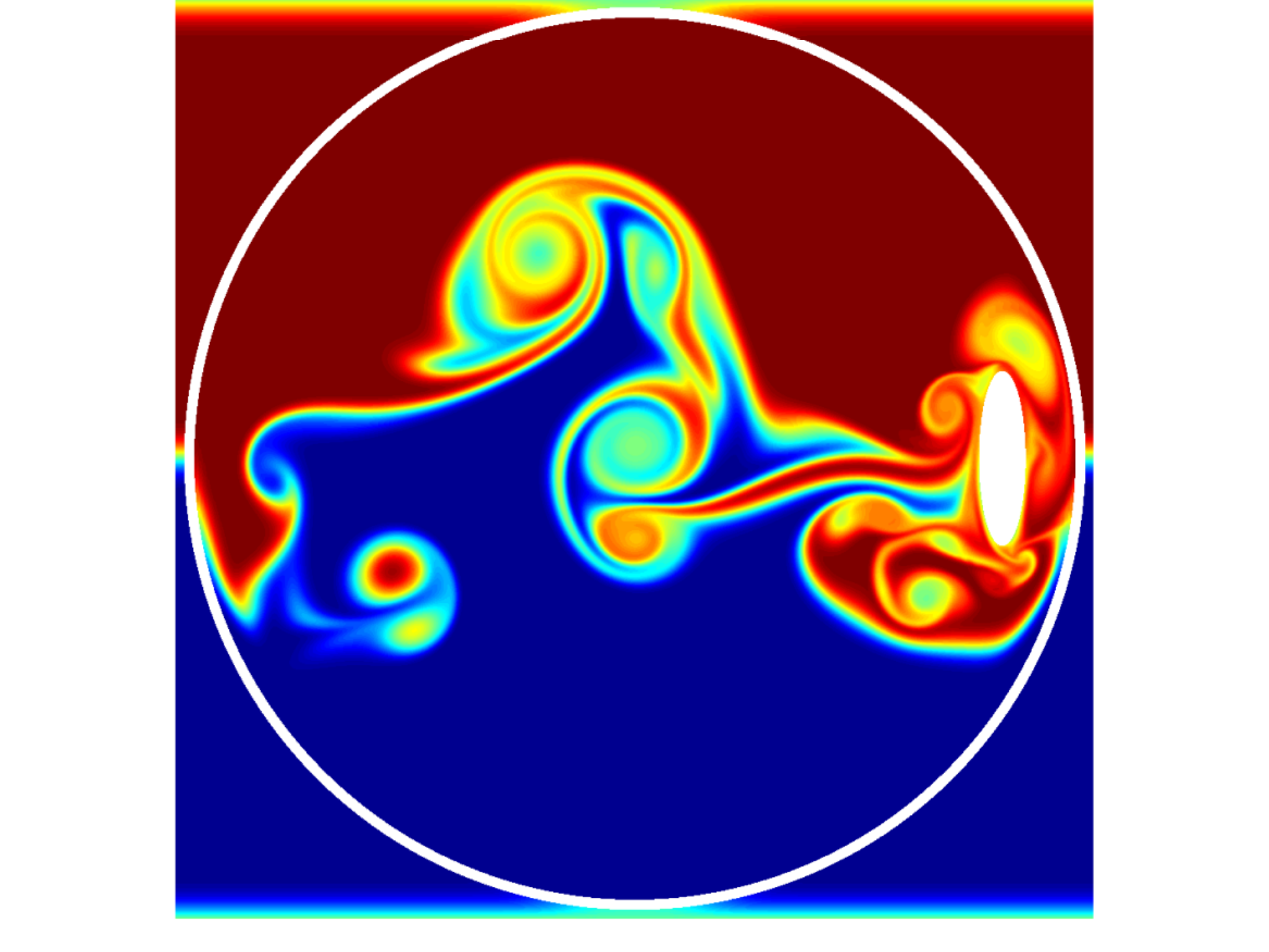} \\
    \includegraphics[width=0.5\textwidth]{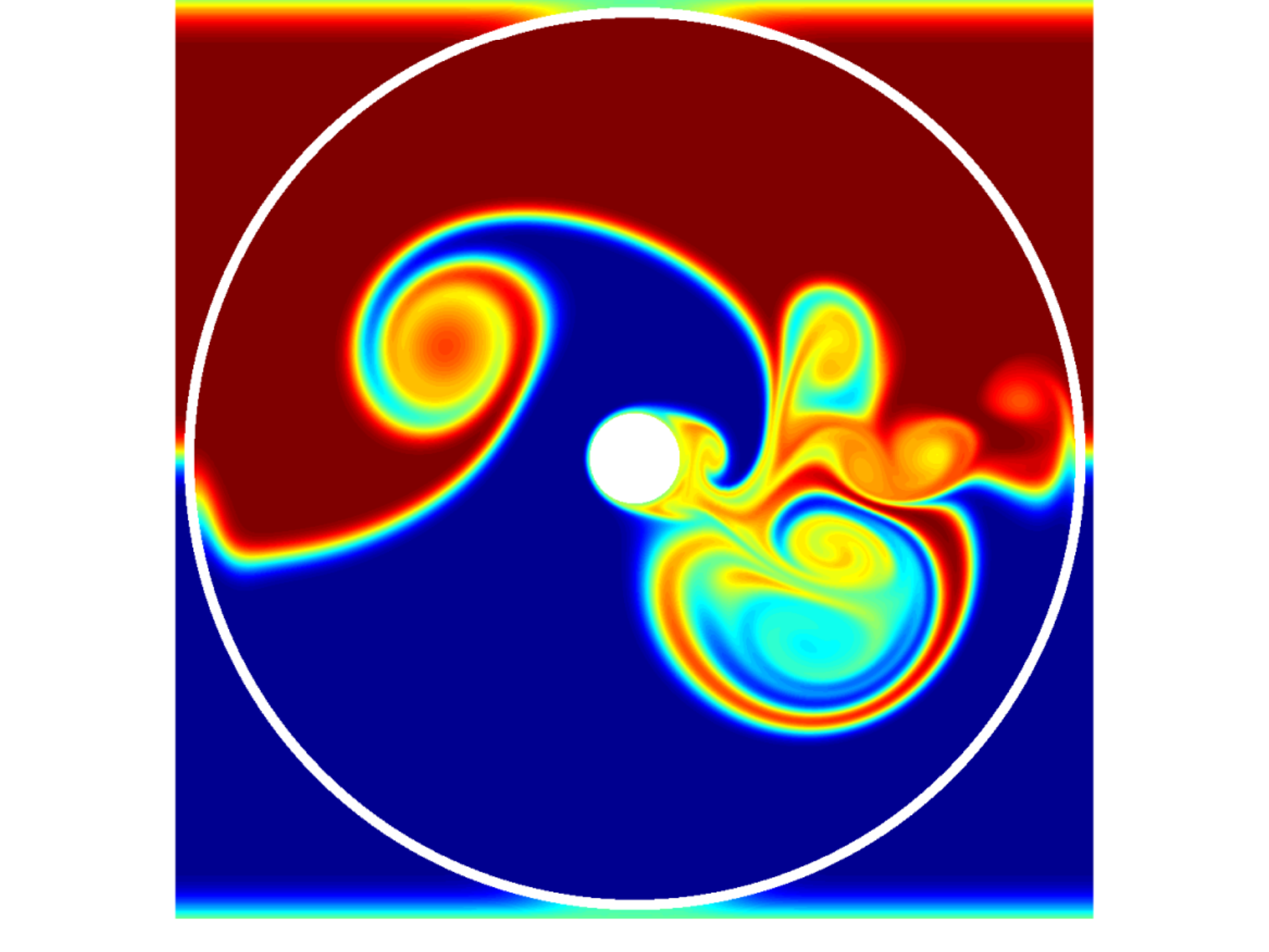} &
    \includegraphics[width=0.5\textwidth]{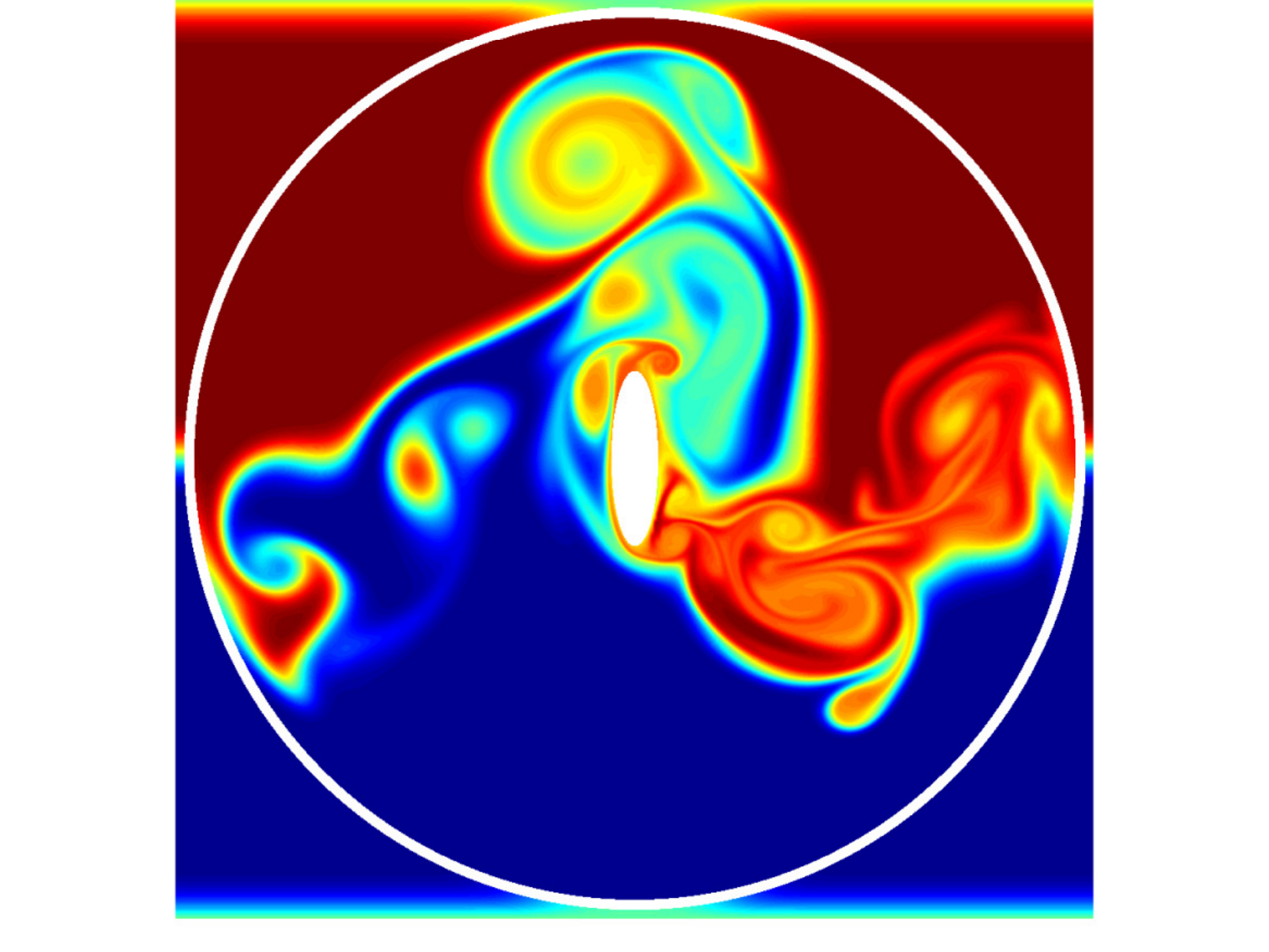} \\
    \includegraphics[width=0.5\textwidth]{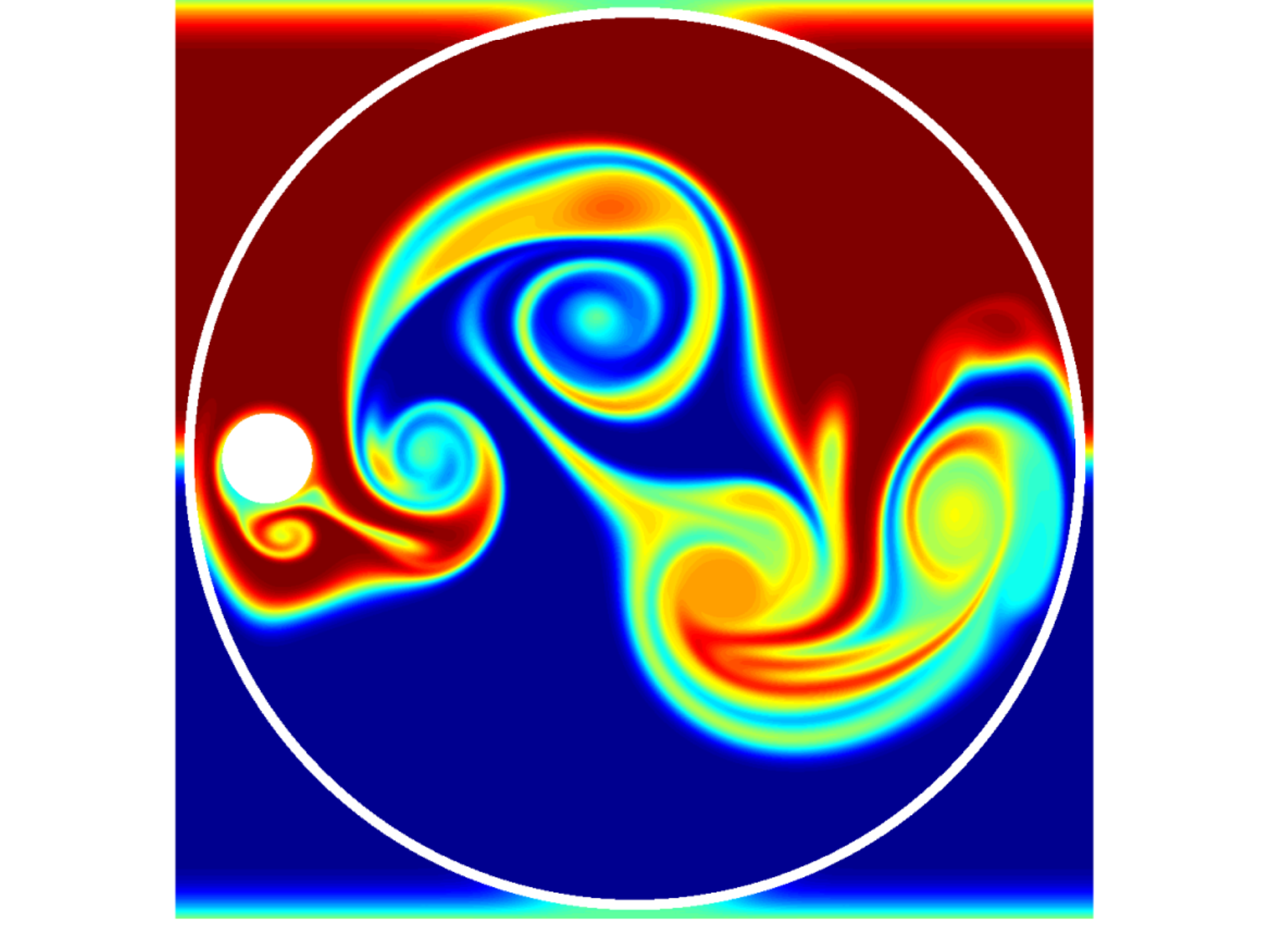} &
    \includegraphics[width=0.5\textwidth]{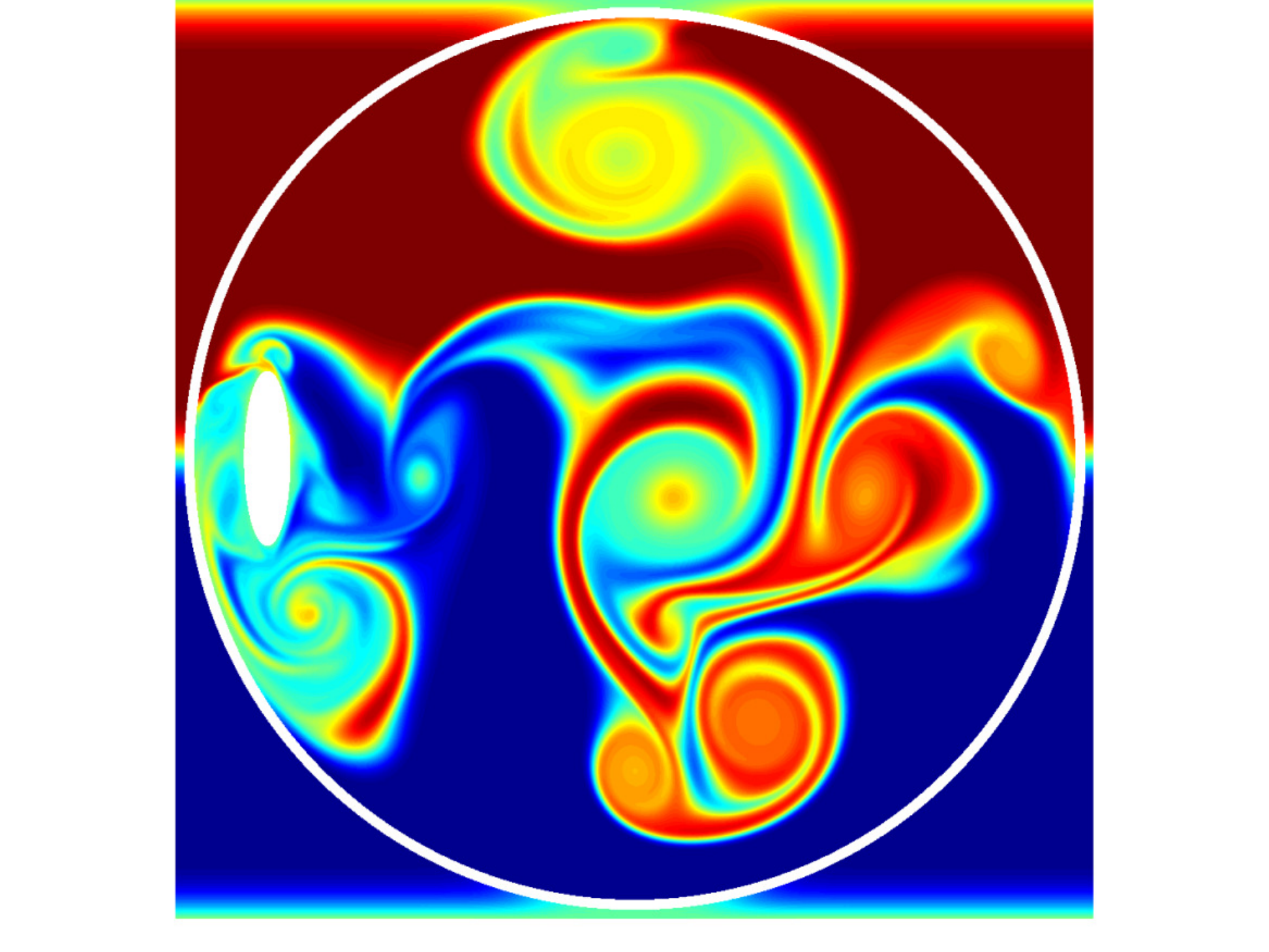}
  \end{tabular}
  \caption{\label{CosPic} Case 4: mixing optimization using one
    horizontally moving, rotating stirrer. Left column: unoptimized
    configuration, with snapshots at $t = 8, 16, 24, 32$
    (top to bottom). Right column: after five direct-adjoint
    optimizations, with snapshots at $t = 8, 16, 24, 32$
    (top to bottom). For videos of these scenarios please refer to
    {\tt{CosBefore.mp4}} and {\tt{CosAfter.mp4}} for the left and
    right column, respectively.}
\end{figure}

\section{\label{sec:concl}Conclusions}

A computational framework has been presented that allows the efficient
optimization of mixing strategies of binary fluids by moving
stirrers. Among the control parameters for the optimization, the speed
along a given path, the rotation speed of the stirrers and the
geometry of the stirrers have been considered, but more complex setups
are within the range of possibilities using the presented
approach. The framework uses Brinkman penalization to embed and
describe the stirrer elements and to track their interaction with the
surrounding fluid. A resulting PDE-constrained optimization problem is
then recast into a direct-adjoint (or primal-dual) formulation, which
is subsequently solved iteratively, employing a checkpointing
technique due the linkage of the direct and adjoint
problem. Particular attention has been paid to the explicit
formulation of path and shape gradients based on the penalized
formulation. These gradients are then used to optimize the mixing
efficiency (in our case, the variance of the passive scalar field),
while observing the user-specified control-energy bounds.

Four test cases have illustrated the feasibility and flexibility of
the presented approach: (i) the optimization of one simple stirrer,
(ii) a double configuration, (iii) a layout with five stirrers, and
(iv) a stirrer moving along user-specified path. In all cases, a
significant improvement in mixing efficiency could be accomplished,
and the optimization algorithm showed notable robustness in finding a
more optimal solution.

Despite these promising results, a few challenges remain. As is the
case for any gradient-based optimization applied to a non-convex
problem, only a local minimum can be guaranteed by our algorithm. User
input (as was the case in setup (ii) in the text) is required to nudge
the convergence towards a global optimum. This nudging could be
accomplished more objectively by coupling the above algorithm to an
annealing-type stochastic algorithm; approaches in this direction will
be pursued in a future effort. Furthermore, the algorithm will be
tested in a multi-parameter environment where a far larger control
space (larger than the one considered here) will be treated; in
particular, the convergence behavior will be monitored in this case,
and methods, such as control-space splitting, will be investigated as
means to accelerate convergence. Finally, three-dimensional layouts
will be considered. The code is highly parallelized and scales well to
many cores; good performance for large-scale optimizations of binary
mixing problems is thus expected.

\bigskip

\section*{Acknowledgements}

We wish to thank the Research Computing Service at Imperial College
London for providing the resources and support for this research.

\appendix

\section{Arithmetic with Hadamard products}

For the sake of clarity and completeness, we will give a brief summary
of rules and relations involving the Hadamard product which has been
used in the formulation of the governing equations and the derivation
of the adjoint equations, and has produced a compact formalism and
notation. In particular, we will consider the steps involved in the
transfer of operators acting on the direct flow variables (such as
velocities, pressure or passive scalar) onto corresponding operators
acting on their adjoint equivalents. While this transfer is rather
straightforward in the matrix-product case, care has to be exercised
when the operator involves Hadamard products.

The Hadamard product, denoted by $\circ,$ of two vectors ${\bf{a}}$
and ${\bf{b}}$ of identical size is defined as
\begin{equation}
  {\bf{c}} = {\bf{a}} \circ {\bf{b}} = {\bf{b}} \circ {\bf{a}} \qquad
  \hbox{ with } \quad c_i = a_i b_i.
\end{equation}
It is defined as the element-wise product of two vectors
and results in a vector ${\bf{c}}$ equal in size to ${\bf{a}}$ or
${\bf{b}}.$

It will be convenient to re-express the Hadamard product of two
vectors in terms of a standard matrix product. To this end, we
introduce the notation ${\bf{a}}^{(D)}$ to indicate a diagonal matrix
with the elements of ${\bf{a}}$ along its main diagonal. With this
notation we can restate the Hadamard product as
\begin{equation}
  {\bf{c}} = {\bf{a}} \circ {\bf{b}} = {\bf{a}}^{(D)} {\bf{b}} =
  {\bf{a}} {\bf{b}}^{(D)}.
\end{equation}
In the derivation of the adjoint equations, we are faced with terms of
the form
\begin{align}
  {\cal{I}} = {\bf{a}}^H \left( {\bf{b}} \circ \left[ {\mathsf{M}}
    {\bf{c}} \right] \right).
\end{align}
Expressions of this type constitute an inner product ${\cal{I}}$ and
arise from the nonlinear terms of the governing equations related to
convective transport, but terms linked to the mask function can also
yield the above example. They appear in the augmented Lagrangian
formulation of the optimization problem. Due to our spatial
discretization, ${\bf{a}}, {\bf{b}}$ and ${\bf{c}}$ are column vectors
of size $n \times 1$ with $n$ as the total number of degrees of
freedom, and ${\mathsf{M}}$ is an $n \times n$ matrix. During the
derivation of the adjoint equations, the vector ${\bf{c}}$ may
represent a first variation of a flow variable, while the vector
${\bf{a}}$ stands for the adjoint variable (see the main text for
details). We seek to isolate this first variation (the vector
${\bf{c}}$) by transferring all operators acting on it onto the
adjoint variable represented by ${\bf{a}}$ while preserving the inner
product. Using the alternative formulation of the Hadamard product
based on diagonal matrices, we obtain

\begin{subeqnarray}
  {\bf{a}}^H \left( {\bf{b}} \circ \left[ {\mathsf{M}} {\bf{c}}
    \right] \right) &=& {\bf{a}}^H \left( {\bf{b}}^{(D)} {\mathsf{M}}
  {\bf{c}} \right), \\
  &=& {\bf{a}}^H \left( \left( {\mathsf{M}}^H {\bf{b}}^{(D)} \right)^H
  {\bf{c}} \right), \\
  &=& \left( {\mathsf{M}}^H {\bf{b}}^{(D)} {\bf{a}} \right)^H {\bf{c}},
  \\
  &=& \left( {\mathsf{M}}^H \left[ {\bf{b}} \circ {\bf{a}} \right]
  \right)^H {\bf{c}}.
\end{subeqnarray}
Using this simple rule we are able to efficiently manipulate most
expressions in our adjoint derivations. We note that in the case of
${\mathsf{M}}$ being an identity matrix, our relation simply reduces
to
\begin{equation}
  {\bf{a}}^H \left( {\bf{b}} \circ {\bf{c}} \right) = \left( {\bf{b}}
  \circ {\bf{a}} \right)^H {\bf{c}} = \left( {\bf{a}}^H \circ
        {\bf{b}}^H \right) {\bf{c}}.
\end{equation}

\section{Explicit $\chi^{\dag}$-derivation}

As a representative example of the full adjoint derivation, we will
more explicitly perform the steps to arrive at the expression for the
adjoint mask function $\chi^{\dag},$ i.e.,
equation~(\ref{Optimal:Chi}). These steps are illustrative of the
remaining part of the adjoint formalism and involve the critical
concepts that also apply to the momentum equations. The formulation as
a spatially discretized problem, the incorporation of the boundary
conditions via penalization and the use of Hadamard products to
describe nonlinear terms aid in making the derivation less unwieldy
and error-prone as in attempts on the continuous equations.

The expression for $\chi^{\dag}$ results, as an optimality condition,
from a first variation of the augmented Lagrangian with respect to the
mask functions $\chi.$ We have

\begin{eqnarray}
  \int_0^{T^{F}}\left(\frac{\partial \mathcal{L}}{\partial
    \chi_i}\right)\delta \chi_i \ \text{d} t &=&
  \int_0^{T^{F}}\left(\frac{\partial \mathcal{J}}{\partial
    \chi_i}\right)\delta \chi_i \ \text{d} t \nonumber \\
  &-& \int_0^{T^{F}} \bm{u}^{\dag,H}_j \frac{\delta\chi_i}{C_{\eta}} \circ
  \bm{u}_j - \bm{u}^{\dag,H}_j \frac{\delta\chi_i}{C_{\eta}} \circ
  (\bm{u}_{s,i})_j \ \text{d} t \nonumber \\
  &-& \int_0^{T^{F}} p^{\dag,H} {\mathsf{A}}_j \left[
    \frac{\delta\chi_i}{C_{\eta}} \circ \bm{u}_j - \frac{\delta\chi_i}{C_{\eta}}
    \circ (\bm{u}_{s,i})_j \right] \ \text{d} t \nonumber \\
  &-& \int_0^{T^{F}} \theta^{\dag,H} \biggl[-\delta\chi_i \circ \bm{u}_j
    \circ[ {\mathsf{A}}_j \theta] + \delta\chi_i \circ
    (\bm{u}_{s,i})_j \circ [ {\mathsf{A}}_j \theta] \nonumber \\
  && -{\mathsf{A}}_j \left( \left[ Pe^{-1}(-\delta\chi_i) +
      \kappa\delta\chi_i \right] \circ {\mathsf{A}}_j \theta
    \right)\biggr]\text{d} t \nonumber \\
  &-& \int_0^{T^{F}} \chi^{\dag,H}_i \delta\chi_i \ \text{d} t.
\end{eqnarray}
The task is to isolate the first variation $\delta \chi_i$ from all
terms and transfer any operator acting on it to act on the remaining
terms. In this effort, we take advantage of the relation for the
Hadamard product, explained in the previous appendix. Continuing from
above, we obtain

\begin{eqnarray}
  &=& \lambda\int_0^{T^{F}}((\bm{u}_{s,i})_j\circ \delta \chi_i)^H
  \mathsf{R}_i((\bm{u}_{s,i})_j\circ \chi_i) + ((\bm{u}_{s,i})_j\circ
  \chi_i)^H \mathsf{R}_i((\bm{u}_{s,i})_j\circ \delta \chi_i)
  \ \text{d} t\nonumber \\
  &-& \int_0^{T^{F}} \frac{\bm{u}^{\dag,H}_j \circ
    \bm{u}_j^H-\bm{u}^{\dag,H}_j \circ \left( \bm{u}_{s,i}
    \right)^H_j}{C_{\eta}} \delta \chi_i \ \text{d} t \nonumber \\
  &-& \int_0^{T^{F}} \left[ \frac{ \left[ {\mathsf{A}}_j^H p^{\dag}
        \right]^H \circ \bm{u}_j^H - \left[ {\mathsf{A}}_j^H p^{\dag}
        \right]^H \circ \left( \bm{u}_{s,i}
      \right)^H_j}{C_{\eta}}\right] \delta\chi_i \ \text{d} t
  \nonumber \\
  &-& \int_0^{T^{F}} \biggl[ -\theta^{\dag,H} \circ \left( \bm{u}_j \circ
    \left[ {\mathsf{A}}_j \theta \right] \right)^H \delta\chi_i +
    \theta^{\dag,H} \circ \left( \left(\bm{u}_{s,i} \right)_j \circ
    \left[ {\mathsf{A}}_j \theta \right] \right)^H \delta\chi_i
    \nonumber \\
  && -\left[ {\mathsf{A}}_j^H \theta^{\dag} \right]^H
    \left[Pe^{-1}\left( -\delta\chi_i \right) + \kappa\delta\chi_i
      \right] \circ {\mathsf{A}}_j \theta \biggr]\text{d} t \nonumber \\
  &-& \int_0^{T^{F}} \chi_i^{\dag,H} \delta\chi_i \ \text{d} t.
\end{eqnarray}
Next, we collect matching terms and gather them under a single
integral. We thus get

\begin{eqnarray}
  &=&
  \int_0^{T^{F}}\Biggl(\left[2\lambda\mathsf{R}_i((\bm{u}_{s,i})_j\circ
    \chi_i)\right]\circ (\bm{u}_{s,i})_j \nonumber \\
  && -\frac{\bm{u}^{\dag}_j \circ
    \left(\bm{u}_j-\left(\bm{u}_{s,i}\right)_j\right) + \left[
      {\mathsf{A}}_j^H p^{\dag} \right] \circ \left( \bm{u}_j -
    \left(\bm{u}_{s,i} \right)_j\right)}{C_{\eta}} \nonumber \\
  && +\theta^{\dag} \circ \left( \bm{u}_j \circ \left[ {\mathsf{A}}_j
    \theta \right] \right) - \theta^{\dag} \circ \left(
  \left(\bm{u}_{s,i}\right)_j \circ \left[ {\mathsf{A}}_j \theta
    \right] \right) \nonumber \\
  && +\left( \kappa-Pe^{-1}\right) \left[ {\mathsf{A}}_j^H
    \theta^{\dag} \circ {\mathsf{A}}_j \theta \right] - \chi^{\dag}
  \Biggr)^H \delta\chi_i \ \text{d} t
\end{eqnarray}
We recall that this integral is identically zero, which implies that
the integrand must vanish. We are then able to explicitly express
$\chi_i^\dag$ in terms of our other variables; we obtain

\begin{eqnarray}
  \chi^{\dag}_i &=& \left[2\lambda\mathsf{R}_i((\bm{u}_{s,i})_j\circ
    \chi_i)\right]\circ (\bm{u}_{s,i})_j \nonumber \\
  &-& \frac{\bm{u}^{\dag}_j\circ( \bm{u}_j-(\bm{u}_{s,i})_j)+
    [{\mathsf{A}}_j^H p^{\dag}] \circ
    (\bm{u}_j-(\bm{u}_{s,i})_j)}{C_{\eta}} \nonumber \\
  &+& (\theta^{\dag} \circ [ {\mathsf{A}}_j \theta]) \circ (\bm{u}_j -
  (\bm{u}_{s,i})_j).
\end{eqnarray}
Some simple manipulations leave us with the result from the main text,

\begin{eqnarray}
  \chi^{\dag}_i &=& \left[2\lambda\mathsf{R}_i((\bm{u}_{s,i})_j\circ
    \chi_i)\right]\circ (\bm{u}_{s,i})_j+ \left[ \theta^{\dag} \circ [
      {\mathsf{A}}_j \theta] - \frac{\Pi^{\dag}_j}{C_{\eta}} \right]
  \circ (\bm{u}_j -(\bm{u}_{s,i})_j) \nonumber \\
  &+& (\kappa-Pe^{-1}) {\mathsf{A}}_j^H \theta^{\dag} \circ
  {\mathsf{A}}_j \theta.
\end{eqnarray}
The system of adjoint evolution equations, as well as other optimality
conditions, are derived in an analogous manner.

\section*{References}

\bibliography{MixingPaperFinal.bib}

\end{document}